\documentclass[a4paper,fleqn,usenatbib]{mnras}

\usepackage{newtxtext,newtxmath}

\usepackage[T1]{fontenc}
\usepackage{ae,aecompl}

\usepackage{graphicx}	
\usepackage{amsmath}	

\def\mrpd{\hbox{mrad\,d$^{-1}$}}
\def\rpd{\hbox{rad\,d$^{-1}$}}

\def\chisq{\hbox{$\chi^2$}}

\def\chisqr{\hbox{$\chi^2_{\rm r}$}}
\def\msun{\hbox{${\rm M}_{\odot}$}}

\def\me{\hbox{${\rm M}_{\oplus}$}}
\def\re{\hbox{${\rm R}_{\oplus}$}}

\def\rsun{\hbox{${\rm R}_{\odot}$}}
\def\lsun{\hbox{${\rm L}_{\odot}$}}

\def\mstar{\hbox{$M_{\star}$}}
\def\rstar{\hbox{$R_{\star}$}}
\def\lstar{\hbox{$L_{\star}$}}
\def\teff{\hbox{$T_{\rm eff}$}}
\def\logg{\hbox{$\log g$}}

\def\vD{\hbox{$v_{\rm D}$}}

\def\ms{\hbox{m\,s$^{-1}$}}
\def\kmsss{\hbox{km$^2$\,s$^{-2}$}}

\def\kms{\hbox{km\,s$^{-1}$}}
\def\gpcc{\hbox{g\,cm$^{-3}$}}
\def\vsini{\hbox{$v \sin i$}}

\def\mic{\hbox{$\mu$m}}

\def\emr{}
\def\ems{}
\def\Bl{\hbox{$B_{\rm \ell}$}}
\def\Bd{\hbox{$B_{\rm d}$}}

\def\degr{\hbox{$^\circ$}}

\def\omeq{\hbox{$\Omega_{\rm eq}$}}
\def\dom{\hbox{$d\Omega$}}

\def\Prot{\hbox{$P_{\rm rot}$}}

\newcommand{\hei}{\hbox{He$\;${\sc i}}}

\newcommand{\pab}{\hbox{Pa${\beta}$}}

\title[Magnetic field and multiplanet system of AU~Mic]{The magnetic field and multiple planets of the young dwarf AU~Mic} 
\author[J.-F.~Donati et al.]{J.-F.~Donati$^{1}$\thanks{E-mail: jean-francois.donati@irap.omp.eu},
           P.I.~Cristofari$^{1}$, B.~Finociety$^{1}$, B.~Klein$^{2,1}$, C.~Moutou$^1$, E.~Gaidos$^{3}$, 
\newauthor C.~Cadieux$^4$, E.~Artigau$^{4}$, A.C.M.~Correia$^{5,6}$, G.~Bou\'e$^{6}$, N.J.~Cook$^{4}$, A.~Carmona$^{7}$, 
\newauthor L.T.~Lehmann$^{1}$, J.~Bouvier$^{7}$, E.~Martioli$^{8,10}$,  J.~Morin$^{9}$, P Fouqu\'e$^{1}$, X.~Delfosse$^{7}$, 
\newauthor R.~Doyon$^{4}$, G.~H\'ebrard$^{10}$, S.H.P.~Alencar$^{11}$, J.~Laskar$^{6}$, L.~Arnold$^{12}$, P.~Petit$^{1}$, 
\newauthor \'A.~K\'osp\'al$^{13}$, A.~Vidotto$^{14}$, C.P.~Folsom$^{15}$ and the SLS collaboration 
\vspace{1mm}
\\ 
$^1$ Univ.\ de Toulouse, CNRS, IRAP, 14 avenue Belin, 31400 Toulouse, France \\ 
$^2$ Sub-department of Astrophysics, Department of Physics, University of Oxford, Oxford, OX1 3RH, UK \\ 
$^3$ Department of Earth Sciences, University of Hawai’i at Manoa, 1680 East-West Road, Honolulu, HI, 96822, USA \\ 
$^4$ Universit\'e de Montr\'eal, D\'epartement de Physique, IREX, Montr\'eal, QC H3C 3J7, Canada \\ 
$^5$ CFisUC, Departamento de F\'isica, Universidade de Coimbra, 3004-516 Coimbra, Portugal \\ 
$^6$ IMCCE, CNRS, Obs.\ de Paris, Sorbonne Univ., 77 av. Denfert-Rochereau, 75014 Paris, France \\ 
$^7$ Univ.\ Grenoble Alpes, CNRS, IPAG, 38000 Grenoble, France \\ 
$^8$ Laborat\'{o}rio Nacional de Astrof\'{\i}sica, Rua Estados Unidos 154, 37504-364, Itajub\'{a}, MG, Brazil \\ 
$^9$ LUPM, Univ.\ de Montpellier, CNRS, F-34095 Montpellier, France \\ 
$^{10}$ Institut d’Astrophysique de Paris, CNRS, Sorbonne Univ., 98 bis bd Arago, 75014 Paris, France \\ 
$^{11}$ Departamento de F\'{\i}sica -- ICEx -- UFMG, Av. Ant\^onio Carlos, 6627, 30270-901 Belo Horizonte, MG, Brazil\\  
$^{12}$ Canada-France-Hawaii Telescope, 65-1238 Mamalahoa Hwy., Kamuela, HI 96743, USA \\ 
$^{13}$ Konkoly Observatory, Research Centre for Astronomy and Earth Sciences, Konkoly-Thege Mikl\'os \'ut 15-17, 1121 Budapest, Hungary \\ 
$^{14}$ Leiden Observatory, Leiden University, Niels Bohrweg 2, 2333 CA Leiden, the Netherlands \\ 
$^{15}$ Tartu Observatory, University of Tartu, Observatooriumi 1, 61602, Tõravere, Estonia 
}

\date{Submitted 2023 Feb 10 -- Accepted 2023 Apr 17 } 

\pubyear{2023}

\begin{document}

\label{firstpage}
\pagerange{\pageref{firstpage}--\pageref{lastpage}}
\maketitle

\begin{abstract}
In this paper we present an analysis of near-infrared spectropolarimetric and velocimetric data of the young M dwarf AU~Mic, collected with SPIRou at the Canada-France-Hawaii telescope from 2019 to 2022, mostly within 
the SPIRou Legacy Survey.  With these data, we study the large- and small-scale magnetic field of AU~Mic, detected through the unpolarized and circularly-polarized Zeeman signatures of spectral lines.  
We find that both are modulated with the stellar rotation period (4.86~d), and evolve on a timescale of months under differential rotation and intrinsic variability.  
The small-scale field, estimated from the broadening of spectral lines, reaches $2.61\pm0.05$~kG.  The large-scale field, inferred with Zeeman-Doppler imaging from Least-Squares Deconvolved profiles of circularly-polarized 
and unpolarized spectral lines, is mostly poloidal and axisymmetric, with an average intensity of $550\pm30$~G.  We also find that surface differential rotation, as derived from the large-scale field, is 
$\simeq$30\% weaker than that of the Sun.  
We detect the radial velocity (RV) signatures of transiting planets b and c, although dwarfed by activity, and put an upper limit on that of candidate planet d, putatively causing 
the transit-timing variations of b and c.   We also report the detection of the RV signature of a new candidate planet (e) orbiting further out with a period of $33.39\pm0.10$~d, i.e., near the 4:1 
resonance with b.  The RV signature of e is detected at 6.5$\sigma$ while those of b and c show up at $\simeq$4$\sigma$, yielding masses of $10.2^{+3.9}_{-2.7}$ and $14.2^{+4.8}_{-3.5}$~\me\ for b and c, and a minimum mass 
of $35.2^{+6.7}_{-5.4}$~\me\ for e.  
\end{abstract}




\section{Introduction}
\label{sec:int}

The formation of stars and their planets has become a very popular forefront topic of modern astrophysics, following the discovery of thousands of exoplanetary systems and the availability 
of many new powerful instruments capable of characterizing them, such as the JWST most recently.  The goal of such studies is to investigate the surprising diversity of the exoplanetary systems 
detected around low-mass stars, and in particular to better understand the formation and evolution of planetary systems like ours.  Studying newly born planetary systems and their 
pre-main-sequence (PMS) host stars is essential in this respect, the first evolutionary steps being those for which we currently have no more than weak observational constraints to guide theoretical models.  

So far, very few multiple planetary systems younger than 50~Myr have been reported around low-mass stars, two of which detected with transit photometry, namely V1298~Tau \citep[hosting 
4 transiting planets,][]{David19} and AU~Mic \citep[with 2 known transiting warm Neptunes,][]{Plavchan20, Martioli21, Szabo21, Szabo22}, then further monitored with precision Doppler velocimetry \citep{Klein21, Suarez22, Zicher22}.  
Given that young low-mass stars are usually quite active and strongly magnetic as a result of their short rotation periods (and convective envelopes), investigating their planets mandatorily requires characterization of the 
magnetic activity of the star so that the impact of this activity can be taken into account, and filtered out from the radial velocity (RV) curves in which planetary signatures hide.  
Constraining the large-scale magnetic fields of PMS stars is also essential for further documenting the parent dynamo processes that are able to amplify and sustain these fields, for 
investigating star-disc interactions and angular momentum evolution for stars whose accretion disc is still present \citep[e.g.,][]{Zanni13, Blinova16}, and for 
studying potential star-planet interactions that may occur if the planets orbit within the Alfven radius of their host stars \citep[e.g.,][]{Strugarek15}.  In this paper, we focus on the 
second and the brightest of these 2 stars, i.e., AU~Mic.  

AU~Mic is an active M1 dwarf that belongs to the $\beta$~Pic moving group \citep[aged $\simeq$20~Myr,][]{Mamajek14,Miret20}.  Hosting an extended debris 
disc with moving features \citep{Kalas04, Boccaletti15, Boccaletti18} and 2 known transiting warm Neptunes \citep{Plavchan20, Martioli21}, it is an ideal target for studying the 
formation and evolution of young planets and their atmospheres \citep{Hirano20}.  
{\emr Several studies focused on estimating the masses of both planets, the 2 
latest ones yielding $M_b=14.3\pm7.7$~\me\ (where \me\ notes the Earth mass) and $M_c=34.9\pm10.8$ \citep{Klein22}, and $M_b=11.7\pm5.0$~\me\ and $M_c=22.2\pm6.7$ \citep{Zicher22}.  
Given the large transit-timing variations (TTVs) of up to $\pm$10~min reported for planets b and c
\citep{Szabo22}, it is likely that the planetary system of AU~Mic includes more, yet undetected, bodies.  A new candidate Earth-mass planet (dubbed d), putatively 
located between b and c, was recently proposed to account for the reported TTVs \citep{Wittrock23}.  } 
Besides, AU~Mic is known for its intense activity and strong 
magnetic field \citep{Kochukhov20, Klein21}, making it a prime target for studying dynamos of largely convective stars, magnetized winds and star-planet 
interactions \citep{Kavanagh21, Klein22, Alvarado22}, 
or escaping planetary atmospheres \citep{Carolan20}.  

In this paper, we report extended near-infrared (nIR) high-resolution spectropolarimetric observations of AU~Mic with SPIRou at the 3.6-m Canada-France-Hawaii Telescope (CFHT) atop Maunakea in 
Hawaii, from early 2019 to mid 2022.  After outlining our observations and data reduction in Sec.~\ref{sec:obs}, we briefly revisit the main parameters of AU~Mic in Sec.~\ref{sec:par}, 
compute the longitudinal and small-scale magnetic fields and their modulation with time in Sec.~\ref{sec:bls}, carry out Zeeman-Doppler Imaging (ZDI) of our 
spectropolarimetric data at the main observing epochs in Sec.~\ref{sec:zdi}, study and model RV variations in Sec.~\ref{sec:lbl}, and investigate several activity proxies in Sec.~\ref{sec:act}.  
We finally summarize and discuss our results, and conclude by suggesting follow-up studies in Sec.~\ref{sec:dis}.

\section{SPIRou observations}
\label{sec:obs}

AU~Mic was intensively observed from early 2019 to mid 2022 with the SPIRou nIR spectropolarimeter / high-precision velocimeter \citep{Donati20} at CFHT, 
mostly within the SPIRou Legacy Survey (SLS), a Large Programme of 310~nights with SPIRou focussing on planetary systems around nearby M dwarfs on the one 
hand, and on the study of magnetized star / planet formation on the other.  SPIRou collects spectra covering the entire 0.95--2.50~\mic\ wavelength range in a single exposure,
at a resolving power of 70,000, and for any given polarization state.  
A total of 235 circular polarization sequences on 194 different nights were collected on AU~Mic with SPIRou over this 3-yr 
period, 181 in the framework of the SLS itself, 38 within the Director's Discretionary Time PI program of Baptiste Klein \citep[run ID 19AD97 and 19BD97, with results published in][]{Klein21}, 15 
within the PI program of Eric Gaidos (run ID 20AH93) and 1 within the PI program of Julien Morin (run ID 19AF26).  As outlined in \citet{Donati20}, each SPIRou polarization sequence consists 
of 4 sub-exposures (except for one featuring 2 sub-exposures only due to bad weather).  Each sub-exposure is associated with a different orientation of the Fresnel rhomb retarders \cite[to remove 
systematics in polarization spectra to first order, see][]{Donati97b}, yielding one unpolarized (Stokes $I$) and one circularly polarized (Stokes $V$) spectrum.  A series of 29 such spectra were 
also collected during the transit of AU~Mic b on 2019 June 16, thanks to which \citet{Martioli20} demonstrated, via the Rossiter-McLaughlin effect, that the planet orbit is prograde and lies in the equatorial plane 
of the host star (within 15\degr) and in the plane of the debris disk \citep{Boccaletti18}.  

With 10 of the 235 spectra collected in bad weather conditions and featuring much lower signal-to-noise ratio (SNR), we are left with a total of 225 Stokes $I$ and $V$ spectra of AU~Mic collected on 
188 different nights.  Exposure times for most sequences are $\simeq$800~s, except for a few of them (e.g., those collected during the transit of AU~Mic b) that were shorter (from 380 to 490~s).  
Peak SNRs range from 407 to 954 (median 785).  The full log of our SPIRou observations is provided in Appendix~\ref{sec:appA} (see Table~\ref{tab:log} provided as supplementary material) .  


All data were processed with a new version of \texttt{Libre ESpRIT}, the nominal reduction pipeline of ESPaDOnS at CFHT, adapted for SPIRou \citep{Donati20}.  These reduced spectra 
were used in particular for the spectropolarimetric analyses outlined in Secs.~\ref{sec:bls} and \ref{sec:zdi}.  Least-Squares Deconvolution \citep[LSD,][]{Donati97b} was then applied to all reduced 
spectra, using a line mask constructed from the VALD-3 database \citep{Ryabchikova15} for an effective temperature \teff=3750~K and a logarithmic surface gravity \logg=4.5 adapted to young early M dwarfs like AU~Mic 
(see Sec.~\ref{sec:par}), and selecting lines of relative depths larger than 3 percent only (for a total of $\simeq$1500 lines, featuring an average Land\'e factor of 1.2).  
The noise levels $\sigma_V$ in the resulting Stokes $V$ LSD profiles range from 0.73 to 1.73 (median 0.96)  in units of $10^{-4} I_c$ where $I_c$ notes the continuum intensity.  
We also applied LSD to our spectra using 2 sub-masks of our main line mask, 
the high-Land\'e mask including lines with Land\'e factors larger than 1.5 (average Land\'e factor 1.7), and the low-Land\'e mask including lines with Land\'e factors lower than 1.0 (average Land\'e factor 0.7), 
both sub-masks featuring more or less the same number of lines ($\simeq$300) and the same average wavelength as the main mask ($\simeq$1700~nm).  

In parallel, our data were also processed with \texttt{APERO} (version 0.7.275), the nominal SPIRou reduction pipeline \citep{Cook22}, currently better optimised in terms of RV precision than \texttt{Libre ESpRIT}.  The reduced 
spectra were then analysed by the line-by-line (LBL) technique \citep[version 0.45,][]{Artigau22} to compute precise RVs for 185 of the 188 nightly-averaged observations collected on AU~Mic\footnote{Three spectra suffered from 
an instrumental issue that affected the SPIRou RV reference module, yielding no precise RV estimates at these dates.}, with a median RV error bar of 3.8~\ms\ (see Table~\ref{tab:log}).  
Our RV data were also corrected for an overall trend, called Zero Point and coming from both the instrument and the reduction.  It is inferred from a Gaussian Process Regression (GPR) applied to the RV curves of a dozen 
RV standard stars regularly monitored with SPIRou, and whose amplitude is small (RMS of a few \ms) with respect to the measured RV variations of AU~Mic.  
Finally, in addition to RVs, the LBL analysis produces other diagnostics, in particular one 
measuring the variations in the average width of line profiles with respect to the median spectrum, which serves as an activity proxy \citep[called differential line width or dLW in][]{Artigau22} linked 
to the Zeeman broadening of unpolarized spectra (see Sec.~\ref{sec:bls}).  These data were used for the RV analysis detailed in Sec.~\ref{sec:lbl}, and in the following section on activity proxies (Sec.~\ref{sec:act}).   


\section{Fundamental parameters of AU~Mic}
\label{sec:par}

AU~Mic (Gl~803, HD~197481, HIP~102409, $V=8.627$, $J=5.436$) is a bright M1 PMS dwarf located at a distance of $9.714\pm0.002$~pc from us \citep{Gaia21}, with a rotation 
period of 4.86~d \citep{Plavchan20, Klein21, Zicher22}, typical of its young age \citep[$\simeq$20~Myr,][]{Mamajek14, Miret20}.  As usual for young active M dwarfs, it exhibits 
photometric fluctuations caused by surface brightness inhomogeneities and flares, with an amplitude of a few percent \citep[e.g.,][]{Plavchan20, Martioli21}.  Its mean $V-I$ color \citep[equal to $2.034\pm0.060$~mag,][]{Kiraga12} 
yields $\teff=3700\pm70$~K \citep{Pecaut13}, consistent with previous estimates \citep[][quoting $3742\pm83$, $3642\pm22$, $3700-3800$ and $3755\pm69$~K respectively]{Gaidos14, Malo14, Afram19, Maldonado20}.  
It implies that AU~Mic has a bolometric magnitude of $7.22\pm0.02$~mag and therefore a logarithmic luminosity with respect to the Sun $\log \lstar/\lsun=-0.99\pm0.01$,  
in agreement with, e.g., \citet{Malo14} and \citet{Cifuentes20}.  The inferred radius is  $\rstar=0.78\pm0.04$~\rsun.  

The most recent interferometric measurements suggest a radius of $0.862\pm0.052$~\rsun\ \citep[when taking into account limb darkening,][]{Gallenne22}, slightly larger 
though still compatible with the previous value within 1$\sigma$.  Given the rotation period of $4.86\pm0.01$~d (the error bar indicating temporal variability rather than precision), 
these two \rstar\ estimates translate into line-of-sight-projected rotation velocities at the equator of $\vsini=8.1\pm0.2$ and $9.0\pm0.5$~\kms\ respectively \citep[AU~Mic being seen almost equator 
on,][]{Martioli20,Martioli21}.  From a fit to infrared spectral lines (including magnetic broadening), \citet{Kochukhov20} find 
$\vsini=9.2\pm0.1$~\kms, consistent with the interferometric radius, although the authors mention that a smaller \vsini\ (of 8.1~\kms) is also possible when assuming a 
larger macroturbulence velocity (both being hard to determine independently).  This further argues in favour of a radius in the range 0.80--0.85~\rsun\ for AU~Mic.  

We used our spectra to redetermine the parameters of AU~Mic with the tool designed for 
characterizing SPIRou spectra of M dwarfs \citep{Cristofari22a, Cristofari22b}.  Since magnetic fields significantly contribute to the 
width of spectral lines in stars as active and magnetic as AU~Mic \citep{Lopez-Valdivia21}, we implemented polarized radiative transfer in the modeling \citep[see][and references therein for more information on the method]{Cristofari23} 
and carried out the analysis on the median SPIRou spectrum of AU~Mic, including the effect of small-scale magnetic fields as in \citet{Kochukhov20}.  In practice, we computed a grid 
of model spectra for different atmospheric parameters (\teff, \logg,  metallicity ${\rm [M/H]}$, abundance of $\alpha$ elements relative to Fe $[\alpha/{\rm Fe}]$) 
and magnetic strengths (0, 2, 4, 6, 8 and 10~kG, assuming in each case a radial field of equal strength over the star), 
and ran a Monte Carlo Markov Chain (MCMC) process to find the atmospheric parameters and combination of magnetic spectra that best match the profiles of 
selected atomic and molecular lines with various magnetic sensitivities (including the Ti lines used by \citealt{Kochukhov20} and the field insensitive CO lines at 2.3~\mic).  

\begin{figure}
\centerline{\includegraphics[scale=0.6,bb=20 20 380 390]{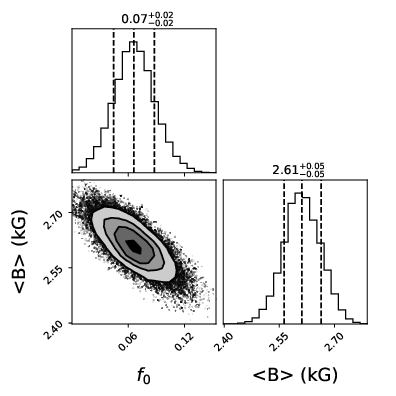}} 
\caption[]{Magnetic parameters of AU~Mic, derived by fitting our median SPIRou spectrum using the atmospheric modeling approach of \citet{Cristofari23},
which incorporates magnetic fields as well as a MCMC process to determine optimal parameters and their error bars.  We find that AU~Mic hosts a small-scale 
magnetic field of <$B$>~$=2.61\pm0.05$~kG, whereas the relative area of non-magnetic regions at the surface of the star is $f_0=0.07\pm0.02$.} 
\label{fig:bmag}
\end{figure}

For the main atmospheric parameters, we find that $\teff=3665\pm31$~K, $\logg=4.52\pm0.05$, ${\rm [M/H]}=0.12\pm0.10$ and $[\alpha/{\rm Fe}]=0.00\pm0.04$.  
For the magnetic properties, we infer that the mean small-scale field at the surface of the star is <$B$>~$=2.61\pm0.05$~kG whereas the optimal coefficient associated with the 
non-magnetic spectrum $f_0$, i.e., the relative stellar surface area featuring no magnetic fields, is $f_0=0.07\pm0.02$ (see Fig.~\ref{fig:bmag}).  
Most of the reconstructed field concentrates within the 2 and 4~kG bins (with respective filling factors $f_2=59$\% and $f_4=33$\%), whereas 
the 6, 8 and 10~kG bins ($f_6=1$\%, $f_8=f_{10}=0$) can be ignored \citep{Cristofari23}.  We stress that taking magnetic fields into account 
is important to derive reliable atmospheric parameters of stars as magnetic as AU~Mic, with potential overestimates, especially in \logg, 
when the effect of magnetic fields is neglected \citep{Cristofari23}.  
The magnetic field at the times of our observations is stronger and covers a larger 
fraction of the star than at the time of the observations of \citet{Kochukhov20}.  

Comparing the temperature and luminosity derived above (implying together a radius of $0.79\pm0.02$~\rsun) with the evolutionary models of 
\citet[][assuming solar metallicity and including overshoot]{Siess00} 
yields a mass of $\mstar=0.43\pm0.03$~\msun, a radius of $\rstar=0.74\pm0.03$~\rsun\ and an age of 13$\pm$2~Myr, which is lower than the age of the $\beta$~Pic moving group.  
Comparing now to the \citet{Baraffe15} models gives a better agreement, with $\mstar=0.55\pm0.05$~\msun, $\rstar=0.78\pm0.02$~\rsun\ and an age of $17\pm4$~Myr.  
Using the Dartmouth models \citep[assuming again solar metallicity and including overshoot,][]{Dotter08} further improves the match with the measured parameters, yielding $\mstar=0.59\pm0.04$~\msun, 
$\rstar=0.79\pm0.02$~\rsun\ and an age of $20\pm5$~Myr.  All predicted radii are smaller than, though still reasonably close to, the interferometric one.  In fact, active M dwarfs have repeatedly been reported to exhibit 
inflated radii with respect to theoretical models, possibly under the effect of magnetic fields \citep{Chabrier07, Morales10, Feiden16}, although there is no consensus on this point yet \citep[e.g.,][]{Morrell19}.   

We assume $\mstar=0.60\pm0.04$~\msun\ and $\rstar=0.82\pm0.02$~\rsun\ for AU~Mic in the rest of the paper, implying $\vsini=8.5\pm0.2$.  The corresponding \logg\ ($4.39\pm0.05$) is slightly smaller 
that the one we measured, further arguing in favour of the Dartmouth models which yield a larger mass and thus a larger \logg\ ($4.42\pm0.05$) than the two others.  Part of the discrepancy may also come from the fact that 
the standard atmospheric models we used to fit our SPIRou spectra may not be well adapted for young stars as strongly active and magnetic as AU~Mic.  
We finally note that AU~Mic is predicted to be still fully convective by the models of \citet{Siess00}, but to have already developed 
a small radiative core (of approximate mass and radius 0.2~\mstar\ and 0.3~\rstar) in the models of \citet{Baraffe15} and \citet{Dotter08}.  

Parameters used in the following sections are listed in Table~\ref{tab:par}.  Rotation cycles are computed using 
a rotation period of 4.86~d, and an arbitrary reference barycentric Julian date of BJD0~$=2459000$.  (Note that a different BJD0 was used in \citealt{Klein21}, 
causing phases in our study to be 0.394 rotation cycle larger for profiles common to both studies.)   

\begin{table}
\caption[]{Parameters of AU~Mic used in / derived from our study} 
\hspace{-6mm}
\begin{tabular}{ccc}
\hline
distance (pc)        & $9.714\pm0.002$ & \citet{Gaia21} \\
$V$ (mag)            & $8.627\pm0.052$ & \citet{Kiraga12}\\ 
$V-I$ (mag)          & $2.034\pm0.060$ & \citet{Kiraga12}\\ 
$J$ (mag)            & $5.436\pm0.017$ & \citet{Cutri03}\\ 
BC$_J$ (mag)         & $1.72\pm0.01$   & \citet{Pecaut13}\\ 
$\log(\lstar/\lsun)$ & $-0.99\pm0.01$  & from \teff, $J$, BC$_J$ and distance\\ 
\teff\ (K)           & $3665\pm31$     & \\
\logg\ (dex)         & $4.52\pm0.05$   & $4.39\pm0.05$ from mass and radius\\ 
${\rm [M/H]}$ (dex)  & $0.12\pm0.10$   & \\ 
$[\alpha/{\rm Fe}]$ (dex)  & $0.00\pm0.04$   & \\ 
\mstar\ (\msun)      & $0.60\pm0.04$    & using \citet{Dotter08} \\
\rstar\ (\rsun)      & $0.82\pm0.02$   & \\
age (Myr)            & $23\pm3$        & \citet{Mamajek14} \\ 
                     & $18.5\pm2.4$    & \citet{Miret20} \\
\Prot\ (d)           & $4.86$          & period used to phase data \\ 
\Prot\ (d)           & $4.856\pm0.003$ & period from \Bl\ data \\ 
\Prot\ (d)           & $4.866\pm0.004$ & period from RV data \\ 
\vsini\ (\kms)       & $8.5\pm0.2$     & from \Prot\ and \rstar \\ 
$i$                  & 80\degr         & assumed for ZDI \\ 
                     & $89.5\pm0.4$\degr & orbit of b, \citet{Martioli21}\\ 
<$B$> (kG)           & $2.61\pm0.05$   & \\ 
$f_0$                & $0.07\pm0.02$   & $f_2=59$\%, $f_4=33$\% \\ 
\omeq\ (\rpd)        & $1.299\pm0.002$ & average over 2020 and 2021 \\ 
\dom\ (\mrpd)        & $37\pm7$        & idem \\ 
\hline
\end{tabular}
\label{tab:par}
\end{table}

\section{The longitudinal field and Zeeman broadening of AU~Mic}
\label{sec:bls}

Using the Stokes $V$ and $I$ LSD profiles of AU~Mic computed in Sec.~\ref{sec:obs}, we computed the longitudinal field \Bl, i.e., the line-of-sight-projected 
component of the vector magnetic field averaged over the visible hemisphere, following \citet{Donati97b}.   
The Stokes $V$ LSD signatures of AU~Mic being quite broad, the first moment is computed over a domain of $\pm45$~\kms\ about the line center, whereas the equivalent width of the 
Stokes $I$ LSD profiles is simply estimated through a Gaussian fit (and found to be $\simeq$2~\kms).  We also computed a null polarization check called $N$ \citep{Donati97b}, and derived  
a mean longitudinal field from this check, which is expected to be equal to 0 within the error bars and to yield a reduced chi-square \chisqr\ close to 1.  

We thus obtained 225 \Bl\ points over the full 3-yr timespan of our observations (see Table~\ref{tab:log} for a complete log), as well as an equal number of values from the $N$ profiles.  
The corresponding \chisqr\ from \Bl\ values and its equivalent from the $N$ profiles are respectively equal to 500 and 0.95 over the whole series, confirming that the field is 
detected and that the error bars are consistent with photon noise.  {\emr We find that \Bl\ ranges from -240 to 260~G, i.e., about an order of magnitude smaller than the average small-scale field 
<$B$> (see Sec.~\ref{sec:par}).}  We then carried out a quasi-periodic (QP) GPR fit to the \Bl\ curve, with the covariance function $c(t,t')$ set to 
\begin{eqnarray}
c(t,t') = \theta_1^2 \exp \left( -\frac{(t-t')^2}{2 \theta_3^2} -\frac{\sin^2 \left( \frac{\pi (t-t')}{\theta_2} \right)}{2 \theta_4^2} \right) 
\label{eq:covar}
\end{eqnarray}
where $\theta_1$ is the amplitude (in G) of the Gaussian Process (GP), $\theta_2$ its recurrence period (very close to \Prot),
$\theta_3$ the evolution timescale on which the \Bl\ curve changes shape (in d), and $\theta_4$ a smoothing parameter 
setting the amount of allowed harmonic complexity.  To these 4 hyper parameters, we added a fifth one called $\theta_5$ that describes the 
additional uncorrelated noise that is needed to obtain the QP GPR fit to the \Bl\ data (denoted $y$) featuring the highest likelihood $\mathcal{L}$, defined by: 
\begin{eqnarray}
2 \log \mathcal{L} = -n \log(2\pi) - \log|C+\Sigma+S| - y^T (C+\Sigma+S)^{-1} y
\label{eq:llik}
\end{eqnarray}
where $C$ is the covariance matrix for all observing epochs, $\Sigma$ the diagonal variance matrix associated with $y$, $S=\theta_5^2 I$ the contribution of the additional white noise 
with $I$ the identity matrix, and $n$ the number of data points.  
Coupling this with a MCMC run to explore the parameter domain, we can determine the optimal set of hyper parameters and their posterior 
distributions / error bars.  

\begin{figure*}
\centerline{\includegraphics[scale=0.49,angle=-90]{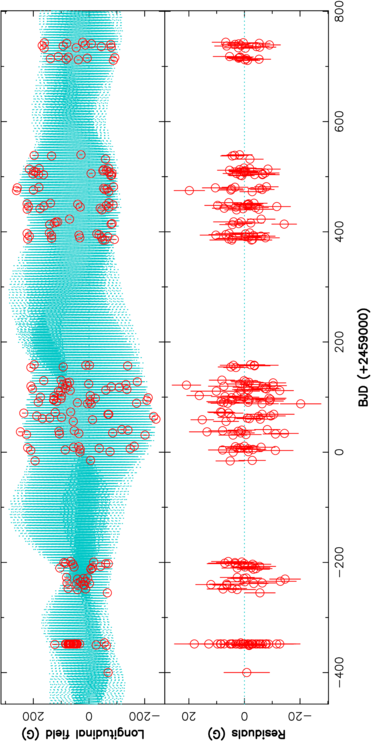} \vspace{2mm}}
\centerline{\includegraphics[scale=0.49,angle=-90]{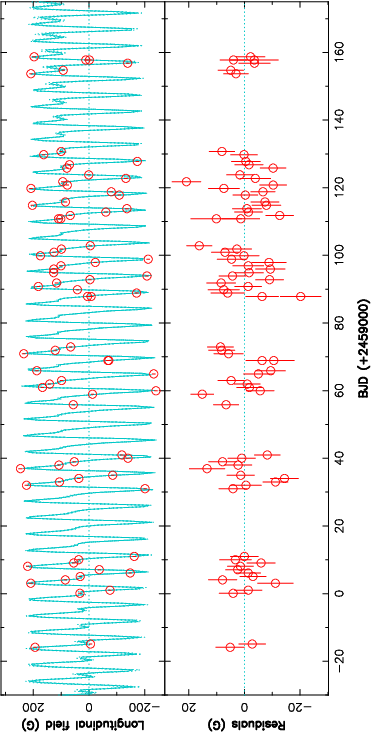} \vspace{2mm}}
\centerline{\includegraphics[scale=0.49,angle=-90]{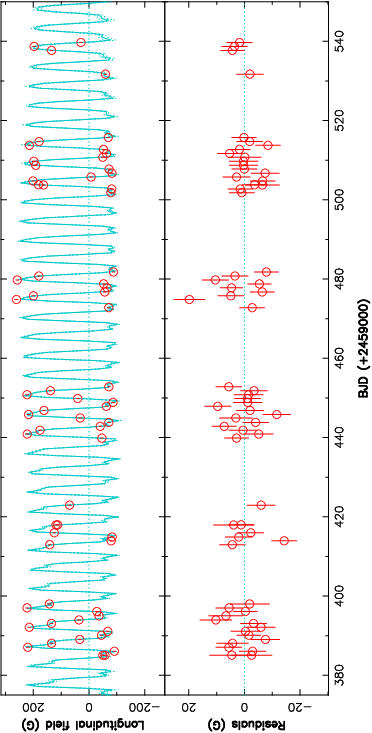}}
\caption[]{Longitudinal magnetic field \Bl\ of AU~Mic (red dots) over our observing period, and QP GPR fit to the data (cyan).  The residuals are shown in the bottom plot of each panel. 
The top panel show the whole data set, whereas the lower 2 panels present a zoom on the 2020 and 2021 data respectively. The RMS of the residuals is 6.2~G, slightly larger than the 
average error bar of 5.2~G, yielding $\chisqr=1.4$, whereas the \chisqr\ with respect to the weighted average is 442.}  
\label{fig:gpb}
\end{figure*}

The result of the fit is shown in Fig.~\ref{fig:gpb} (top panel), with a zoom on seasons 2020 and 2021 also provided in the medium and bottom 
panels.  The fitted GP parameters and error bars are listed in the top section of Table~\ref{tab:gpr}.  
All parameters are well defined, in particular the recurrence period $\theta_2$, found to be equal to $4.856\pm0.003$~d \citep[i.e., close to the 
estimates of][]{Plavchan20, Klein21, Cale21, Zicher22} and the evolution timescale, which we measure at $80\pm12$~d, i.e., half the duration of a typical observing season.  
The data are fitted to a RMS level of 6.2~G, slightly larger than the average error bar on our \Bl\ measurements (of 5.2~G), yielding $\chisqr=1.4$.  
The \Bl\ data are thus not fitted down to the photon-noise level, suggesting an additional source of noise, e.g., intrinsic variability caused by 
activity (e.g., stochastic changes in the large-scale field, flares), modeled by the GP models with $\theta_5$ being significantly different from 0.  
The season-to-season variations of \Bl\ are quite obvious, with the modulation shrinking to a minimum in October 2019 (BJD 2458800) and reaching a maximum in the following 
season (BJD 2459100).  Moreover, as expected from the rather short evolution timescale, the fitted \Bl\ curve also evolves significantly within each season, as can be seen 
on, e.g., the middle panel of Fig.~\ref{fig:gpb}.  Running GPR on the 2020 and 2021 data subsets further suggests that evolution was faster in 2020 ($\theta_3=73\pm11$~d) than in 2021 ($\theta_3=140\pm40$~d).  

\begin{table}
\caption[]{Results of our MCMC modeling of the \Bl\ (first 5 rows) and <$B$> (last 5 rows) curves of AU~Mic.  For each hyper parameter, we list the fitted value along with the corresponding 
error bar, as well as the assumed prior.  The knee of the modified Jeffreys prior is set to $\sigma_{B}$, i.e., the median error bars of our \Bl\ and <$B$> estimates (i.e., 5.2 and 20~G respectively).  
For the recurrence period $\theta_2$, using a uniform prior yields the same result.  
For the evolution timescale $\theta_3$, the log Gaussian prior is set to 100~d (within a factor of 2), a typical value for early M dwarfs.  } 
\center{\scalebox{0.9}{\hspace{-4mm}
\begin{tabular}{cccc}
\hline
Parameter   & Name & value & Prior   \\
\hline
GP amplitude (G)     & $\theta_1$  & $100\pm12$       & mod Jeffreys ($\sigma_{B}$) \\
Rec.\ period (d)     & $\theta_2$  & $4.856\pm0.003$  & Gaussian (4.86, 0.1) \\
Evol.\ timescale (d) & $\theta_3$  & $80\pm12$        & log Gaussian ($\log$ 100, $\log$ 2) \\
Smoothing            & $\theta_4$  & $0.43\pm0.04$    & Uniform  (0, 3) \\
White noise (G)      & $\theta_5$  & $6.6\pm0.9$      & mod Jeffreys ($\sigma_{B}$) \\
\hline
GP amplitude (kG)    & $\theta_1$  & $0.16\pm0.03$    & mod Jeffreys ($\sigma_{B}$) \\
Rec.\ period (d)     & $\theta_2$  & $4.859\pm0.004$  & Gaussian (4.86, 0.1) \\
Evol.\ timescale (d) & $\theta_3$  & $153\pm18$       & log Gaussian ($\log$ 100, $\log$ 2) \\
Smoothing            & $\theta_4$  & $0.71\pm0.10$    & Uniform  (0, 3) \\
White noise (kG)     & $\theta_5$  & $0.01\pm0.01$    & mod Jeffreys ($\sigma_{B}$) \\
\hline
\end{tabular}}} 
\label{tab:gpr}
\end{table}

\begin{figure*}
\centerline{\includegraphics[scale=0.49,angle=-90]{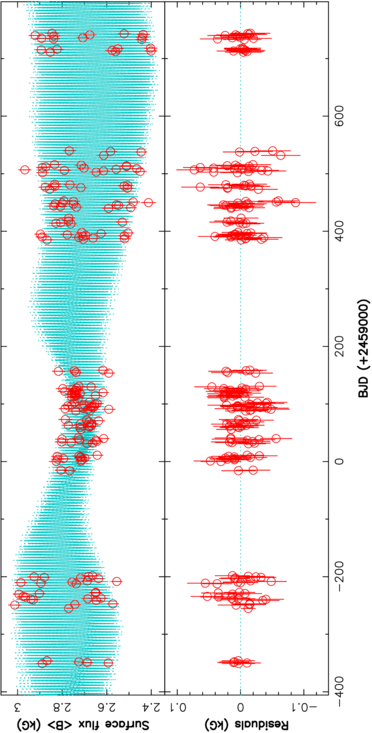} \vspace{2mm}}
\centerline{\includegraphics[scale=0.49,angle=-90]{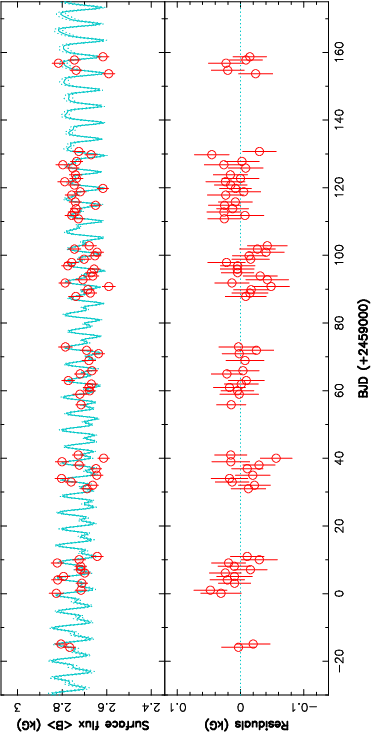} \vspace{2mm}}
\centerline{\includegraphics[scale=0.49,angle=-90]{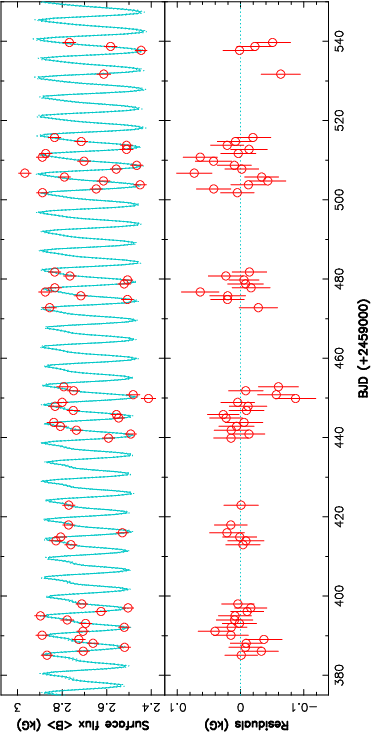}}
\caption[]{Same as Fig.~\ref{fig:gpb} for the small-scale magnetic flux <$B$> at the surface of AU~Mic.  The RMS of the residuals is 0.02~kG, yielding $\chisqr=0.78$, whereas the \chisqr\ with respect to 
the weighted average is 22.4. }  
\label{fig:gpbs}
\end{figure*}

In parallel to the \Bl\ analysis, we used the new tool of \citet{Cristofari23}, with which we analysed the nightly medians of our Stokes $I$ spectra (see Sec.~\ref{sec:par}), 
to derive the small-scale field at the stellar surface at each observing epoch and investigate its rotational modulation 
(freezing all non-magnetic parameters to the values derived from the median spectrum, even though magnetic and non magnetic parameters are mostly uncorrelated).  The 
derived values of <$B$> are listed in Table~\ref{tab:log}.  As for \Bl, we carried out a QP GPR on the <$B$> values, resulting in a \chisqr\ of 0.17, i.e., much lower than 1.  It reflects that the 
retrieved formal error bars are absolute error bars (including systematics) rather than relative ones, thereby underestimating the precision at which night-to-night variations are measured.  
{\emr To derive relative error bars, we simply rescaled the formal ones using the dispersion of the residuals, thereby ensuring that the \chisqr\ of the QP GPR is close to 1 while $\theta_5$ remains 
consistent with 0.}   
We obtained the results shown in Fig.~\ref{fig:gpbs} and the hyper-parameters listed in the bottom section of Table~\ref{tab:gpr}, yielding now $\chisqr=0.78$.  

We find that <$B$> is clearly modulated by the rotation cycle, with a recurrence period of $4.859\pm0.004$~d, 
i.e., marginally larger than the one derived from \Bl.  The semi-amplitude of the modulation is small, less than 0.1~kG in 2020 and reaching a maximum of 0.25~kG in 2022.  
We note that the modulation of <$B$> is smallest when that of \Bl\ is largest and vice versa on our 4 observing seasons.  Finally, the evolution timescale is twice longer for <$B$> than for \Bl.  
Over our 4 seasons of observations, <$B$> decreases from about 2.80~kG down to about 2.65~kG (see Fig.~\ref{fig:gpbs}) and is on average slightly larger than that estimated from the median spectrum 
(see Sec.~\ref{sec:par}).  More specifically, this weakening shows up as a decrease of the fitted 4~kG coefficient  
$f_4$, and a corresponding increase of the 2~kG coefficient $f_2$ (both being strongly anti-correlated, with a correlation coefficient $R=-0.90$).  Whereas $f_4$ correlates 
with <$B$> ($R=0.70$) and varies from 0.32 to 0.18, $f_2$ is anti-correlated with <$B$> ($R=-0.80$) and varies from 0.60 to 0.75.  We also note that, as for \Bl, the 
evolution of <$B$> as derived by GPR is faster in 2020 ($\theta_3=90\pm30$~d) than in 2021 ($\theta_3=300\pm100$~d).  
These results illustrate that both \Bl\ and <$B$>, probing different characteristics of the field, are quite useful and very complementary to analyse the magnetic properties of active stars like AU~Mic. 

We also looked at the Stokes $I$ LSD profiles computed with the high-Land\'e and the low-Land\'e masks.  We find that the profiles from the first set are clearly broader than those from the 
second set as a result of Zeeman broadening, with the median full-width-at-half-maximum (FWHM) of the high-Land\'e and low-Land\'e LSD profiles being respectively equal to 31.9 and 22.5~\kms.  
The amount of quadratic differential broadening between the 2 sets of profiles is found to be $22.6\pm1.0$~\kms\ on average, with no clear evolution with time nor modulation with rotation phase.  
The FWHM of the high-Landé LSD profiles, dominated by Zeeman broadening, is weakly modulated by the rotation period and exhibits long-term variations of up to 2~\kms\ over the full observing period.  
A GPR fit to the data yields a period of $4.85\pm0.02$~d and a semi-amplitude of up to 1~\kms.  
A similar behaviour of about half the amplitude is observed on the FWHM of low-Landé LSD profiles.  The overall trends on the FWHM of the LSD profile from 
high-Land\'e lines mimic those on <$B$>, i.e., a small decrease over the 4 seasons and a minimum modulation amplitude in 2020.  The correlation factor between FWHMs and <$B$> is found to be $R=0.60$, 
suggesting that FWHMs can be used as an alternate proxy for <$B$>, albeit with a loss of precision.  


\section{Magnetic field and differential rotation of AU~Mic}
\label{sec:zdi}

Using time-series of Stokes $V$ and $I$ LSD profiles, one can model the large-scale magnetic field at the surface of AU~Mic, along with constraints on the small-scale field.  
This is achieved with ZDI, a tomographic imaging tool that inverts phase-resolved sets of LSD profiles into maps of the large-scale vector field \citep[e.g.,][]{Donati06b, Klein21}.  
In the particular case of AU~Mic, Stokes $I$ LSD profiles are significantly broadened by magnetic fields and can be used to further constrain the magnetic map and give insights 
on the small-scale field.  

\subsection{Zeeman-Doppler Imaging}
\label{sec:zd0}

In practice, ZDI proceeds iteratively, starting from a null magnetic field and adding information as it explores the parameter space using conjugate gradient techniques.  At each 
iteration, ZDI compares the synthetic Stokes profiles of the current magnetic image with observed ones, and loops until it reaches the requested level of agreement with the data (i.e., a 
given \chisqr).  As the problem is ill-posed and features an infinite number of solutions of variable complexity, we choose the simplest one, i.e., the solution with minimum information or 
maximum entropy that matches the data at the requested level \citep[e.g.,][]{Skilling84}.  The surface of the star is decomposed into 3000 grid cells.  
Local Stokes $I$ and $V$ profiles in each grid cell are computed using Unno-Rachkovsky's equation of the polarized radiative transfer equation in a plane-parallel Milne Eddington atmosphere 
\citep{Landi04}, then integrated over the visible surface of the star at each observed rotation phase (assuming a linear center-to-limb darkening law for the continuum, with a coefficient of 0.3) 
to yield the synthetic profiles corresponding to the reconstructed image.  The mean wavelength and Land\'e factor of our LSD profiles are 1700~nm and 1.2.  

The magnetic field at the surface of the star is described through a spherical harmonics (SH) expansion, using the formalism of \citet{Donati06b} in which the poloidal and toroidal components 
of the vector field are expressed with 3 sets of complex SH coefficients, $\alpha_{\ell,m}$ and $\beta_{\ell,m}$ for the poloidal component, and $\gamma_{\ell,m}$ for the toroidal component\footnote{After a few years, 
the expressions of \citet{Donati06b} were modified to achieve a more consistent description of the field, with $\beta_{\ell,m}$ being replaced by $\alpha_{\ell,m}+\beta_{\ell,m}$ in the equations of the meridional 
and azimuthal field components \citep[see, e.g.,][]{Lehmann22, Finociety22}.}, where $\ell$ and $m$ note the degree and order of the corresponding SH term in the expansion.  

ZDI can also model brightness inhomogeneities at the surface of the star, simultaneously with large-scale magnetic fields.  In the particular case of AU~Mic, we find that the 
distortions, and especially the broadening, of the LSD Stokes $I$ profiles are dominated by magnetic effects, with only a small impact of surface brightness inhomogeneities (in agreement 
with the small amplitude of photometric variations, of order of a few percent).  For instance, the Doppler width of the local profile \vD\ that is needed to reproduce the average Stokes $I$ profile 
of AU~Mic with minimal Zeeman broadening is found to be $\vD=5.3$~\kms\ (assuming $\vsini=8.5$~\kms, see Sec.~\ref{sec:par}), whereas this parameter is typically equal to $\vD\simeq3$~\kms\ for weakly-active, slowly-rotating M dwarfs of similar spectral type.  The large difference in FWHM between the average Stokes $I$ LSD profiles associated with high-Land\'e and low-Land\'e lines (see Sec.~\ref{sec:bls}) further confirms 
that this excess broadening is mostly of magnetic origin.  In practice, we find that the Stokes $I$ and $V$ profiles of AU~Mic can be entirely explained by magnetic field variations 
at the surface of the star.  Consistently reproducing the FWHMs of the Stokes $I$ LSD profiles associated with high-Land\'e and low-Land\'e lines for small-scale fields of about 2.5~kG \citep[see 
Sec.~\ref{sec:bls} and][]{Kochukhov20} requires setting $\vD=3.5$~\kms, on the high side of what is observed for weakly-active M dwarfs.  

Given this, we chose to carry-out 2 different sets of complementary magnetic reconstructions.  We first focus on Stokes $V$ LSD profiles only and assume $\vD=5.3$~\kms, i.e., the 
Doppler width of the local profile that enables to reproduce Stokes $I$ LSD profiles with minimal Zeeman broadening.  We further assume that only a fraction $f_V$ of each grid cell (called filling 
factor of the large-scale 
field, equal for all cells) actually contributes to Stokes $V$ profiles, with a magnetic flux over the cells equal to $B_V$ (and thus a magnetic field within the magnetic portion of the cells equal 
to $B_V/f_V$).  This approach, {\ems called `Stokes $V$ analysis' below}, allows one to model the large-scale field and its temporal evolution over the 3~yr of our observations.  It is similar to the study of \citet{Klein21} 
in this respect (regarding \vD\ and $f_V$ in particular), except for simultaneous brightness imaging that has very little impact on the reconstructed magnetic map and that we therefore left out of the 
process.  

In a second step, we model both Stokes $V$ and Stokes $I$ LSD profiles, this time assuming $\vD=3.5$~\kms, i.e., the value that yields the 
observed FWHMs of the Stokes $I$ LSD profiles of high-Land\'e et low-Land\'e lines 
when the Zeeman broadening of a $\simeq$2.5~kG small-scale field is taken into account.  
This approach, {\ems called `Stokes $I$ \& $V$ analysis' below}, leads to a significantly stronger reconstructed field (than in the Stokes $V$ analysis), which should now be consistent with both the large-scale field constraints 
provided by Stokes $V$ profiles and the small-scale field ones coming from Stokes $I$ data.  In this case, we further assume that a fraction $f_I$ of each grid cell (called filling factor of the small-scale 
field, again equal for all cells) hosts small-scale fields of strength $B_V/f_V$ (i.e., with a small-scale magnetic flux over the cell equal to $B_I = B_V f_I/f_V$).  This simple model implies in 
particular that the small-scale field locally scales up with the large-scale field (with a scaling factor of $f_I/f_V$), which is likely no more than a rough approximation.  
That the small-scale field is modulated with rotation (see Sec.~\ref{sec:bls}) actually suggests that it may indeed spatially correlate with the large-scale field to some degree.  
In practice, we carried out ZDI for various values of $f_I$ and $f_V$, and selected the pair that fits the data best.  In both cases, we assumed $i=80\degr$ for the inclination of the rotation axis 
to the line of sight, i.e., slightly lower than the inclination of the orbital planes of planets b and c (see Table~\ref{tab:par}), to reduce mirroring effects of the imaging process between the upper and lower hemispheres.  

Both methods have their own pros and cons.  The {\ems Stokes $V$ analysis} makes it possible to optimally fit the Stokes $V$ profiles, yielding the minimal large-scale field (and variations 
with time) required by the data, but is not able to account for the observed small-scale field.  It is well adapted to investigate the temporal evolution of the large-scale field, 
either from season to season, or under the effect of surface differential rotation (DR) within a given season.  
{\ems The  Stokes $I$ \& $V$ analysis} is better suited to study the small-scale field and presumably yields a more accurate estimate of the large-scale field as well, but is less optimal to monitor 
its temporal changes.   
We did not attempt to model low-level brightness inhomogeneities whose impact on the Stokes $I$ and $V$ profiles is quite small, nor to model changes in the Stokes $I$ 
profiles caused by magnetic fields with pseudo-brightness features inducing similar profile distortions \citep[as done in][]{Klein21}.  We report the results of both approaches in Secs.~\ref{sec:zd1} 
and \ref{sec:zd2}.  

\begin{figure*}
\centerline{\large\bf 2019\raisebox{0.3\totalheight}{\includegraphics[scale=0.45,angle=-90]{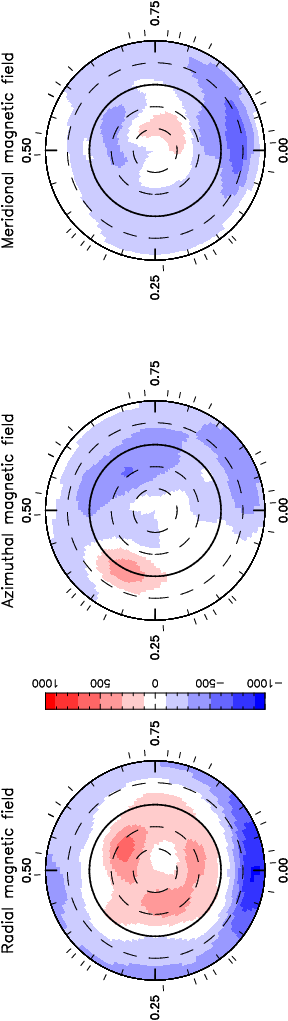}}\vspace{1mm}}
\centerline{\large\bf 2020\raisebox{0.3\totalheight}{\includegraphics[scale=0.45,angle=-90]{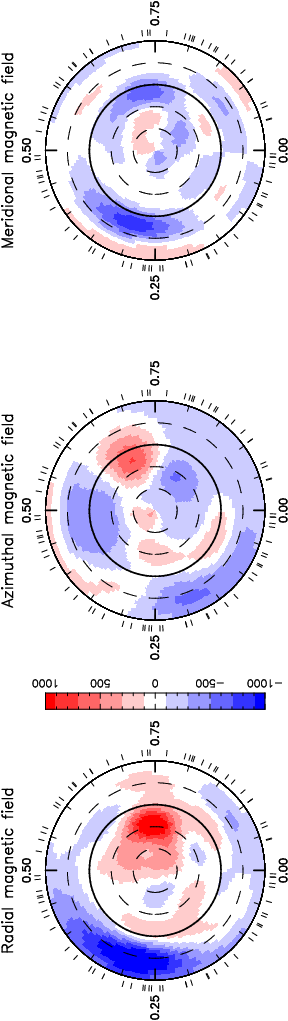}}\vspace{1mm}} 
\centerline{\large\bf 2021\raisebox{0.3\totalheight}{\includegraphics[scale=0.45,angle=-90]{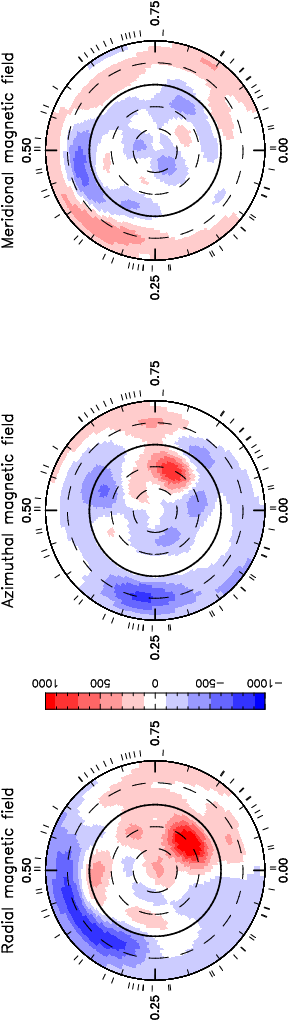}}\vspace{1mm}} 
\centerline{\large\bf 2022\raisebox{0.3\totalheight}{\includegraphics[scale=0.45,angle=-90]{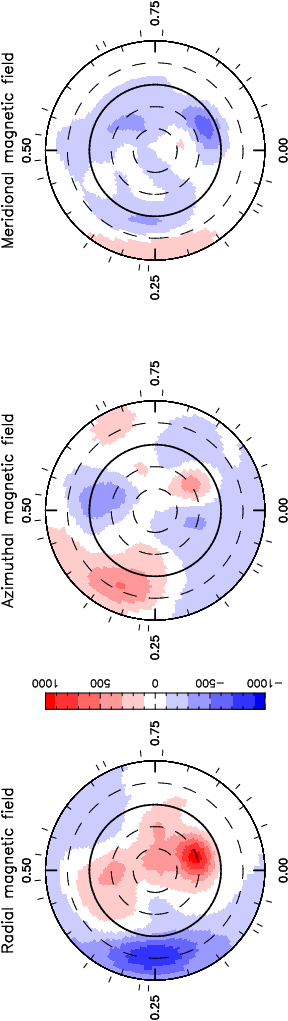}}\vspace{1mm}} 
\caption[]{Reconstructed maps of the large-scale field of AU~Mic (left, middle and right columns for the radial, azimuthal and 
meridional components in spherical coordinates, in G), for season 2019 Sep-Nov, 2020 Apr-Nov, 2021 Jun-Nov and 2022 May-Jun (top to bottom row respectively), derived 
from the Stokes $V$ LSD profiles of Fig.~\ref{fig:fit} using ZDI (see Sec.~\ref{sec:zd1}).  The maps are shown in a flattened polar projection down to latitude  
$-60$\degr, with the north pole at the center and the equator depicted as a bold line.  Outer ticks indicate phases of observations.
Positive radial, azimuthal and meridional fields respectively point outwards, counterclockwise and polewards.  Each image shows the evolving magnetic map (as a result of DR) 
at mid-time throughout the season.  } 
\label{fig:map}
\end{figure*}

Finally, we also looked for signatures of DR at the surface of the star, in the same way as in previous studies, i.e., by assuming a 2-parameter DR law similar to that of the 
Sun, with the rotation rate at latitude $\theta$ being given by $\Omega(\theta)=\omeq - \dom \sin^2 \theta$, where \omeq\ and \dom\ respectively stand for the rotation rate at the equator and the 
difference in rotation rate between the equator and the pole.  We then look for the pair of DR parameters that provides the best fit to the data at given image information content.  To 
ensure maximum sensitivity, we concentrate on Stokes $V$ data only that are best suited for diagnosing temporal variability of the large-scale field.  Results are reported in Sec.~\ref{sec:zd3}.  

Since the timescale on which the large-scale field of AU~Mic evolves ($\simeq$80~d, see Sec.~\ref{sec:bls}) is much shorter than our overall observing window of 3~yr, we chose to divide our 
dataset into 4 different subsets, each more or less corresponding to one of our observing season (2019 Sep-Nov, 2020 Apr-Nov, 2021 Jun-Nov, 2022 May-Jun) containing respectively 28, 78, 65 and 20 spectra, 
and covering time slots of 57, 175, 155 and 33~d (about 12, 36, 32 and 7 rotation cycles), i.e., 0.4 to 2.2$\times$ the evolution timescale of the longitudinal magnetic field $\theta_3$ (80~d, 
see Sec.~\ref{sec:bls}).  
The median time shifts between these 4 successive seasons, equal to 298, 391 and 265~d (about 61, 80 and 55 rotation cycles), are 3.3--4.9$\times$ larger than $\theta_3$.  
The few spectra collected early in 2019 April and June (at 6 main epochs, see Table~\ref{tab:log}), not providing enough phase coverage by themselves, were excluded from this analysis.  

Images shown in  Secs.~\ref{sec:zd1} and \ref{sec:zd2} are derived with DR parameters of $\omeq=1.299$~\rpd\ and $\dom= 0.037$~\rpd, an average of the values we infer from our 
2020 and 2021 data sets (see Sec.~\ref{sec:zd3}).  

\subsection{\ems{Stokes $V$ analysis of LSD profiles}}
\label{sec:zd1}

The sets of Stokes $V$ LSD profiles collected with SPIRou for the 4 main seasons outlined previously, along with the ZDI fit to the data (assuming $\vD=5.3$~\kms\ for the Doppler width of the local 
profile), are presented in Fig.~\ref{fig:fit} (as supplementary material), 
whereas the corresponding reconstructed images are shown in Fig.~\ref{fig:map}.  As in \citet{Klein21}, we find that $f_V\simeq0.2$ provides the best fit to the data for all epochs.  Error bars of 
Stokes $V$ LSD profiles (see Table~\ref{tab:log}) had to be increased by 40--60 percent for ZDI to be able to reach a unit \chisqr, as with \Bl\ for which GPR also diagnosed the presence of additional 
uncorrelated noise (see Sec.~\ref{sec:bls}).  The required increase in error bars is larger for our 2 longest seasons (2020 and 2021) and smaller for the shortest ones (2019 and 2022), further 
confirming that it likely reflects intrinsic variability from the intense activity of AU~Mic.  

Taking into account the phase shift mentioned in Sec.~\ref{sec:par}, our 2019 magnetic image resembles that of \citet{Klein21}, with a radial magnetic field reaching 700~G at phase 0.6 \citep[phase 
0.2 in][]{Klein21} at mid latitudes, along with consistent patches of azimuthal field of different polarities.  Although both studies use the same data (except for one low-SNR observation, marked with an 
'x' in Table~\ref{tab:log}, left out from the first analysis and whose impact on the reconstructed image is insignificant), the two images are not exactly identical, the new one being reconstructed 
for a slightly larger \vsini\ (8.5~\kms\ instead of 7.8~\kms) and a different pair of DR parameters (see Sec.~\ref{sec:zd3}).  As a result, the field we reconstruct is slightly smaller than (though 
still consistent with) that of \citet{Klein21}, with a quadratically-averaged large-scale magnetic flux over the stellar surface equal to <$B_V$>$\simeq$380~G.  The large-scale field is mostly poloidal 
and axisymmetric, with the poloidal component enclosing 75\% of the reconstructed field energy, 75\% of which in axisymmetric modes.  The dipole component reaches a strength of $\Bd=430$~G at the pole, 
and is inclined at 15\degr\ to the rotation axis towards phase 0.65.  

\begin{figure*}
\centerline{\large\bf 2020\raisebox{0.3\totalheight}{\includegraphics[scale=0.45,angle=-90]{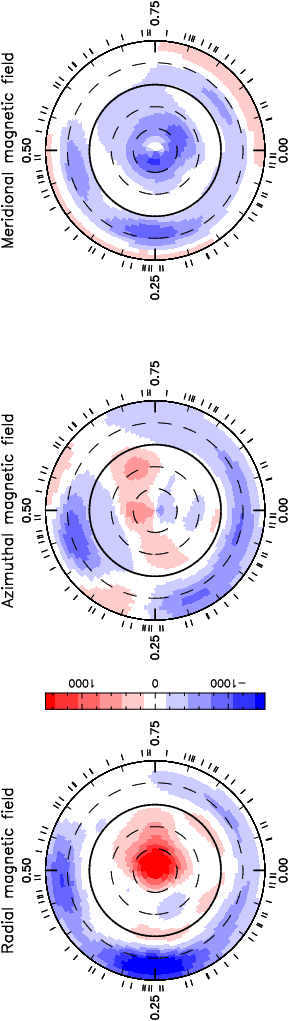}}} 
\caption[]{Same as second top row of Fig.~\ref{fig:map} (2020 Apr-Nov), with the field now reconstructed using both Stokes $I$ \& $V$ LSD profiles (see Sec.~\ref{sec:zd2}).  Note the different color scale 
to depict magnetic fluxes. }  
\label{fig:mapI}
\end{figure*}

The 2020 and 2021 images are reconstructed from data sets covering a 3$\times$ longer time span and are thus much better constrained, especially regarding DR (see Sec.~\ref{sec:zd3}).  Again, 
we find that AU~Mic hosts a dominantly poloidal large-scale field enclosing 85\% of the reconstructed energy (of which 35\% in axisymmetric modes), and that <$B_V$> ranges from 360~G (in 2021) 
to 400~G (in 2020).  Both magnetic maps feature a strong positive radial field region at a latitude of 30\degr\ (at phases 0.73 in 2020 and 0.91 in 2021) where $B_V$ reaches 1.2~kG, and another 
similar one of opposite polarity in the other hemisphere, located more or less symmetrically from the positive one with respect to the centre of the star.  These radial field regions are 
accompanied by an azimuthal field region of identical polarity, located at a similar latitude and a slightly smaller phase.  Apart from a global phase shift (of 0.17 cycle) that likely reflects 
the effect of DR at a latitude of $\simeq$30\degr\ over the time span that separates both data sets (equal to 80 rotation cycles of AU~Mic), the 2020 and 2021 magnetic maps share obvious 
similarities, featuring dipole components of $\Bd=440$~G and 400~G respectively tilted at 40\degr\ and 50\degr\ to the rotation axis (towards phase 0.75 and 0.85).  Despite this resemblance, 
the detailed maps depart enough from one another to generate \Bl\ curves that are significantly different, as a result of temporal evolution (on a timescale of 80~d, see Sec.~\ref{sec:bls} and 
Fig.~\ref{fig:gpb}). 

Although derived from the sparsest of our 4 data sets, the 2022 magnetic image of AU~Mic is largely similar to the previous ones, with a 1~kG positive radial field region reconstructed at 
mid latitude (phase 0.96), but with the main negative radial field region no longer showing up at the antipodes of the positive one.  The azimuthal field regions accompanying the radial field 
ones are also much weaker than in the previous 2 images.  We find that <$B_V$>, equal to 330~G, is weaker than in the previous seasons, and so is the dipole component $\Bd=380$~G (tilted at 
35\degr\ towards phase 0.80).  Once more, the poloidal component largely dominates the large-scale field topology, enclosing 90\% of the reconstructed magnetic energy, 45\% of which in axisymmetric 
modes.  

All properties of the reconstructed large-scale magnetic field are summarised in Table~\ref{tab:mag} for our 4 observing seasons.  

\begin{table*}
\caption[]{Properties of the large-scale (columns 2 to 6 and 8 to 10) and small-scale (column 7) magnetic field of AU~Mic for our 4 observing seasons.  
Columns 2 to 5 correspond to the ZDI Stokes $V$ analysis (assuming $f_V=0.2$ and $\vD=5.3$~\kms, see Sec.~\ref{sec:zd1}), whereas columns 6 to 10 summarize the results of 
the ZDI Stokes $I$ \& $V$ analysis (assuming $f_I=0.9$, $f_V=0.2$ and $\vD=3.5$~\kms, see Sec.~\ref{sec:zd2}).  Columns 3 and 8 list the polar strengths of the 
dipole component \Bd, columns 4 and 9 the tilts of the dipole component to the rotation axis, and columns 5 and 10 the amount of magnetic energy reconstructed in the poloidal component of the 
field and in the axisymmetric modes of this component.  Error bars on field values and percentages are typically equal to 5--10\%. }  
\center{\scalebox{1.0}{\hspace{0mm}
\begin{tabular}{cccccccccc}
\hline
         & \multicolumn{4}{c}{Stokes $V$ analysis} & \multicolumn{5}{c}{Stokes $I$ \& $V$ analysis}    \\
         & \multicolumn{4}{c}{($f_V=0.2$, $\vD=5.3$~\kms)} & \multicolumn{5}{c}{($f_I=0.9$, $f_V=0.2$, $\vD=3.5$~\kms)}    \\
\hline
Season   & <$B_V$> & \Bd  & tilt & poloidal / axisym & <$B_V$> & <$B_I$> & \Bd  & tilt & poloidal / axisym \\
         &  (G)    &   (G)       & (\degr) & (\%)  &  (G)    & (kG)    &   (G)       &  (\degr) & (\%) \\ 
\hline
2019 Sep-Nov & 380 & 430 & 15 & 75 / 75 & 550 & 2.5 & 650 & 10 & 85 / 90 \\ 
2020 Apr-Nov & 400 & 440 & 40 & 85 / 35 & 570 & 2.6 & 660 & 25 & 90 / 75 \\ 
2021 Jun-Nov & 360 & 400 & 50 & 85 / 35 & 530 & 2.4 & 650 & 25 & 90 / 75 \\ 
2022 May-Jun & 330 & 380 & 35 & 90 / 45 & 520 & 2.3 & 660 & 20 & 90 / 80 \\ 
\hline
\end{tabular}}}
\label{tab:mag}
\end{table*}

\subsection{\ems{Stokes $I$ \& $V$ analysis of LSD profiles}}
\label{sec:zd2}

We now analyse Stokes $I$ and $V$ LSD profiles together, setting now $\vD=3.5$~\kms\ for the Doppler width of the local profile.  It allows us to infer constraints 
on the small-scale and large-scale magnetic fields $B_I$ and $B_V$ simultaneously, under the assumption that the first scales up with the second with a fixed factor of $f_I/f_V$ (see 
Sec.~\ref{sec:zd0}).  In practice, we find that $f_I=0.9$ provides good results, in rough agreement with the results of Sec.~\ref{sec:par}, implying a scaling factor of the small-scale 
to large-scale field equal to $f_I/f_V=4.5$.  This is also consistent with previous results, indicating that active M dwarfs like AU~Mic are able to trigger large-scale fields <$B_V$> 
whose strength reaches up to 30 percent that of small-scale fields <$B_I$> \citep{Morin10, Kochukhov21}.  

With this approach, we find that the large-scale field of AU~Mic is significantly stronger than that derived in Sec.~\ref{sec:zd1}, including in particular a more intense, mostly axisymmetric, 
dipole component.  This is consistent with the small-scale field <$B$> directly measured from magnetically sensitive lines, and from the observed differential broadening between the high-Land\'e 
and low-Land\'e Stokes $I$ LSD profiles.~\ref{sec:bls}).  
Since small-scale fields are assumed to scale-up with large-scale fields in our simple model, ZDI 
has no other choice than increasing the large-scale field as well to generate the adequate level of magnetic broadening in the Stokes $I$ LSD profiles.  Despite this additional constraint, ZDI is still 
able to fit the Stokes $V$ LSD profiles at the same time as the Stokes $I$ LSD profiles, adding to the large-scale field a nearly axisymmetric dipole component that contributes no more than 
marginally to the Stokes $V$ profiles (as a result of the geometrical configuration, with the star being close to equator-on for an Earth-based observer).  

The properties of the reconstructed magnetic fields are summarized in columns 6 to 10 of Table~\ref{tab:mag} for our 4 seasons.  In average, we find that <$B_V$> now reaches fluxes of 520--570~G, 
i.e., 1.5$\times$ larger than when fitting Stokes $V$ LSD profiles only.  It implies small-scale field fluxes of <$B_I$>=2.3--2.6~kG, in agreement with the results of Secs.~\ref{sec:par} and 
\ref{sec:bls}, as well as with those of \citet{Kochukhov20}.  The difference with the results of Sec.~\ref{sec:zd1} also shows up on the inferred dipole component, now 1.6$\times$ larger than in 
the Stokes $V$-only reconstruction, with the poloidal component enclosing 75--90\% of the reconstructed magnetic energy.  As the added dipole is mostly axisymmetric, the tilt of the overall dipole 
component to the rotation axis is smaller (10-25\degr) and the poloidal component is mostly axisymmetric (75--90\% in terms of magnetic energy).  We show one example reconstruction for 
season 2020 in Fig.~\ref{fig:mapI}, the inferred maps for the 3 other seasons looking similar.  

We stress that, although our new set of magnetic maps are able to fit both Stokes $I$ and $V$ data, they may still not match the real ones as the imaging problem is ill-posed.  This is especially true 
for AU~Mic, whose almost equator-on orientation and small \vsini\ both contribute to the problem degeneracy.  For instance, adding small-scale tangled fields 
more or less evenly at the surface of AU~Mic without modifying the large-scale field reconstructed from Stokes $V$ profiles only, may also provide a comparable fit to the Stokes $I$ and $V$ profiles, 
but with a magnetic topology that does not have a fixed amount of small-scale to large-scale field ratio over the surface.  This second model would however generate very little rotational 
modulation of the small-scale field, thereby contradicting our <$B$> measurements from magnetically sensitive lines (see Sec.~\ref{sec:bls}).  Further constraining the imaging process would require 
collecting LSD profiles for Stokes $Q$ and $U$ LSD profiles, in addition to Stokes $V$ and $I$ profiles, as previously suggested by \citet{Kochukhov20}.  Fig.~\ref{fig:lqu} (provided as supplementary 
material) shows for instance that 
Stokes $Q$ and $U$ signatures (not measured in our campaign) would be detectable and allow one to unambiguously differentiate between the maps of Secs.~\ref{sec:zd1} and \ref{sec:zd2}.  

\subsection{Differential rotation from Stokes $V$ LSD profiles}
\label{sec:zd3}

Last but not least, one can study the amount of latitudinal DR shearing the surface of AU~Mic from the recurrence periods of the Stokes $V$ signatures associated with the magnetic features 
reconstructed at different latitudes.  As the goal is to diagnose subtle evolution of the Stokes $V$ signatures with time, the most reliable approach is to focus on Stokes $V$ LSD profiles 
only (as in Sec.~\ref{sec:zd1}), even though the large-scale field itself may actually be stronger than what Stokes $V$ LSD profiles alone indicate (see Sec.~\ref{sec:zd2}).  
As outlined in Sec.~\ref{sec:zd0}, this is achieved by reconstructing magnetic maps at given information content (i.e., at given magnetic energy) over a given grid of DR parameters.  We then 
fit the resulting \chisqr\ map with a 2D paraboloid, the location of the minimum and the paraboloid curvature at this location respectively providing the optimal DR parameters and the associated 
error bars \citep{Donati03b}.  

We first find that data sets corresponding to seasons 2019 and 2022 do not span long enough a time slot and include too few profiles to yield reliable DR estimates, especially for a star like 
AU~Mic whose \vsini\ is on the low side, and that exhibits a high level of intrinsic variability, even on its large-scale field (see Sec.~\ref{sec:bls}).  More specifically, we find that the 
derived \chisq\ maps for both epochs are noisy, showing low-level fluctuations with no obvious minimum over the grid of DR parameters, possibly as a result of short-term intrinsic 
variability distorting magnetic maps and preventing the DR signal to build-up in a consistent way.  For season 2019 and with the same data, \citet{Klein21} was able to locate a minimum in the 
\chisqr\ map outside our grid of DR parameters, corresponding to an unexpectedly strong DR for an M dwarf like AU~Mic.  We thus suspect this early estimate to rather reflect the impact of intrinsic variability 
rather than the shearing effect of DR.  

\begin{figure*}
\centerline{ \includegraphics[scale=0.3,angle=-90]{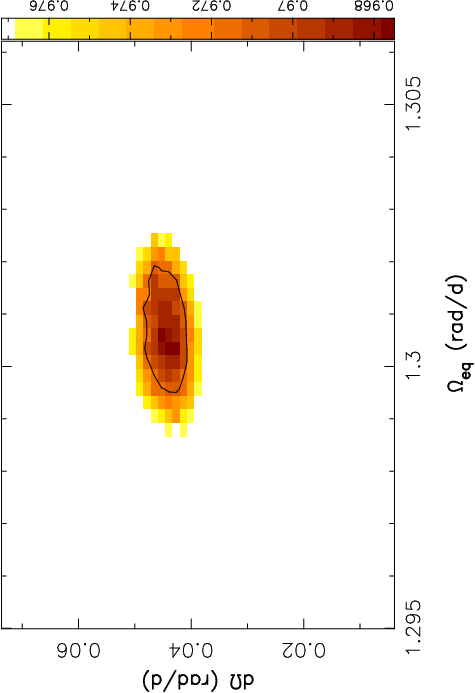}\hspace{3mm}\includegraphics[scale=0.33,angle=-90]{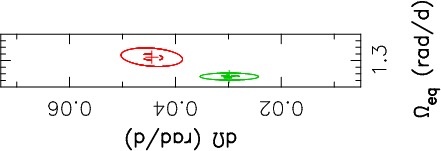}\hspace{0mm}\includegraphics[scale=0.3,angle=-90]{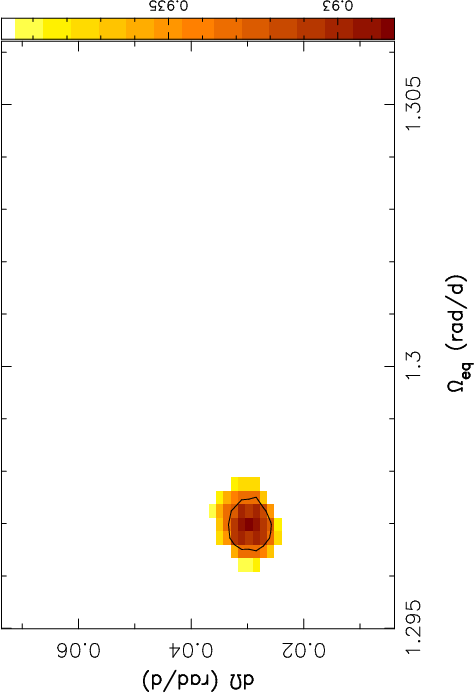}} 
\caption[]{Measuring DR at the surface of AU~Mic by minimizing the \chisqr\ of the ZDI fit to our Stokes $V$ LSD profiles over a grid of DR parameters (see Sec.~\ref{sec:zd3}), for season 2020 Apr-Nov 
(left panel) and 2021 Jun-Nov (right panel).  In both panels, the black contour depicts the 3$\sigma$ confidence interval for the pair of DR parameters. The central panel shows the 1$\sigma$ and 3$\sigma$ 
confidence ellipses for both seasons (in red for 2020 and green for 2021) together on the same plot.  }  
\label{fig:drp}
\end{figure*}

The data sets we collected over the 2020 and 2021 seasons, respectively spanning 36 and 32 rotation cycles and gathering 78 and 65 Stokes $V$ LSD profiles, are much better suited for deriving 
reliable DR estimates.  We find that $\omeq=1.3006\pm0.0004$~\rpd\ and $\dom=0.0445\pm0.0014$~\rpd\ for season 2020, and $\omeq=1.2970\pm0.0002$~\rpd\ and $\dom=0.0298\pm0.0013$~\rpd\ for season 
2021, with the two estimates differing by more than 3$\sigma$ between both epochs suggesting that DR may be varying at the surface of AU~Mic (see Fig.~\ref{fig:drp}), on a timescale similar to 
that on which the large-scale field evolves.  
We however note that the error bars on DR parameters are bigger in 2020 than in 2021 despite the larger number of points in the data set and the similar reconstructed magnetic images (see Fig.~\ref{fig:map}).  
We suspect that it again reflects the impact of intrinsic variability (e.g., from stochastic changes of the magnetic maps), that happened to be larger in 2020 than in 2021 judging from the decay time and residuals inferred by GPR  
from the \Bl\ data (see Sec.~\ref{sec:bls} and Fig.~\ref{fig:gpb}), and which presumably broadened the 2D \chisqr\ paraboloid in 2020 more than in 2021.  Splitting the data set of each season into 2 subsets and 
carrying out the same process on each subset indeed yields noisy \chisqr\ maps and discrepant DR parameters, consistent with our previous conclusion that extensive observations collected over a 
full season are needed to obtain a reliable DR measurement.  

Given that DR measurements of AU~Mic are apparently quite sensitive to intrinsic variability, the difference between the values inferred from our 2020 and 2021 Stokes $V$ data may not be so 
significant, despite their differing by more than 3$\sigma$.  This is why all magnetic images presented in Secs.~\ref{sec:zd1} and \ref{sec:zd2} were derived using average DR parameters from 
both seasons (i.e., $\omeq=1.299$~\rpd\ and $\dom= 0.037$~\rpd).  These average parameters imply that the equator of AU~Mic rotates in about 4.84~d whereas its pole rotates in about 4.98~d, with a 
typical timescale of 170~d for the equator to lap the pole by one complete cycle.  This timescale is about twice longer than the evolution timescale derived from the GP fit to the \Bl\ data (see 
Table~\ref{tab:par}), indicating that DR itself is likely not the main contributor to the overall large-scale field distortion with time.  

In this context, the nominal period of AU~Mic (of 4.86~d, used to phase our data, see Table~\ref{tab:par}) corresponds to a latitude of 24\degr\ whereas that derived from the GP fit to the \Bl\ data 
($4.856\pm0.003$~d) corresponds to latitudes in the range 20--24\degr, and that derived from the GP fit of the <$B$> data ($4.859\pm0.004$~d) corresponds to latitudes in the range 22--26\degr.  
Similarly, the phase shift of surface features at latitude 30\degr\ over a timescale of 391~d (i.e., the time shift 
between seasons 2020 and 2021) is expected to be +0.19, slightly larger than, though still comparable to, the observed phase drift of the main radial and azimuthal field features reconstructed at 
at both epochs at this latitude (of order +0.15, see middle rows of Fig.~\ref{fig:map}).

\section{The multi-planet system of AU~Mic}
\label{sec:lbl}

We analysed the 185 RV points derived by the LBL technique \citep{Artigau22} from our nightly observations of AU~Mic (see Table~\ref{tab:log}), looking for the RV signatures of the 2 known transiting warm Neptunes  
hiding within the dominant activity signal modulated by the rotation period.  Our rich data set also allows us to investigate potential RV signatures of additional planets in the AU~Mic system, either more distant 
ones that may not be transiting, {\emr or small inner ones like the candidate Earth-mass planet (dubbed d) recently proposed by \citet{Wittrock23}, potentially located between b and c and putatively causing the large TTVs 
reported for both \citep{Szabo22}. } 
For planets b and c, the orbital periods and transit times are known with high precision (see Table~\ref{tab:pla}), leaving us with the RV semi-amplitude $K_b$ and $K_c$ to be determined \citep[assuming circular orbits, 
consistent with the results of][]{Zicher22}.  {\emr For candidate planet d, also assumed to be on a circular orbit, we choose the most likely solution of \citet{Wittrock23}, associated with a period of $12.7381\pm0.0013$~d 
and a conjunction BJD of $2458333.32\pm0.36$~d (both parameters fixed in our modeling), which leaves us with only $K_d$ to be adjusted.}   
For each additional planet to be considered, it adds 3 more free parameters to the problem, the orbital period $P_i$, the RV semi-amplitude $K_i$ and the date of conjunction BJD$_i$ 
(assuming again circular orbits).  Although TTVs are quite significant for the actual transits of b and c \citep{Szabo22}, we do not take them into account in our RV modeling, as they still 
correspond to very small phase shifts (of about 0.3 percent of an orbital cycle for the innermost planet).  The default transit times and orbital periods that we use for planets b and c in our analysis are those 
of \citet{Szabo22} that minimize the amplitude of TTVs over all transits observed so far.  

\begin{figure*}
\centerline{\includegraphics[scale=0.6,angle=-90]{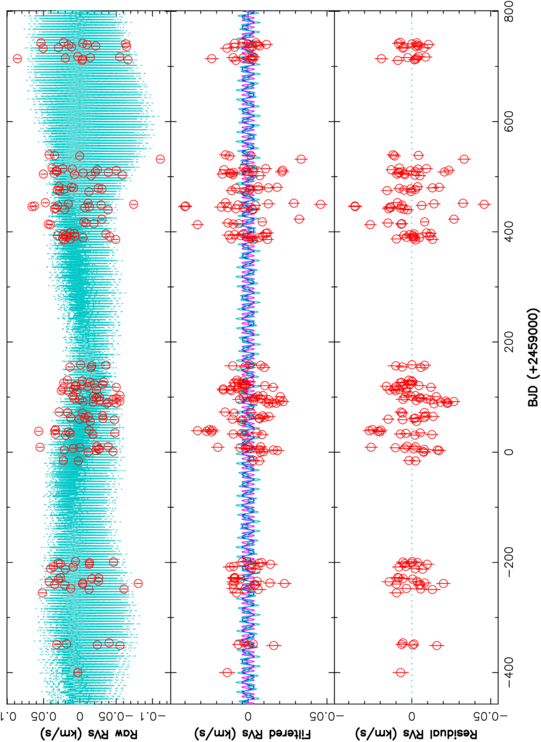}} 
\caption[]{Raw (top), filtered (middle) and residual (bottom) RVs of AU~Mic (red dots) over our observing period.  The top plot shows the MCMC fit to the data, including a QP GPR modeling of the activity and 
the RV signatures of planets b and c (cyan), whereas the middle plot shows the planet RV signatures (pink, blue and cyan for planets b, c and b+c respectively) once activity is filtered out. 
{\emr The RMS of the residuals is 11.1~\ms, 2.9$\times$ larger than the median error bars (3.8~\ms) on individual RV points.}  } 
\label{fig:rvr}
\end{figure*}

In practice, we use a GPR with a QP kernel to model activity, coupled to a MCMC process to determine the posterior distributions of the planet parameters and of the GP hyper parameters.  We run 
different joint models, one featuring planets b and c only which we take as a reference, plus others, either without b and c, or with additional planets whose RV 
signatures may also be present in the data.  The marginal logarithmic likelihood $\log \mathcal{L}_M$ of a given solution is computed using the approach of \citet{Chib01} as described in \citet{Haywood14}, and the 
significance of the RV signatures of additional non-transiting planets is estimated from the difference in $\log \mathcal{L}_M$ (i.e., the logarithmic Bayes Factor $\log$~BF) with respect to our reference model.   

\begin{table*}
\caption[]{{\emr MCMC results for the 4 studied cases (no planet, b+c, b+c+e, b+c+e+d).  For each case, we list the recovered GP and planet parameters with their error bars, as well as the priors used whenever 
relevant.  The last 4 rows give the \chisqr\ 
and the RMS of the best fit to our RV data, as well as the associated marginal logarithmic likelihood $\log \mathcal{L}_M$ and marginal logarithmic likelihood variation $\Delta \log \mathcal{L}_M$ with respect to the 
reference case (b+c).  }}  
\center{\scalebox{1.0}{\hspace{0mm}
\begin{tabular}{cccccc}
\hline
Parameter          & No planet            & b+c                  & b+c+e                & b+c+e+d              &   Prior \\ 
\hline
$\theta_1$ (\ms)   & \emr $30.1^{+3.9}_{-3.4}$ & \emr $30.3^{+3.9}_{-3.5}$ & \emr $31.6^{+4.7}_{-4.1}$ & \emr $31.5^{+4.7}_{-4.1}$ & mod Jeffreys ($\sigma_{\rm RV}$) \\  
$\theta_2$ (d)     & \emr $4.865\pm0.005$      & \emr $4.863\pm0.006$      & \emr $4.865\pm0.005$      & \emr $4.865\pm0.005$      & Gaussian (4.86, 0.1) \\ 
$\theta_3$ (d)     & \emr $139^{+23}_{-20}$    & \emr $136^{+23}_{-20}$    & \emr $136^{+22}_{-19}$    & \emr $136^{+22}_{-19}$    & log Gaussian ($\log$ 140, $\log$ 1.5) \\ 
$\theta_4$         & \emr $0.29\pm0.04$        & \emr $0.29\pm0.04$        & \emr $0.37\pm0.06$        & \emr $0.37\pm0.06$        & Uniform  (0, 3) \\ 
$\theta_5$ (\ms)   & \emr $14.2\pm1.0$         & \emr $13.5\pm1.0$         & \emr $12.5\pm0.9$         & \emr $12.7\pm0.9$         & mod Jeffreys ($\sigma_{\rm RV}$) \\ 
\hline
$K_b$ (\ms)        &                      & \emr $4.1^{+1.8}_{-1.2}$  & \emr $4.5^{+1.7}_{-1.2}$  & \emr $4.6^{+1.7}_{-1.2}$  & mod Jeffreys ($\sigma_{\rm RV}$) \\ 
$P_b$ (d)          &                      & 8.4631427            & 8.4631427            & 8.4631427            & fixed from \citet{Szabo22} \\ 
BJD$_b$ (2459000+) &                      & $-669.61584$         & $-669.61584$         & $-669.61584$         & fixed from \citet{Szabo22} \\ 
$M_b$ (\me)        &                      & \emr $9.3^{+4.1}_{-2.7}$  & \emr $10.2^{+3.9}_{-2.7}$ & \emr $10.4^{+3.9}_{-2.7}$ & derived from $K_b$, $P_b$ and \mstar \\ 
\hline
$K_c$ (\ms)        &                      & \emr $4.0^{+1.7}_{-1.2}$  & \emr $4.8^{+1.6}_{-1.2}$  & \emr $5.1^{+1.6}_{-1.2}$  & mod Jeffreys ($\sigma_{\rm RV}$) \\ 
$P_c$ (d)          &                      & 18.85882             & 18.85882             & 18.85882             & fixed from \citet{Szabo22} \\ 
BJD$_c$ (2459000+) &                      & 454.8973             & 454.8973             & 454.8973             & fixed from \citet{Szabo22} \\ 
$M_c$ (\me)        &                      & \emr $11.8^{+5.1}_{-3.5}$ & \emr $14.2^{+4.8}_{-3.5}$ & \emr $15.1^{+4.8}_{-3.5}$ & derived from $K_c$, $P_c$ and \mstar \\ 
\hline
$K_e$ (\ms)        &                      &                      & \emr $11.1^{+2.1}_{-1.7}$ & \emr $11.3^{+2.2}_{-1.8}$ & mod Jeffreys ($\sigma_{\rm RV}$) \\  
$P_e$ (d)          &                      &                      & \emr $33.39\pm0.10$       & \emr $33.39\pm0.10$       & Gaussian (33.4, 1.0) \\ 
BJD$_e$ (2459000+) &                      &                      & \emr $117.1\pm0.9$        & \emr $117.1\pm0.9$        & Gaussian (118, 8) \\ 
$M_e$ (\me)        &                      &                      & \emr $35.2^{+6.7}_{-5.4}$ & \emr $35.9^{+6.9}_{-5.8}$ & derived from $K_e$, $P_e$ and \mstar  \\ 
\hline
\emr $K_d$ (\ms)        &                      &                      &                      & \emr $1.1^{+1.1}_{-0.5}$  & \emr mod Jeffreys ($\sigma_{\rm RV}$) \\  
\emr $P_d$ (d)          &                      &                      &                      & \emr 12.73812             & \emr fixed from \citet{Wittrock23} \\ 
\emr BJD$_d$ (2459000+) &                      &                      &                      & \emr $-666.6789$          & \emr fixed from \citet{Wittrock23} \\ 
\emr $M_d$ (\me)        &                      &                      &                      & \emr $2.9^{+2.9}_{-1.3}$ & \emr derived from $K_d$, $P_d$ and \mstar  \\ 
\hline
\chisqr            & \emr 10.4                 & \emr 9.8                  & \emr 8.5                  & \emr 8.5                  &  \\ 
RMS (\ms)          & \emr 11.5                 & \emr 11.1                 & \emr 10.4                 & \emr 10.4                 &  \\ 
$\log \mathcal{L}_M$ & \emr 458.8              & \emr 465.1                & \emr 481.4                & \emr 481.7                &  \\
$\log {\rm BF} = \Delta \log \mathcal{L}_M$ & \emr $-6.3$  & \emr 0.0      & \emr 16.3                 & \emr 16.6                 &  \\
\hline 
\end{tabular}}}
\label{tab:pla}
\end{table*}

{\emr 
When including planets b and c only, we derive similar semi-amplitudes for their RV signatures, equal to $K_b=4.1^{+1.8}_{-1.2}$~\ms\ and $K_c=4.0^{+1.7}_{-1.2}$~\ms\ (see Table~\ref{tab:pla} 
and Fig.~\ref{fig:rvr}), less than the estimates published earlier but nonetheless consistent within about 2$\sigma$ \citep{Klein21,Cale21,Zicher22}.  The RV signatures of both planets show up at a 
level of 3.4$\sigma$.  The GP amplitude reaches $30\pm4$~\ms, about  
7$\times$ larger than the semi-amplitudes of the planet signatures, stressing how intense activity is in the case of as young a star as AU~Mic.  Besides, the RMS of the fit to our RV data is equal 
to 11.1~\ms, 2.9$\times$ larger than the median error bar of our RV measurements (3.8~\ms, see Table~\ref{tab:log}), a likely result of a high level of activity-induced intrinsic variability and 
of potential RV signatures of additional system planets not yet included in the analysis.  } This illustrates how tricky the detection of planet RV signatures can be for very active stars, even in the case of transiting planets 
whose orbital periods and transit times are well documented from high-precision photometry, and how efficient activity filtering needs to be to reliably unveil planet RV signatures.  
{\emr When including planets b and c and compared to a model with no planet, we find that $\log \mathcal{L}_M$ increases by 6.3 while the dispersion of RV residuals and the additional white 
noise parameter $\theta_5$ both decrease (see Table~\ref{tab:pla}), confirming that adding both planets provides a more reliable description of our RV data.  Adjusting the eccentricity of planets b and c along 
with the other parameters yields no more than a small improvement ($\Delta \log \mathcal{L}_M=0.3$) and eccentricities compatible with 0 (with error bars of 0.04 and 0.08 for planets b and c), 
consistent with the results of \citet{Zicher22} and justifying our a priori assumption of circular orbits\footnote{We adjust eccentricities using variables $\sqrt e \cos \omega$ and 
$\sqrt e \sin \omega $ ($e$ being the eccentricity and $\omega$ the angle of periastron) and Gaussian priors (0.0, 0.3) for both, in agreement with the observed distribution of eccentricities 
for multi-planet systems \citep{vanEylen19}.  }.  
} 

\begin{figure*}
\centerline{\includegraphics[scale=0.6,angle=-90]{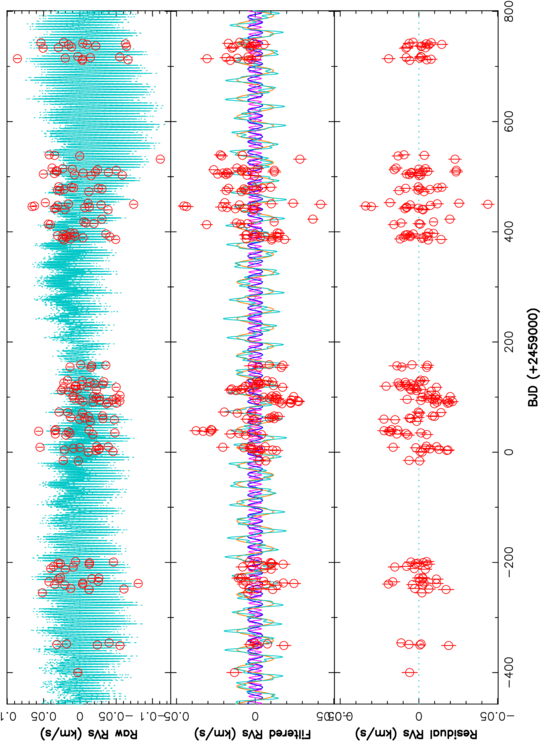}} 
\caption[]{Same as Fig.~\ref{fig:rvr} but including candidate planet e (orange and cyan curves for e and b+c+e in the middle plot) in the MCMC modeling.  {\emr The RMS of the residuals is now 10.4~\ms.} } 
\label{fig:rv2}
\end{figure*}

\begin{figure*}
\centerline{\includegraphics[scale=0.35,angle=-90]{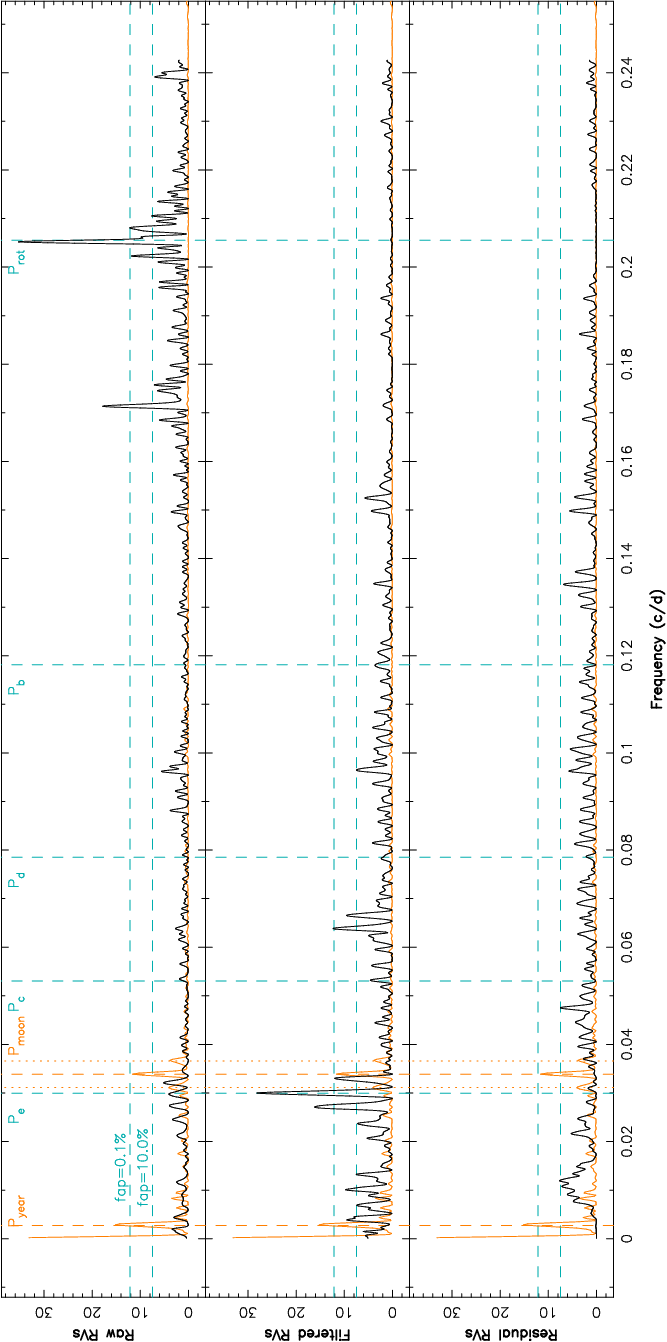}} 
\caption[]{Periodogram of the raw (top), filtered (middle) and residual (bottom) RVs when including candidate planet e (plus b and c) in the MCMC modeling.  
{\emr The cyan vertical dashed lines trace the rotation period of the star and the orbital periods of planets b and c, and of candidate planets d and e, } 
while the horizontal dashed line indicate the 10 and 0.1\% FAP levels in the periodogram of our RV data.    
The orange curve depicts the window function, whereas the orange vertical dashed and dotted line outline the 1-yr period, the synodic period of the Moon and its 1-yr aliases.  } 
\label{fig:per}
\end{figure*}

\begin{figure}
\includegraphics[scale=0.45,angle=-90]{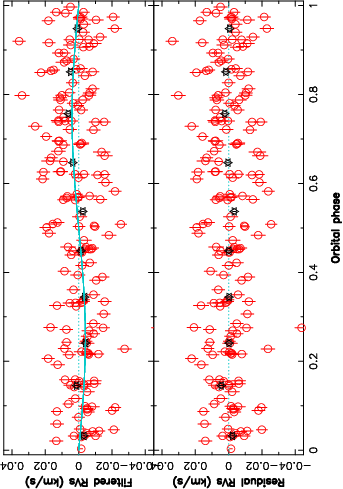}
\includegraphics[scale=0.45,angle=-90]{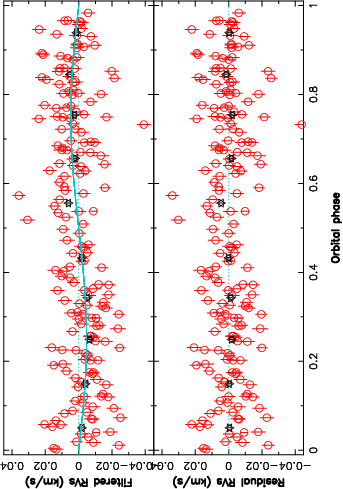}
\includegraphics[scale=0.45,angle=-90]{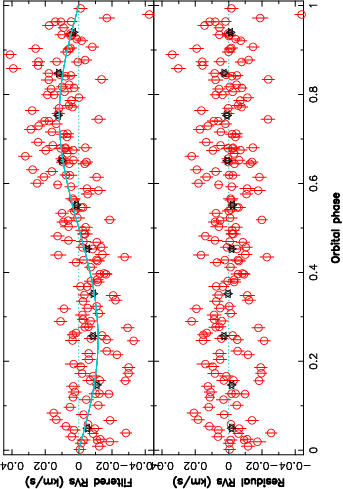}
\caption[]{Filtered (top plot) and residual (bottom plot) RVs for transiting planets b (top panel) and c (middle panel), and for candidate planet e (bottom panel) of 
AU~Mic.  The red dots are the individual RV points with the respective error bars, whereas the black stars are average RVs over 0.1 phase bins.  
{\emr As in Fig.~\ref{fig:rv2}, the dispersion of RV residuals is 10.4~\ms.}   } 
\label{fig:rvf}
\end{figure}

We explored the possibility of an additional system planet whose RV signatures would still be hiding in our data.  
By looking at the periodogram of the filtered RV data, we find residual power at a period of 33.4~d that may hint at the presence of a candidate planet (dubbed planet e), which would be located further out close to a 
4:1 resonance with planet b and 7:1 resonance with the rotation period of the host star.  
{\emr When candidate planet e is taken into account in the modeling and MCMC searches again for the most likely combination of activity and planet signatures, the 33.4~d peak dominates the periodogram of the filtered 
RVs at a level that corresponds to a false alarm probability (FAP) that the signal is spurious of order $10^{-10}$.  
We find an orbital period of $P_e=33.39\pm0.10$~d and a semi-amplitude of $K_e=11.1^{+2.1}_{-1.7}$~\ms\ for candidate planet e, i.e., about 2.5$\times$ larger than those of planets b and c.  
The RV signal is detected at a level of 6.5$\sigma$, with $\log \mathcal{L}_M$ increasing by 16.3 with respect to the reference case with planets b and c only, and the RMS of the RV residuals 
and the additional white noise $\theta_5$ both decreasing.  With this model, the semi-amplitudes associated with planets b and c slightly increase, reaching now $K_b=4.5^{+1.7}_{-1.2}$~\ms\ and 
$K_c=4.8^{+1.6}_{-1.2}$~\ms, implying a $\simeq$4$\sigma$ detection.  
The corresponding fit and periodogram are shown in Figs.~\ref{fig:rv2} and \ref{fig:per}, whereas the phase-folded filtered RV data for planets b, c and e are presented in Fig.~\ref{fig:rvf} 
(both before and after binning on phase bins of 10 percent of their orbital cycles).  No peak crosses the 10\% FAP threshold in the residual RVs (see Fig.\ref{fig:per} bottom plot).  
We note that the derived period for candidate planet e is close to a 1-yr alias of the synodic Moon period (32.1~d, visible in the periodogram of the window function, see Fig.~\ref{fig:per}), but 
distant from it by more than the FWHM of the periodogram peak, ensuring they are distinct.  Adjusting eccentricities of all 3 planets yields again values consistent with 0 (error bars of 
0.04, 0.07 and 0.09 for planets b, c and e respectively) and no more than a small increase in marginal logarithmic likelihood ($\Delta \log \mathcal{L}_M=0.8$), confirming that the model with circular 
orbits is enough to describe our RV data.  

Although the periodogram of the RV residuals shows no obvious signal, we investigated whether our RV data could suggest the presence of additional planets beside candidate planet e, in particular the candidate planet d 
suggested by \citet{Wittrock23} that may orbit between b and c and potentially explain the large TTVs reported for b and c.  When including all 4 planets in the modeling (assuming again circular orbits), we find very 
similar results for b, c and e (see Table~\ref{tab:pla}), and a very small semi-amplitude of $K_d=1.1^{+1.1}_{-0.5}$~\ms, which in fact amounts to a mere 3$\sigma$ upper limit on $K_d$ of 4.4~\ms.  As expected, 
the marginal logarithmic likelihood increases very little ($\Delta \log \mathcal{L}_M=0.3$) when d is included in the modeling, implying that our data provide no evidence for its existence (within the quoted 3$\sigma$ 
upper limit).  } 

We finally note that the parameters of the fitted GP are similar for all cases, except for the additional white noise ($\theta_5$) which decreases when adding planets b, c and e  in the model.  
We find that the timescale on which the activity jitter evolves is longer than that on which \Bl\ changes but similar to that associated with <$B$>.  
We also obtain that the rotation period derived from the activity jitter ($\theta_2=4.865\pm0.005$~d) is marginally longer than the nominal one used to phase the data (4.86~d, see Table~\ref{tab:par}) and those derived 
from \Bl\ and <$B$> measurements (respectively equal to $4.856\pm0.003$ and $4.859\pm0.004$~d).  These periods indicate that the center of gravity of the surface features generating the activity jitter is located 
within the range of latitude 25--29\degr\ (see Sec.~\ref{sec:zd3}).  We point out that 
the smoothing factor $\theta_4$ derived from RV data is smaller than that inferred from \Bl\ and <$B$>, as expected from the fact that RV is proportional to the derivative of the integrated flux 
(at first order), whereas \Bl\ and <$B$> simply grow as the integral of the magnetic field vector / strength over the visible hemisphere.  The periodogram of the filtered RVs (see Fig.\ref{fig:per}, middle plot) clearly 
demonstrates that activity filtering using GPR is quite efficient despite the long observing window, with no excess power remaining at the rotation period, harmonics and aliases.  


\section{Activity proxies of AU~Mic}
\label{sec:act}

The main goal of this section is to investigate how the various activity proxies correlate with RVs, to find out whether one can be used to achieve a filtering of the activity jitter that is as accurate as, or even 
more efficient than, that achieved through GPR (see Sec.~\ref{sec:lbl}).  

The first obvious activity proxy to investigate is the small-scale field <$B$>, already studied in Sec.~\ref{sec:bls} and reported to correlate well with RVs in the particular case of the Sun \citep{Haywood22}.  
We find that RVs of AU~Mic correlate poorly with \Bl\ ($R=0.33$) and <$B$> ($R=0.12$), but nicely with the first time derivative of <$B$> (computed from the GPR fit to <$B$>, $R=0.78$).  
This is similar to what was reported by \citet{Suarez20} for Proxima, where RVs are found not to correlate with the FWHM of the cross-correlation profile, presumably linked to <$B$> via magnetic broadening, but 
with its first time derivative.  This is also consistent with our finding that <$B$> correlates reasonably well with the FWHM of the Stokes $I$ profiles of high-Land\'e lines in AU~Mic ($R=0.60$, see Sec.~\ref{sec:bls}), 
but not with what was reported by \citet{Klein21} in the particular case of the small late-2019 subset (where RVs correlated well with FWHMs).  
Altogether, it indicates that the activity jitter of M dwarfs like Proxima and AU~Mic reflects the RV impact of brightness or magnetic features at the surface of the star, rather than the signature of inhibited convective 
blue-shift, expected to be much smaller for M dwarfs than for the Sun.  

However, trying to directly filter out RVs of AU~Mic using the first time derivative of <$B$> as a proxy to predict the activity jitter is less efficient than the GPR filtering outlined in Sec.~\ref{sec:lbl}.  
We find that the RV residuals are still modulated with rotation, albeit with an amplitude of about 45\% that of the original jitter.  The predicted RV jitter within each season is indeed not entirely consistent with the 
observed RVs, either in amplitude or in phase pattern, leaving residuals that still dominate the RV signatures from the 3 planets.  {\emr Applying the same 3-planet plus QP GPR modeling as 
that of Sec.~\ref{sec:lbl} on the filtered RVs yields results for the parameters of b, c and e and RV residuals that are very similar to those obtained in Sec.~\ref{sec:lbl}.}  
It confirms that the RV signatures detected for all 3 planets are not a spurious artifact 
induced by activity, but brings no improvement in the RMS of RV residuals.  We note that the periodogram of <$B$> only features a strong peak at the rotation period of AU~Mic, and virtually no power at the 
orbital periods of the planets.  

Looking now at the LBL activity proxy dLW, we find that it is modulated by rotation, with a QP GPR fit yielding a period of $4.88\pm0.03$~d, consistent with AU~Mic's reference period of 4.86~d.  
The dispersion with respect to the fit is 3$\times$ larger than the median error bar of individual points (a likely result of intrinsic variability), whereas the modulation amplitude 
derived by GPR is only 25\% larger than the dispersion.  It is less useful in this respect than <$B$> or even the FWHM of the Stokes $I$ profiles of high-Land\'e lines, both exhibiting 
clearer rotational modulation (see Sec.~\ref{sec:bls}).  This difference likely comes from the fact that dLW, computed from all spectral lines (including molecular features, on average less sensitive to magnetic fields), 
is less appropriate for diagnosing small-scale magnetic fields than <$B$> or the FWHM of the Stokes $I$ LSD profiles of atomic lines with high Land\'e factors.  
We also note that RVs poorly correlate with dLW, as already reported for <$B$> and FWHMs.    

Since dLW is measured on individual lines with LBL, one can carry out a weighted principal component analysis \citep[wPCA,][]{Delchambre15} on the time series corresponding to all individual lines, to find out how 
temporal variations of line widths differ from line to line, as a likely impact of small-scale fields (Cadieux et al., in prep.).  When applied to AU~Mic, we find that the strongest wPCA component $W_1(t)$ is enough to 
explain most of the line-to-line differences, and that $W_1$ is clearly modulated with rotation, with a QP GPR yielding a recurrence period of $4.859\pm0.003$~d and a decay time of $141\pm15$~d, fully consistent with 
those derived for <$B$> (see lower section of Table~\ref{tab:gpr}).  
Furthermore, we find that that $W_1$ is strongly correlated with <$B$> ($R=0.96$). Unsurprisingly, using $W_1$ (instead of <$B$>) to filter RVs yields results very similar to those outlined above, i.e., an activity 
jitter reduced in amplitude by a factor of 2 compared to the original one, but still dominating the planet signatures.  The RV planet signatures derived from the filtered RVs are again consistent with 
the values listed in Table~\ref{tab:pla}.  

We also measured equivalent width variations (EWVs) of the 1083~nm \hei\ triplet and 1282~nm \pab\ line, tracing stellar activity and potentially star-planet interactions \citep{Klein22}, and whose profiles are shown 
in Fig.~\ref{fig:hpl} (supplementary material).  
We proceeded as in \citet{Finociety21,Finociety23}, computing a median spectrum by which all spectra are divided, and fitting a Gaussian profile of fixed width and position (in the stellar rest 
frame) to the residual spectra.  The derived values and error bars are listed in Table~\ref{tab:log}, with a few flares detected in \hei\ but not in \pab.  
We find that both indices are modulated with the stellar rotation period.  Fitting the EWVs with GPR yields periods of $4.86\pm0.01$~d for both the \hei\ and \pab\ lines, when setting the decay 
time to 100~d (i.e., a value close to that derived from \Bl\ data, see Sec.~\ref{sec:bls}).  In both cases, the excess white noise, quantifying the intrinsic variability of both activity indicators, is 
significantly larger than the formal photon-noise error bars (median of 0.030 and 0.015~\kms\ for \hei\ and \pab), by an order of magnitude or more, reaching 0.5 and 
0.1~\kms\ for \hei\ and \pab.  It confirms again the strong variability that AU~Mic triggers at all times, especially in the \hei\ line for which the semi-amplitude of the 
modulation is only about 70\% the size of the excess white noise (whereas both are comparable in strength for \pab).  While \pab\ EWVs correlate reasonably well with <$B$> ($R=0.60$), it is not 
the case for \hei\ EWVs that are much more dispersed, presumably as a result of intrinsic variability and chromospheric activity.  We also find that the \pab\ EWVs slowly decrease with time like <$B$> does, 
suggesting that AU~Mic may be progressing towards activity minimum along its putative cycle \citep{Ibanez19}.  

As for <$B$>, the periodogram of the \pab\ EWVs is featureless apart from the main peak at \Prot\ (with a FAP well below the 0.1\% threshold).  In particular, no signal shows up at the orbital 
periods of the planets.  The periodogram of the \hei\ EWVs is much more noisy, with again a main peak at \Prot\ (FAP of 0.1\%).  
Multiple peaks are also present at a FAP level of 10\% or higher, including at frequencies close to the orbital periods of transiting planets b and c.  As these peaks do not correspond to a safe detection, 
and since similar peaks are also present at different periods throughout the periodogram, we conclude that their proximity with the 
orbital frequencies of planets b and c is no more than a coincidence.  

All activity proxies discussed here exhibit rotational modulation, though with different levels of significance.  However, even the best of them, i.e., <$B$> and $W_1$ 
(whose first time derivatives correlate well with RVs), are moderately successful at filtering the dominant activity jitter from the RV curve of AU~Mic.

\section{Summary and discussion}
\label{sec:dis}

Our paper presents a detailed study of the young active star AU~Mic based on 235 high-resolution unpolarized and circularly polarized spectra collected with SPIRou at CFHT from early 2019 to mid 2022,  
covering a timespan of 1,144~d over 4 successive seasons.  

We revisited the main parameters of AU~Mic, including its surface magnetic flux, using the median SPIRou spectrum, and found that $\teff=3665\pm31$~K, $\logg=4.52\pm0.05$, 
${\rm [M/H]}=0.12\pm0.10$  and $[\alpha/{\rm Fe}]=0.00\pm0.04$.  These parameters are consistent with a mass and radius of $0.60\pm0.04$~\msun\ and $0.82\pm0.02$~\rsun, respectively, at an age of $\simeq$20~Myr, 
in the context of the \citet{Baraffe15} and \citet{Dotter08} evolutionary models, except for the estimated \logg\ that is larger (by about 3$\sigma$) than that expected from the mass and radius.  Given the nominal 
rotation period of AU~Mic (4.86~d), the projected rotation velocity \vsini\ is $8.5\pm0.2$~\kms, consistent with previous literature estimates.  Both evolutionary models further suggest that AU~Mic already developed 
a small radiative core.  The small-scale magnetic field <$B$>, adjusted with the atmospheric parameters on the median spectrum from a set of nIR lines with known Zeeman patterns \citep{Cristofari23}, is 
equal to $2.61\pm0.05$~kG, again consistent with previous literature measurements and typical to field strengths of active M dwarfs \citep[e.g.,][]{Kochukhov20, Reiners22}.  We find that modeling Zeeman broadening 
is important to derive accurate atmospheric parameters for strongly magnetic stars like AU~Mic.  

As in \citet{Klein21}, the large-scale magnetic field of AU~Mic is detected through circularly polarized Zeeman signatures of atomic lines.  The longitudinal field \Bl, probing the large-scale 
field, exhibits obvious rotational modulation with a period of $4.856\pm0.003$~d.  The corresponding pattern features a semi-amplitude ranging from 100 (in 2019) to 250~G (in 2020) and evolves on a timescale of $80\pm12$~d, 
typical to largely- or fully-convective M dwarfs whose large-scale fields are usually stable over a few months \citep[e.g.,][]{Morin08a, Morin08b, Hebrard16}.  
The small-scale field <$B$> also exhibits rotational modulation though weaker than that of \Bl, with a different pattern (minimum and maximum amplitude in 2020 and 2021 respectively) and a slightly longer period of 
$4.859\pm0.004$~d.  The evolution timescale is about twice longer for <B> than for \Bl.  
The FWHM of the Stokes $I$ LSD profiles of atomic lines with high Land\'e factors comes as an alternate option for measuring <$B$>, albeit with a loss of precision. 

Applying ZDI on the rotationally modulated sets of LSD Stokes $V$ profiles of each observing season, and setting the Doppler width of the local profile to $\vD=5.3$~\kms\ (to match the shape of the average LSD Stokes $I$ 
profile with minimal Zeeman broadening), we find that the large-scale field of AU~Mic reaches $\simeq$400~G, is mainly poloidal and axisymmetric, and with a dipole component of $\simeq$400~G tilted at 15--50\degr\ with 
respect to the rotation axis depending on the season.  Optimal results are obtained for a filling factor of the large-scale field $f_V\simeq0.2$.  If we include Stokes $I$ LSD profiles in the fitting procedure as well and 
further assume that the local small-scale field scales up with the local large-scale field (with the filling factor of the small-scale field set to $f_I\simeq0.9$ and the Doppler width of the local profile to a more 
conventional value of $\vD=3.5$~\kms), we find that a stronger large-scale field of $\simeq$550~G and a small-scale field of $\simeq$2.5~kG are needed to simultaneously reproduce LSD Stokes $V$ and $I$ profiles.  
The reconstructed field is even more poloidal and axisymmetric, with a dipole component of $\simeq$650~G only moderately tilted with respect to the rotation axis (by 10\degr\ to 25\degr).  This magnetic topology is 
consistent with those of fully or largely convective main-sequence M dwarfs, whose convective zone is deeper than about half the stellar radius \citep{Donati09}, whereas the ratio of magnetic flux between small and large 
scales ($f_V/f_I=0.22$) agrees with previous measurements \citep{Morin08b, Kochukhov21}.  

Whereas {\ems the Stokes $I$ \& $V$ analysis} gives a more reliable description of the overall strength and geometry of the average large-scale and small scale fields at 
the surface of AU~Mic, the {\ems Stokes $V$ analysis} better accounts for the seasonal evolution of the large-scale field.  The difference in large-scale field strength 
between these two approaches reflects that some of the large-scale magnetic topology of AU~Mic (and in particular the axisymmetric component) may remain undetected through Stokes $V$ data only, as a result of the almost 
perpendicular orientation of the stellar rotation axis with respect to the line of sight (assuming that the orbital plane of the planets coincides with the equatorial plane of the star).  Collecting Stokes $Q$ and $U$ 
(in addition to Stokes $V$) data of AU~Mic would help in this respect, as these Zeeman signatures should be detectable in AU~Mic and could be used to efficiently differentiate between both magnetic configurations (see 
Fig.~\ref{fig:lqu}).  An accurate estimate of the large-scale dipole field of AU~Mic is also needed for studies of potential interactions between the host star  and its close-in planets \citep{Kavanagh21, Klein22}, 
{\emr or for space weather simulations in the system \citep{Carolan20, Alvarado22, Mesquita22}}.  

From the temporal evolution of Stokes $V$ profiles in 2020 and 2021, we retrieve the average amount of latitudinal differential rotation shearing the surface of AU~Mic, and find that the rotation rate at the equator \omeq\ 
and the difference in rotation rate between the equator and pole \dom\ are equal to 1.299 and 0.037~\rpd, corresponding to rotation periods at the equator and pole of 4.84 and 4.98~d, respectively, and to a latitudinal 
shear equal to about two thirds that at the surface of the Sun.  This amount of DR is typical to that found on partly convective low-mass stars \citep[e.g.,][]{Hebrard16}.  As the values derived at each epoch 
($\omeq=1.3006\pm0.0004$ and $\dom=0.0445\pm0.0014$~\rpd\ in 2020 and $\omeq=1.2970\pm0.0002$ and $\dom=0.0298\pm0.0013$~\rpd\ in 2021) differ by more than 3$\sigma$, we can conclude that differential rotation at the 
surface of AU~Mic is likely varying with time.  We nonetheless caution that the unusual amount of intrinsic variability observed for this star may have induced part of this discrepancy, making it essential to monitor the 
star over at least several months to minimize its impact and ensure we can derive reliable estimates of the DR parameters.  

RVs inferred with the LBL technique \citep{Artigau22} from the SPIRou spectra of AU~Mic processed with \texttt{APERO} \citep{Cook22} show an overall scatter of 31~\ms\ RMS over the full observing period, with some seasons 
(e.g., 2020) exhibiting a smaller than average dispersion (24~\ms).  This is a factor of about 4 smaller than the average scatter at visible wavelengths (127~\ms\ RMS) reported  by \citet{Zicher22}, illustrating how 
efficient SPIRou is to obtain precise RVs of active M dwarfs \citep[e.g.,][]{Carmona23}.  
These RV variations are strongly modulated with a period of $4.866\pm0.004$~d, marginally longer than that on which \Bl\ and <$B$> are 
modulated.  We further obtain that RVs are well correlated, neither with \Bl\ nor with <$B$> but with the first time derivative of <$B$> ($R=0.78$), similar to what was reported for Proxima \citep[using the FWHM of the 
cross-correlation function as a proxy for <$B$>,][]{Suarez20}.  Using this correlation to filter RVs from the activity jitter improves the situation but leaves a significant fraction (about 45\%) of the jitter.  A QP GPR fit 
to the RVs provides a more efficient filtering, leaving essentially no signal at the rotation period and its harmonics and aliases.  The typical evolution timescale of the activity jitter is found to be $136\pm21$~d, larger 
than that of \Bl\ but consistent with that of <$B$>.  Although the dLW activity index provided by LBL does not correlate well with <$B$>, we find that the strongest wPCA component $W_1(t)$ of all per-line dLW time series 
is strongly correlated with <$B$> in AU~Mic ($R=0.96$) and can thereby serve as a much better activity proxy than dLW itself (Cadieux et al., in prep.).  Filtering RVs using the first time derivative of $W_1$ yields results 
similar to (though not better than) those achieved with <$B$>.  {\emr Machine Learning is apparently a promising alternative to GPR for filtering activity \citep{Perger23}, with the advantage of being based on a physical background.} 

{\emr Modeling the RV signatures of transiting planets b and c \citep[with the orbital periods and TESS transit times set to the values quoted in][]{Szabo22} at the same time as the activity jitter yields semi-amplitudes of 
$K_b=4.1^{+1.8}_{-1.2}$~\ms\ and $K_c=4.0^{+1.7}_{-1.2}$~\ms\ for planets b and c, with a residual RV scatter of 11.1~\ms\ RMS.  This is consistent with previous results from optical data \citep[where the activity jitter is 
4$\times$ larger,][]{Zicher22} for planet b, but smaller by a factor of 2 for planet c.  We suspect that the difference mainly reflects residuals when filtering the strong activity jitter from optical data, that resulted in 
relatively large error bars on the semi-amplitudes of both planets (of 2.5~\ms, i.e., 1.5--2$\times$ larger than ours).  When looking at residuals in the periodogram of filtered RVs, we find excess power at a period of 33.4~d 
that hints at the presence of another planet in the system, dubbed candidate planet e.  Fitting orbital parameters of planet e along with those of planets b and c and the hyper-parameters of the GP describing activity yields 
$K_e=11.1^{+2.1}_{-1.7}$~\ms\ and $P_e=33.39\pm0.10$~d, and slightly larger semi-amplitudes for planets b and c ($K_b=4.5^{+1.7}_{-1.2}$~\ms, $K_c=4.8^{+1.6}_{-1.2}$~\ms), with the RV signatures detected at a 6.5$\sigma$ level 
for planet e and $\simeq$4$\sigma$ level for b and c.  We find that including candidate planet e gives a logarithmic Bayes factor of 16.3 for this model with respect to that featuring planets b and c only (plus activity), 
indicating that the detection is reliable, and induces a decrease in the residual RV scatter (now 10.4~\ms\ RMS).  Fitting eccentricities of all 3 planets yields values compatible with zero (with error bars of 0.04, 0.07 and 
0.09 for planets b, c and e respectively) along with a non-significant increase in marginal logarithmic likelihood, confirming that circular orbits for all 3 planets is the default model to be used here.  } 
We stress that the period of candidate planet e is close to a 1-yr alias of the Moon synodic period (32.1~d, see window function in Fig.~\ref{fig:per}) but distant from it by 13$\times$ the error bar on the derived 
period, ensuring that both peaks are distinct.  We nonetheless caution that this proximity is suspicious, hence why we choose to refer to planet e as a candidate planet at this stage.  We also note that candidate planet e is 
close to a 4:1 resonance with planet b, and to a 7:1 resonance with the rotation of the star.  

\begin{figure}
\centerline{\includegraphics[scale=0.3]{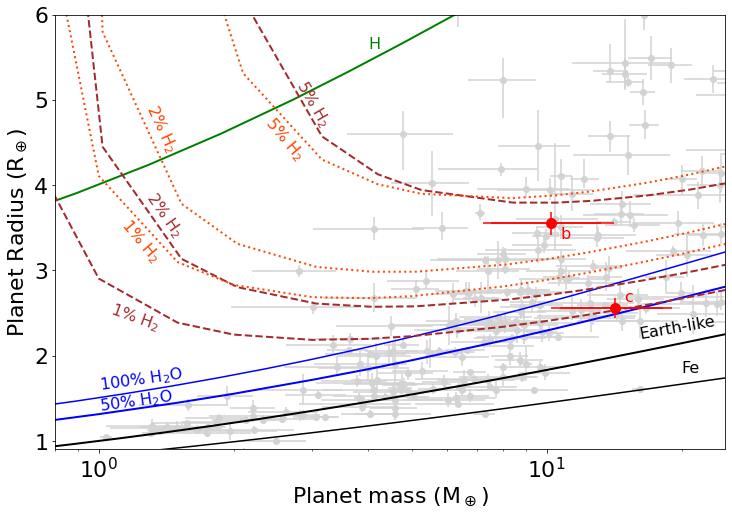}} 
\caption[]{Mass-radius diagram for exoplanets whose mass and radius are known with a relative precision better than 30\% (grey points).  AU~Mic b and c are shown with red filled circles and corresponding error bars, and 
are expected to contract with age \citep[e.g.,][]{Owen13}.  Theoretical models of \citet{Zeng16, Zeng19} corresponding to various inner planet structures / compositions are depicted with 
black (100\% iron and Earth-like), green (100\% H) and blue (100\% and 50\% H$_2$O envelope) full lines.  Models with a 1\%, 2\% and 5\% H$_2$ atmosphere with either an Earth-like (brown dashes) or a 50\% water (orange 
dashes) interior are also shown.  }
\label{fig:mrp}
\end{figure}

{\emr The derived semi-amplitudes yield masses of $10.2^{+3.9}_{-2.7}$, $14.2^{+4.8}_{-3.5}$ and $35.2^{+6.7}_{-5.4}$~\me\ for planets b, c and e, respectively.  The corresponding densities for 
planets b and c \citep[whose radii are $3.55\pm0.13$ and $2.56\pm0.12$~\re, where \re\ notes the Earth radius][]{Szabo22} are $1.26^{+0.68}_{-0.43}$ and $4.7^{+2.5}_{-1.6}$~\gpcc, i.e., about 3.7$\times$ 
larger for c than for b} whereas the opposite is usually observed in evolved multi-planet systems \citep[but not for all, e.g.,][]{Leleu21} as a result of the difference in equilibrium temperatures \citep[$593\pm21$ 
and $454\pm16$~K for b and c respectively,][$\simeq$370~K for candidate planet e using similar assumptions]{Martioli21}.  However, the observed density contrast also reflects the fact that both planets did not yet 
complete their contraction process, in particular planet b that is still rather inflated \citep{Zicher22}.  Both planets are not expected to evolve in the same way as a result of cooling, contraction 
or photo-evaporation of the H/He atmosphere, given the difference in mass and distance from the star.  In particular, given its higher equilibrium temperature {\emr possibly boosted by induction heating \citep{Kislyakova18}}, 
and its marginally lower mass, b is expected to increase its density more than c, hence reducing the density contrast between both.  Predicting the evolution of b and c from their observed positions in a radius versus 
mass diagram (see Fig.~\ref{fig:mrp}) requires  calculations with, e.g., the MESA models \citep{Owen13}, as in \citet{Zicher22}.  The inflation of planet b makes it an ideal target for future atmospheric characterization, 
{\emr especially given its strong level of irradiation that should also extend its atmosphere and provide opportunities of investigating deeper atmospheric layers \citep{GarciaMunoz21}}.  
{\emr Whereas planet e alone should place the planetary system of AU~Mic in the 'ordered' category recently defined by \citet{Mishra23}, planet d and e should put it in the 'mixed' class.  In both cases, AU~Mic will bring 
valuable observational constraints on young planetary system architectures and their expected evolution with time.} 

\begin{figure*}
\centerline{\includegraphics[scale=0.5,bb=40 10 800 300]{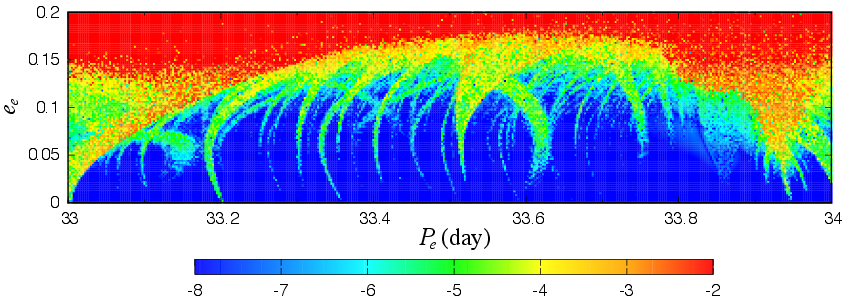}} 
\caption[]{{\emr Period versus eccentricity stability diagram for candidate planet e of the b+c+e AU~Mic system}. The phase space of the system is explored by varying the period (in d) and eccentricity of candidate planet e about the value 
derived in this work (Table~\ref{tab:pla}).  For each initial condition, the system is integrated over 5~kyr with the symplectic integrator SABAC4 \citep{Laskar01}.  A stability criterion is derived from the frequency 
analysis of the mean longitude \citep{Laskar93}.  The chaotic behaviour is quantified by the variation in the mean motion, with the color scale coding the decimal logarithm of the mean motion variation \citep{Correia10}. 
The red areas corresponds to unstable orbits, and the blue ones to orbits that are presumably stable on a Gyr timescale.} 
\label{fig:stab}
\end{figure*}

Assuming it is located in the same orbital plane as transiting planets b and c and given the impact parameters of b and c \citep[respectively equal to $0.17\pm0.11$ and $0.58\pm0.13$~\rstar,][]{Szabo22}, candidate planet e 
is probably not transiting or only through a grazing transit given the distance at which it orbits, i.e., $0.171\pm0.005$~au, as opposed to $0.0645\pm0.0013$ and $0.1101\pm0.0022$~au for planets b and c.  No transit at 
this period has been reported yet, although the limited monitoring windows of TESS for AU~Mic (28~d) would make it more difficult to detect (if it happens) than those of planets b and c.  
{\emr We note that the derived epoch of transit (or of conjunction) implies that if planet e is transiting, several transits should have been observed in TESS data (in sector \#1 at BJD $2458449.3\pm2.9$ and sector \#27 at 
BJD $2459050.3\pm1.1$) and potentially during CHEOPS \#2 and \#3 visits as well \citep[at BJDs $2459083.7\pm1.0$ and $2459117.1\pm0.9$,][]{Szabo21}. 
Since these data show no evidence of a transit, we conclude that candidate planet e is unlikely to transit. } 

By studying the stability of the b+c+e system, we find that candidate planet e can be stable on a Gyr timescale for the orbital periods and modest eccentricities allowed by the RV data (see Fig.~\ref{fig:stab}).  
The best fit orbital solution (see Table~\ref{tab:pla}) is not in exact resonance with any of the other planets, implying that planet e is not able to significantly perturb the orbits of planets b and c nor 
to generate the TTVs of up to $\pm$10~min reported for planet b \citep{Szabo22}.  By scanning a range of values about those derived in our study for the parameters of planet e (mass, period, eccentricity, and 
argument of periastron) to investigate whether a reasonable fit to the observed TESS, Spitzer and CHEOPS transit epochs was achievable, we obtain that no combination is able to reproduce transit data, even for 
orbital periods corresponding to a 4:1 resonance with planet b.  

{\emr We also investigated whether candidate planet d, suggested to be orbiting between b and c and potentially causing the reported TTVs \citep{Wittrock23}, is detectable in our data.  
We find no clear evidence of its presence and derive a 3$\sigma$ upper limit on the amplitude of its RV signature ($K_d<4.4$~\ms) and on its minimum mass ($M_d<11.4$~\me), consistent with the 
prediction \citep[$\simeq$1~\me,][]{Wittrock23}.  
Besides, we confirm that b+c+d is stable for the 12.74-d period and low-eccentricity orbit of planet d suggested by \citet[][]{Wittrock23}, whereas b+c+d+e, though less stable, should still be able to survive 
on long timescales (see Fig.~\ref{fig:stab2} in supplementary material).}  {\ems We also verified that the simulated TTVs are much more consistent with the transit timings of planets b and c once candidate planet d is included 
\citep[see Fig.~\ref{fig:ttv} in supplementary material, in agreement with][]{Wittrock23}}.  

Finally, we studied the \pab\ and \hei\ lines and in particular the temporal modulation of their equivalent widths (EWVs) over our 4 observing seasons.  We find that these lines exhibit clear modulation with periods 
that are consistent with the rotation period of the star, despite showing both a significant level of intrinsic variability.  EWVs are clearer for \pab\ (than for \hei) and are found to correlate reasonably well, 
for this line, with our estimates of the small-scale field <$B$>.  Both \pab\ EWVs and <$B$> slowly decrease with time over the 3~yr of our observing campaign, suggesting that AU~Mic may be progressing towards activity 
minimum along its putative cycle \citep{Ibanez19}.  As opposed to \citet{ Klein22}, we find no clear evidence that either line is modulated at the period of the system planets.  The low-significance peaks (FAP of 
$\simeq$10\%) observed close to the orbital periods of planets b and c in the periodogram of \hei\ EWVs are similar in strength to many others and cannot reliably be considered as a sign of, e.g., star-planet 
interaction between AU~Mic and its inner planets.  

We will pursue our spectropolarimetric monitoring of AU~Mic with SPIRou for another 2~yr within the context of the new SPICE Large Programme, that was allocated 175 nights of CFHT time from late 2022 to mid 2024, and 
whose aim is to consolidate and enhance the results of the SLS.  The first goal is to firmly ascertain the existence of candidate planets e and possibly d, and improve at the same time the precision on the mass estimates 
of all  planets, a rather tricky task  given the extreme activity level of AU~Mic, even in the nIR where the RV activity jitter is 4$\times$ smaller than at visible wavelengths.  With its system of young planets (2 of 
which transiting) that are still likely evolving with time, AU~Mic is an ideal laboratory to test and constrain models of planet formation and evolution, and to investigate the atmospheric composition of young inflated 
planets.  Besides, we also aim at pursuing the magnetic monitoring of AU~Mic on a longer timescale to investigate whether the poloidal and toroidal components of large-scale field vary in a cyclic way 
\citep[e.g.,][]{Ibanez19}, switching sign as they do on the Sun, and to document the changes in the small-scale field as the large-scale field evolves with time.  Given that no clear sign switch was observed yet in 
either field component, the present data suggest that, if there is a cycle, the period is at least 5 years.  We stress that having access to Stokes $Q$ and $U$ observations of AU~Mic with SPIRou for at least a partial 
season, and thereby complement the Stokes $V$ and $I$ data to be collected with SPICE, would further enhance our ability to consistently model the magnetic topology and verify the assumptions on which the results presented 
in this paper rely.  More generally, multi-wavelength, multi-instrument monitoring campaigns of AU~Mic, involving in particular precision photometry from space (TESS, CHEOPS and later-on PLATO), optical and 
nIR high-resolution spectropolarimetry and velocimetry (e.g., ESPaDOnS, SPIRou, CRIRES+ for longer wavelengths), and low-resolution nIR spectroscopy from space with the JWST, especially during transits, would be a 
must to characterize in detail this young multi-planet system of our immediate neighbourhood.  

\section*{Acknowledgements}
This project received funds from the European Research Council (ERC) under the H2020 research \& innovation program (grant agreements 
\#740651 NewWorlds, \#742095 SPIDI, \#865624 GPRV, \#716155 SACCRED, \#817540 ASTROFLOW), the Agence Nationale pour la Recherche (ANR, project ANR-18-CE31-0019 SPlaSH) and 
the Investissements d’Avenir program (project ANR-15-IDEX-02).  
SHPA and EM acknowledge funding from FAPEMIG, CNPq and CAPES, while ACMC acknowledges project funds UIDB/04564/2020, UIDP/04564/2020, and PTDC/FIS-AST/7002/2020. 
We thank F.~M\'enard and an anonymous referee for valuable comments on an earlier version of the manuscript.  

Our study is based on data obtained at the CFHT, operated by the CNRC (Canada), INSU/CNRS (France) and the University of Hawaii.
The authors wish to recognise and acknowledge the very significant cultural role and reverence that the summit of Maunakea has always had 
within the indigenous Hawaiian community.  We are most fortunate to have the opportunity to conduct observations from this mountain.
This work also benefited from the SIMBAD CDS database at URL {\tt http://simbad.u-strasbg.fr/simbad} and the ADS system at URL {\tt https://ui.adsabs.harvard.edu}.

\section*{Data availability}  Most data underlying this paper are part of the SLS, and will be publicly available from the Canadian Astronomy Data
Center by February 2024, while the PI and DDT data are already public.   

\bibliography{aumic-arxiv}

\begin{thebibliography}{}
\makeatletter
\relax
\def\mn@urlcharsother{\let\do\@makeother \do\$\do\&\do\#\do\^\do\_\do\%\do\~}
\def\mn@doi{\begingroup\mn@urlcharsother \@ifnextchar [ {\mn@doi@}
  {\mn@doi@[]}}
\def\mn@doi@[#1]#2{\def\@tempa{#1}\ifx\@tempa\@empty \href
  {http://dx.doi.org/#2} {doi:#2}\else \href {http://dx.doi.org/#2} {#1}\fi
  \endgroup}
\def\mn@eprint#1#2{\mn@eprint@#1:#2::\@nil}
\def\mn@eprint@arXiv#1{\href {http://arxiv.org/abs/#1} {{\tt arXiv:#1}}}
\def\mn@eprint@dblp#1{\href {http://dblp.uni-trier.de/rec/bibtex/#1.xml}
  {dblp:#1}}
\def\mn@eprint@#1:#2:#3:#4\@nil{\def\@tempa {#1}\def\@tempb {#2}\def\@tempc
  {#3}\ifx \@tempc \@empty \let \@tempc \@tempb \let \@tempb \@tempa \fi \ifx
  \@tempb \@empty \def\@tempb {arXiv}\fi \@ifundefined
  {mn@eprint@\@tempb}{\@tempb:\@tempc}{\expandafter \expandafter \csname
  mn@eprint@\@tempb\endcsname \expandafter{\@tempc}}}

\bibitem[\protect\citeauthoryear{{Afram} \& {Berdyugina}}{{Afram} \&
  {Berdyugina}}{2019}]{Afram19}
{Afram} N.,  {Berdyugina} S.~V.,  2019, \mn@doi [\aap]
  {10.1051/0004-6361/201935793}, \href
  {https://ui.adsabs.harvard.edu/abs/2019A&A...629A..83A} {629, A83}

\bibitem[\protect\citeauthoryear{{Alvarado-G{\'o}mez}
  et~al.,}{{Alvarado-G{\'o}mez} et~al.}{2022}]{Alvarado22}
{Alvarado-G{\'o}mez} J.~D.,  et~al., 2022, \mn@doi [\apj]
  {10.3847/1538-4357/ac54b8}, \href
  {https://ui.adsabs.harvard.edu/abs/2022ApJ...928..147A} {928, 147}

\bibitem[\protect\citeauthoryear{{Artigau} et~al.,}{{Artigau}
  et~al.}{2022}]{Artigau22}
{Artigau} {\'E}.,  et~al., 2022, \mn@doi [\aj] {10.3847/1538-3881/ac7ce6},
  \href {https://ui.adsabs.harvard.edu/abs/2022AJ....164...84A} {164, 84}

\bibitem[\protect\citeauthoryear{{Baraffe}, {Homeier}, {Allard}  \&
  {Chabrier}}{{Baraffe} et~al.}{2015}]{Baraffe15}
{Baraffe} I.,  {Homeier} D.,  {Allard} F.,   {Chabrier} G.,  2015, \mn@doi
  [\aap] {10.1051/0004-6361/201425481}, \href
  {http://adsabs.harvard.edu/abs/2015A%26A...577A..42B} {577, A42}

\bibitem[\protect\citeauthoryear{{Blinova}, {Romanova}  \&
  {Lovelace}}{{Blinova} et~al.}{2016}]{Blinova16}
{Blinova} A.~A.,  {Romanova} M.~M.,   {Lovelace} R.~V.~E.,  2016, \mn@doi
  [\mnras] {10.1093/mnras/stw786}, \href
  {http://adsabs.harvard.edu/abs/2016MNRAS.459.2354B} {459, 2354}

\bibitem[\protect\citeauthoryear{{Boccaletti} et~al.,}{{Boccaletti}
  et~al.}{2015}]{Boccaletti15}
{Boccaletti} A.,  et~al., 2015, \mn@doi [\nat] {10.1038/nature15705}, \href
  {https://ui.adsabs.harvard.edu/abs/2015Natur.526..230B} {526, 230}

\bibitem[\protect\citeauthoryear{{Boccaletti} et~al.,}{{Boccaletti}
  et~al.}{2018}]{Boccaletti18}
{Boccaletti} A.,  et~al., 2018, \mn@doi [\aap] {10.1051/0004-6361/201732462},
  \href {https://ui.adsabs.harvard.edu/abs/2018A&A...614A..52B} {614, A52}

\bibitem[\protect\citeauthoryear{{Cale} et~al.,}{{Cale} et~al.}{2021}]{Cale21}
{Cale} B.~L.,  et~al., 2021, \mn@doi [\aj] {10.3847/1538-3881/ac2c80}, \href
  {https://ui.adsabs.harvard.edu/abs/2021AJ....162..295C} {162, 295}

\bibitem[\protect\citeauthoryear{{Carmona} et~al.,}{{Carmona}
  et~al.}{2023}]{Carmona23}
{Carmona} A.,  et~al., 2023, arXiv e-prints, \href
  {https://ui.adsabs.harvard.edu/abs/2023arXiv230316712C} {p. arXiv:2303.16712}

\bibitem[\protect\citeauthoryear{{Carolan}, {Vidotto}, {Plavchan}, {Villarreal
  D'Angelo}  \& {Hazra}}{{Carolan} et~al.}{2020}]{Carolan20}
{Carolan} S.,  {Vidotto} A.~A.,  {Plavchan} P.,  {Villarreal D'Angelo} C.,
  {Hazra} G.,  2020, \mn@doi [\mnras] {10.1093/mnrasl/slaa127}, \href
  {https://ui.adsabs.harvard.edu/abs/2020MNRAS.498L..53C} {498, L53}

\bibitem[\protect\citeauthoryear{{Chabrier}, {Gallardo}  \&
  {Baraffe}}{{Chabrier} et~al.}{2007}]{Chabrier07}
{Chabrier} G.,  {Gallardo} J.,   {Baraffe} I.,  2007, \mn@doi [\aap]
  {10.1051/0004-6361:20077702}, \href
  {https://ui.adsabs.harvard.edu/abs/2007A&A...472L..17C} {472, L17}

\bibitem[\protect\citeauthoryear{{Chib} \& {Jeliazkov}}{{Chib} \&
  {Jeliazkov}}{2001}]{Chib01}
{Chib} S.,  {Jeliazkov} I.,  2001, Journal of the American Statistical
  Association, \href {http://adsabs.harvard.edu/abs/2016arXiv160800962B} {96,
  270}

\bibitem[\protect\citeauthoryear{{Cifuentes} et~al.,}{{Cifuentes}
  et~al.}{2020}]{Cifuentes20}
{Cifuentes} C.,  et~al., 2020, \mn@doi [\aap] {10.1051/0004-6361/202038295},
  \href {https://ui.adsabs.harvard.edu/abs/2020A&A...642A.115C} {642, A115}

\bibitem[\protect\citeauthoryear{{Cook} et~al.,}{{Cook} et~al.}{2022}]{Cook22}
{Cook} N.~J.,  et~al., 2022, \mn@doi [\pasp] {10.1088/1538-3873/ac9e74}, \href
  {https://ui.adsabs.harvard.edu/abs/2022PASP..134k4509C} {134, 114509}

\bibitem[\protect\citeauthoryear{{Correia} et~al.,}{{Correia}
  et~al.}{2010}]{Correia10}
{Correia} A.~C.~M.,  et~al., 2010, \mn@doi [\aap]
  {10.1051/0004-6361/200912700}, \href
  {https://ui.adsabs.harvard.edu/abs/2010A&A...511A..21C} {511, A21}

\bibitem[\protect\citeauthoryear{{Cristofari} et~al.,}{{Cristofari}
  et~al.}{2022a}]{Cristofari22a}
{Cristofari} P.~I.,  et~al., 2022a, \mn@doi [\mnras] {10.1093/mnras/stab3679},
  \href {https://ui.adsabs.harvard.edu/abs/2022MNRAS.511.1893C} {511, 1893}

\bibitem[\protect\citeauthoryear{{Cristofari} et~al.,}{{Cristofari}
  et~al.}{2022b}]{Cristofari22b}
{Cristofari} P.~I.,  et~al., 2022b, \mn@doi [\mnras] {10.1093/mnras/stac2364},
  \href {https://ui.adsabs.harvard.edu/abs/2022MNRAS.516.3802C} {516, 3802}

\bibitem[\protect\citeauthoryear{{Cristofari} et~al.,}{{Cristofari}
  et~al.}{2023}]{Cristofari23}
{Cristofari} P.~I.,  et~al., 2023, \mn@doi [\mnras] {10.1093/mnras/stad865},
  \href {https://ui.adsabs.harvard.edu/abs/2023MNRAS.tmp..815C} {}

\bibitem[\protect\citeauthoryear{{Cutri} et~al.,}{{Cutri}
  et~al.}{2003}]{Cutri03}
{Cutri} R.~M.,  et~al., 2003, VizieR Online Data Catalog, \href
  {https://ui.adsabs.harvard.edu/abs/2003yCat.2246....0C} {p. II/246}

\bibitem[\protect\citeauthoryear{{David}, {Petigura}, {Luger},
  {Foreman-Mackey}, {Livingston}, {Mamajek}  \& {Hillenbrand}}{{David}
  et~al.}{2019}]{David19}
{David} T.~J.,  {Petigura} E.~A.,  {Luger} R.,  {Foreman-Mackey} D.,
  {Livingston} J.~H.,  {Mamajek} E.~E.,   {Hillenbrand} L.~A.,  2019, \mn@doi
  [\apjl] {10.3847/2041-8213/ab4c99}, \href
  {https://ui.adsabs.harvard.edu/abs/2019ApJ...885L..12D} {885, L12}

\bibitem[\protect\citeauthoryear{{Delchambre}}{{Delchambre}}{2015}]{Delchambre15}
{Delchambre} L.,  2015, \mn@doi [\mnras] {10.1093/mnras/stu2219}, \href
  {https://ui.adsabs.harvard.edu/abs/2015MNRAS.446.3545D} {446, 3545}

\bibitem[\protect\citeauthoryear{{Donati} \& {Landstreet}}{{Donati} \&
  {Landstreet}}{2009}]{Donati09}
{Donati} J.,  {Landstreet} J.~D.,  2009, \mn@doi [\araa]
  {10.1146/annurev-astro-082708-101833}, \href
  {http://adsabs.harvard.edu/abs/2009ARA%26A..47..333D} {47, 333}

\bibitem[\protect\citeauthoryear{{Donati}, {Semel}, {Carter}, {Rees}  \&
  {Collier Cameron}}{{Donati} et~al.}{1997}]{Donati97b}
{Donati} J.-F.,  {Semel} M.,  {Carter} B.~D.,  {Rees} D.~E.,   {Collier
  Cameron} A.,  1997, \mnras, \href
  {http://adsabs.harvard.edu/abs/1997MNRAS.291..658D} {291, 658}

\bibitem[\protect\citeauthoryear{{Donati}, {Collier Cameron}  \&
  {Petit}}{{Donati} et~al.}{2003}]{Donati03b}
{Donati} J.-F.,  {Collier Cameron} A.,   {Petit} P.,  2003, \mnras, 345, 1187

\bibitem[\protect\citeauthoryear{{Donati} et~al.,}{{Donati}
  et~al.}{2006}]{Donati06b}
{Donati} J.-F.,  et~al., 2006, \mn@doi [\mnras]
  {10.1111/j.1365-2966.2006.10558.x}, \href
  {http://adsabs.harvard.edu/abs/2006MNRAS.370..629D} {370, 629}

\bibitem[\protect\citeauthoryear{{Donati} et~al.,}{{Donati}
  et~al.}{2020}]{Donati20}
{Donati} J.~F.,  et~al., 2020, \mn@doi [\mnras] {10.1093/mnras/staa2569}, \href
  {https://ui.adsabs.harvard.edu/abs/2020MNRAS.498.5684D} {498, 5684}

\bibitem[\protect\citeauthoryear{{Dotter}, {Chaboyer}, {Jevremovi{\'c}},
  {Kostov}, {Baron}  \& {Ferguson}}{{Dotter} et~al.}{2008}]{Dotter08}
{Dotter} A.,  {Chaboyer} B.,  {Jevremovi{\'c}} D.,  {Kostov} V.,  {Baron} E.,
  {Ferguson} J.~W.,  2008, \mn@doi [\apjs] {10.1086/589654}, \href
  {https://ui.adsabs.harvard.edu/abs/2008ApJS..178...89D} {178, 89}

\bibitem[\protect\citeauthoryear{{Feiden}}{{Feiden}}{2016}]{Feiden16}
{Feiden} G.~A.,  2016, \mn@doi [\aap]
  {10.1051/0004-6361/20152761310.48550/arXiv.1604.08036}, \href
  {https://ui.adsabs.harvard.edu/abs/2016A&A...593A..99F} {593, A99}

\bibitem[\protect\citeauthoryear{{Finociety} \& {Donati}}{{Finociety} \&
  {Donati}}{2022}]{Finociety22}
{Finociety} B.,  {Donati} J.~F.,  2022, \mn@doi [\mnras]
  {10.1093/mnras/stac2682}, \href
  {https://ui.adsabs.harvard.edu/abs/2022MNRAS.516.5887F} {516, 5887}

\bibitem[\protect\citeauthoryear{{Finociety} et~al.,}{{Finociety}
  et~al.}{2021}]{Finociety21}
{Finociety} B.,  et~al., 2021, \mn@doi [\mnras] {10.1093/mnras/stab2778}, \href
  {https://ui.adsabs.harvard.edu/abs/2021MNRAS.508.3427F} {508, 3427}

\bibitem[\protect\citeauthoryear{{Finociety} et~al.,}{{Finociety}
  et~al.}{2023}]{Finociety23}
{Finociety} B.,  et~al., 2023, \mn@doi [\mnras] {10.1093/mnras/stad267}, \href
  {https://ui.adsabs.harvard.edu/abs/2023MNRAS.520.3049F} {520, 3049}

\bibitem[\protect\citeauthoryear{{Gaia Collaboration} et~al.,}{{Gaia
  Collaboration} et~al.}{2021}]{Gaia21}
{Gaia Collaboration} et~al., 2021, \mn@doi [\aap]
  {10.1051/0004-6361/202039657}, \href
  {https://ui.adsabs.harvard.edu/abs/2021A&A...649A...1G} {649, A1}

\bibitem[\protect\citeauthoryear{{Gaidos} et~al.,}{{Gaidos}
  et~al.}{2014}]{Gaidos14}
{Gaidos} E.,  et~al., 2014, \mn@doi [\mnras] {10.1093/mnras/stu1313}, \href
  {https://ui.adsabs.harvard.edu/abs/2014MNRAS.443.2561G} {443, 2561}

\bibitem[\protect\citeauthoryear{{Gallenne}, {Desgrange}, {Milli},
  {Sanchez-Bermudez}, {Chauvin}, {Kraus}, {Girard}  \& {Boccaletti}}{{Gallenne}
  et~al.}{2022}]{Gallenne22}
{Gallenne} A.,  {Desgrange} C.,  {Milli} J.,  {Sanchez-Bermudez} J.,  {Chauvin}
  G.,  {Kraus} S.,  {Girard} J.~H.,   {Boccaletti} A.,  2022, \mn@doi [\aap]
  {10.1051/0004-6361/202244226}, \href
  {https://ui.adsabs.harvard.edu/abs/2022A&A...665A..41G} {665, A41}

\bibitem[\protect\citeauthoryear{{Garc{\'\i}a Mu{\~n}oz}, {Fossati},
  {Youngblood}, {Nettelmann}, {Gandolfi}, {Cabrera}  \& {Rauer}}{{Garc{\'\i}a
  Mu{\~n}oz} et~al.}{2021}]{GarciaMunoz21}
{Garc{\'\i}a Mu{\~n}oz} A.,  {Fossati} L.,  {Youngblood} A.,  {Nettelmann} N.,
  {Gandolfi} D.,  {Cabrera} J.,   {Rauer} H.,  2021, \mn@doi [\apjl]
  {10.3847/2041-8213/abd9b8}, \href
  {https://ui.adsabs.harvard.edu/abs/2021ApJ...907L..36G} {907, L36}

\bibitem[\protect\citeauthoryear{{Haywood} et~al.,}{{Haywood}
  et~al.}{2014}]{Haywood14}
{Haywood} R.~D.,  et~al., 2014, \mn@doi [\mnras] {10.1093/mnras/stu1320}, \href
  {http://adsabs.harvard.edu/abs/2014MNRAS.443.2517H} {443, 2517}

\bibitem[\protect\citeauthoryear{{Haywood} et~al.,}{{Haywood}
  et~al.}{2022}]{Haywood22}
{Haywood} R.~D.,  et~al., 2022, \mn@doi [\apj] {10.3847/1538-4357/ac7c12},
  \href {https://ui.adsabs.harvard.edu/abs/2022ApJ...935....6H} {935, 6}

\bibitem[\protect\citeauthoryear{{H{\'e}brard}, {Donati}, {Delfosse}, {Morin},
  {Moutou}  \& {Boisse}}{{H{\'e}brard} et~al.}{2016}]{Hebrard16}
{H{\'e}brard} {\'E}.~M.,  {Donati} J.~F.,  {Delfosse} X.,  {Morin} J.,
  {Moutou} C.,   {Boisse} I.,  2016, \mn@doi [\mnras] {10.1093/mnras/stw1346},
  \href {https://ui.adsabs.harvard.edu/abs/2016MNRAS.461.1465H} {461, 1465}

\bibitem[\protect\citeauthoryear{{Hirano} et~al.,}{{Hirano}
  et~al.}{2020}]{Hirano20}
{Hirano} T.,  et~al., 2020, \mn@doi [\apjl] {10.3847/2041-8213/aba6eb}, \href
  {https://ui.adsabs.harvard.edu/abs/2020ApJ...899L..13H} {899, L13}

\bibitem[\protect\citeauthoryear{{Iba{\~n}ez Bustos}, {Buccino}, {Flores},
  {Martinez}, {Maizel}, {Messina}  \& {Mauas}}{{Iba{\~n}ez Bustos}
  et~al.}{2019}]{Ibanez19}
{Iba{\~n}ez Bustos} R.~V.,  {Buccino} A.~P.,  {Flores} M.,  {Martinez} C.~I.,
  {Maizel} D.,  {Messina} S.,   {Mauas} P.~J.~D.,  2019, \mn@doi [\mnras]
  {10.1093/mnras/sty3147}, \href
  {https://ui.adsabs.harvard.edu/abs/2019MNRAS.483.1159I} {483, 1159}

\bibitem[\protect\citeauthoryear{{Kalas}, {Liu}  \& {Matthews}}{{Kalas}
  et~al.}{2004}]{Kalas04}
{Kalas} P.,  {Liu} M.~C.,   {Matthews} B.~C.,  2004, \mn@doi [Science]
  {10.1126/science.1093420}, \href
  {https://ui.adsabs.harvard.edu/abs/2004Sci...303.1990K} {303, 1990}

\bibitem[\protect\citeauthoryear{{Kavanagh}, {Vidotto}, {Klein}, {Jardine},
  {Donati}  \& {{\'O} Fionnag{\'a}in}}{{Kavanagh} et~al.}{2021}]{Kavanagh21}
{Kavanagh} R.~D.,  {Vidotto} A.~A.,  {Klein} B.,  {Jardine} M.~M.,  {Donati}
  J.-F.,   {{\'O} Fionnag{\'a}in} D.,  2021, \mn@doi [\mnras]
  {10.1093/mnras/stab929}, \href
  {https://ui.adsabs.harvard.edu/abs/2021MNRAS.504.1511K} {504, 1511}

\bibitem[\protect\citeauthoryear{{Kiraga}}{{Kiraga}}{2012}]{Kiraga12}
{Kiraga} M.,  2012, \actaa, \href
  {https://ui.adsabs.harvard.edu/abs/2012AcA....62...67K} {62, 67}

\bibitem[\protect\citeauthoryear{{Kislyakova}, {Fossati}, {Johnstone}, {Noack},
  {L{\"u}ftinger}, {Zaitsev}  \& {Lammer}}{{Kislyakova}
  et~al.}{2018}]{Kislyakova18}
{Kislyakova} K.~G.,  {Fossati} L.,  {Johnstone} C.~P.,  {Noack} L.,
  {L{\"u}ftinger} T.,  {Zaitsev} V.~V.,   {Lammer} H.,  2018, \mn@doi [\apj]
  {10.3847/1538-4357/aabae4}, \href
  {https://ui.adsabs.harvard.edu/abs/2018ApJ...858..105K} {858, 105}

\bibitem[\protect\citeauthoryear{{Klein} et~al.,}{{Klein}
  et~al.}{2021}]{Klein21}
{Klein} B.,  et~al., 2021, \mn@doi [\mnras] {10.1093/mnras/staa3702}, \href
  {https://ui.adsabs.harvard.edu/abs/2021MNRAS.502..188K} {502, 188}

\bibitem[\protect\citeauthoryear{{Klein} et~al.,}{{Klein}
  et~al.}{2022}]{Klein22}
{Klein} B.,  et~al., 2022, \mn@doi [\mnras] {10.1093/mnras/stac761}, \href
  {https://ui.adsabs.harvard.edu/abs/2022MNRAS.512.5067K} {512, 5067}

\bibitem[\protect\citeauthoryear{{Kochukhov}}{{Kochukhov}}{2021}]{Kochukhov21}
{Kochukhov} O.,  2021, \mn@doi [\aapr] {10.1007/s00159-020-00130-3}, \href
  {https://ui.adsabs.harvard.edu/abs/2021A&ARv..29....1K} {29, 1}

\bibitem[\protect\citeauthoryear{{Kochukhov} \& {Reiners}}{{Kochukhov} \&
  {Reiners}}{2020}]{Kochukhov20}
{Kochukhov} O.,  {Reiners} A.,  2020, \mn@doi [\apj]
  {10.3847/1538-4357/abb2a2}, \href
  {https://ui.adsabs.harvard.edu/abs/2020ApJ...902...43K} {902, 43}

\bibitem[\protect\citeauthoryear{{Landi degl'Innocenti} \& {Landolfi}}{{Landi
  degl'Innocenti} \& {Landolfi}}{2004}]{Landi04}
{Landi degl'Innocenti} E.,  {Landolfi} M.,  2004, {Polarisation in spectral
  lines}.
Dordrecht/Boston/London: Kluwer Academic Publishers

\bibitem[\protect\citeauthoryear{{Laskar}}{{Laskar}}{1993}]{Laskar93}
{Laskar} J.,  1993, \mn@doi [Physica D Nonlinear Phenomena]
  {10.1016/0167-2789(93)90210-R}, \href
  {https://ui.adsabs.harvard.edu/abs/1993PhyD...67..257L} {67, 257}

\bibitem[\protect\citeauthoryear{{Laskar} \& {Robutel}}{{Laskar} \&
  {Robutel}}{2001}]{Laskar01}
{Laskar} J.,  {Robutel} P.,  2001, \mn@doi [Celestial Mechanics and Dynamical
  Astronomy] {10.48550/arXiv.astro-ph/0005074}, \href
  {https://ui.adsabs.harvard.edu/abs/2001CeMDA..80...39L} {80, 39}

\bibitem[\protect\citeauthoryear{{Lehmann} \& {Donati}}{{Lehmann} \&
  {Donati}}{2022}]{Lehmann22}
{Lehmann} L.~T.,  {Donati} J.~F.,  2022, \mn@doi [\mnras]
  {10.1093/mnras/stac1519}, \href
  {https://ui.adsabs.harvard.edu/abs/2022MNRAS.514.2333L} {514, 2333}

\bibitem[\protect\citeauthoryear{{Leleu} et~al.,}{{Leleu}
  et~al.}{2021}]{Leleu21}
{Leleu} A.,  et~al., 2021, \mn@doi [\aap] {10.1051/0004-6361/202039767}, \href
  {https://ui.adsabs.harvard.edu/abs/2021A&A...649A..26L} {649, A26}

\bibitem[\protect\citeauthoryear{{L{\'o}pez-Valdivia}
  et~al.,}{{L{\'o}pez-Valdivia} et~al.}{2021}]{Lopez-Valdivia21}
{L{\'o}pez-Valdivia} R.,  et~al., 2021, \mn@doi [\apj]
  {10.3847/1538-4357/ac1a7b}, \href
  {https://ui.adsabs.harvard.edu/abs/2021ApJ...921...53L} {921, 53}

\bibitem[\protect\citeauthoryear{{Maldonado} et~al.,}{{Maldonado}
  et~al.}{2020}]{Maldonado20}
{Maldonado} J.,  et~al., 2020, \mn@doi [\aap] {10.1051/0004-6361/202039478},
  \href {https://ui.adsabs.harvard.edu/abs/2020A&A...644A..68M} {644, A68}

\bibitem[\protect\citeauthoryear{{Malo}, {Doyon}, {Feiden}, {Albert},
  {Lafreni{\`e}re}, {Artigau}, {Gagn{\'e}}  \& {Riedel}}{{Malo}
  et~al.}{2014}]{Malo14}
{Malo} L.,  {Doyon} R.,  {Feiden} G.~A.,  {Albert} L.,  {Lafreni{\`e}re} D.,
  {Artigau} {\'E}.,  {Gagn{\'e}} J.,   {Riedel} A.,  2014, \mn@doi [\apj]
  {10.1088/0004-637X/792/1/37}, \href
  {https://ui.adsabs.harvard.edu/abs/2014ApJ...792...37M} {792, 37}

\bibitem[\protect\citeauthoryear{{Mamajek} \& {Bell}}{{Mamajek} \&
  {Bell}}{2014}]{Mamajek14}
{Mamajek} E.~E.,  {Bell} C. P.~M.,  2014, \mn@doi [\mnras]
  {10.1093/mnras/stu1894}, \href
  {https://ui.adsabs.harvard.edu/abs/2014MNRAS.445.2169M} {445, 2169}

\bibitem[\protect\citeauthoryear{{Martioli} et~al.,}{{Martioli}
  et~al.}{2020}]{Martioli20}
{Martioli} E.,  et~al., 2020, \mn@doi [\aap] {10.1051/0004-6361/202038695},
  \href {https://ui.adsabs.harvard.edu/abs/2020A&A...641L...1M} {641, L1}

\bibitem[\protect\citeauthoryear{{Martioli}, {H{\'e}brard}, {Correia}, {Laskar}
   \& {Lecavelier des Etangs}}{{Martioli} et~al.}{2021}]{Martioli21}
{Martioli} E.,  {H{\'e}brard} G.,  {Correia} A.~C.~M.,  {Laskar} J.,
  {Lecavelier des Etangs} A.,  2021, \mn@doi [\aap]
  {10.1051/0004-6361/202040235}, \href
  {https://ui.adsabs.harvard.edu/abs/2021A&A...649A.177M} {649, A177}

\bibitem[\protect\citeauthoryear{{Mesquita}, {Rodgers-Lee}, {Vidotto}  \&
  {Kavanagh}}{{Mesquita} et~al.}{2022}]{Mesquita22}
{Mesquita} A.~L.,  {Rodgers-Lee} D.,  {Vidotto} A.~A.,   {Kavanagh} R.~D.,
  2022, \mn@doi [\mnras] {10.1093/mnras/stac1624}, \href
  {https://ui.adsabs.harvard.edu/abs/2022MNRAS.515.1218M} {515, 1218}

\bibitem[\protect\citeauthoryear{{Miret-Roig} et~al.,}{{Miret-Roig}
  et~al.}{2020}]{Miret20}
{Miret-Roig} N.,  et~al., 2020, \mn@doi [\aap] {10.1051/0004-6361/202038765},
  \href {https://ui.adsabs.harvard.edu/abs/2020A&A...642A.179M} {642, A179}

\bibitem[\protect\citeauthoryear{{Mishra}, {Alibert}, {Udry}  \&
  {Mordasini}}{{Mishra} et~al.}{2023}]{Mishra23}
{Mishra} L.,  {Alibert} Y.,  {Udry} S.,   {Mordasini} C.,  2023, \mn@doi [\aap]
  {10.1051/0004-6361/202243751}, \href
  {https://ui.adsabs.harvard.edu/abs/2023A&A...670A..68M} {670, A68}

\bibitem[\protect\citeauthoryear{{Morales}, {Gallardo}, {Ribas}, {Jordi},
  {Baraffe}  \& {Chabrier}}{{Morales} et~al.}{2010}]{Morales10}
{Morales} J.~C.,  {Gallardo} J.,  {Ribas} I.,  {Jordi} C.,  {Baraffe} I.,
  {Chabrier} G.,  2010, \mn@doi [\apj]
  {10.1088/0004-637X/718/1/50210.48550/arXiv.1005.5720}, \href
  {https://ui.adsabs.harvard.edu/abs/2010ApJ...718..502M} {718, 502}

\bibitem[\protect\citeauthoryear{{Morin} et~al.,}{{Morin}
  et~al.}{2008a}]{Morin08a}
{Morin} J.,  et~al., 2008a, \mnras, 384, 77

\bibitem[\protect\citeauthoryear{{Morin} et~al.,}{{Morin}
  et~al.}{2008b}]{Morin08b}
{Morin} J.,  et~al., 2008b, \mn@doi [\mnras]
  {10.1111/j.1365-2966.2008.13809.x}, \href
  {http://adsabs.harvard.edu/abs/2008MNRAS.390..567M} {390, 567}

\bibitem[\protect\citeauthoryear{{Morin}, {Donati}, {Petit}, {Delfosse},
  {Forveille}  \& {Jardine}}{{Morin} et~al.}{2010}]{Morin10}
{Morin} J.,  {Donati} J.,  {Petit} P.,  {Delfosse} X.,  {Forveille} T.,
  {Jardine} M.~M.,  2010, \mn@doi [\mnras] {10.1111/j.1365-2966.2010.17101.x},
  \href {http://adsabs.harvard.edu/abs/2010MNRAS.407.2269M} {407, 2269}

\bibitem[\protect\citeauthoryear{{Morrell} \& {Naylor}}{{Morrell} \&
  {Naylor}}{2019}]{Morrell19}
{Morrell} S.,  {Naylor} T.,  2019, \mn@doi [\mnras] {10.1093/mnras/stz2242},
  \href {https://ui.adsabs.harvard.edu/abs/2019MNRAS.489.2615M} {489, 2615}

\bibitem[\protect\citeauthoryear{{Owen} \& {Wu}}{{Owen} \& {Wu}}{2013}]{Owen13}
{Owen} J.~E.,  {Wu} Y.,  2013, \mn@doi [\apj] {10.1088/0004-637X/775/2/105},
  \href {https://ui.adsabs.harvard.edu/abs/2013ApJ...775..105O} {775, 105}

\bibitem[\protect\citeauthoryear{{Pecaut} \& {Mamajek}}{{Pecaut} \&
  {Mamajek}}{2013}]{Pecaut13}
{Pecaut} M.~J.,  {Mamajek} E.~E.,  2013, \mn@doi [\apjs]
  {10.1088/0067-0049/208/1/9}, \href
  {http://adsabs.harvard.edu/abs/2013ApJS..208....9P} {208, 9}

\bibitem[\protect\citeauthoryear{{Perger} et~al.,}{{Perger}
  et~al.}{2023}]{Perger23}
{Perger} M.,  et~al., 2023, \mn@doi [arXiv e-prints]
  {10.48550/arXiv.2301.12872}, \href
  {https://ui.adsabs.harvard.edu/abs/2023arXiv230112872P} {p. arXiv:2301.12872}

\bibitem[\protect\citeauthoryear{{Plavchan} et~al.,}{{Plavchan}
  et~al.}{2020}]{Plavchan20}
{Plavchan} P.,  et~al., 2020, \mn@doi [\nat] {10.1038/s41586-020-2400-z}, \href
  {https://ui.adsabs.harvard.edu/abs/2020Natur.582..497P} {582, 497}

\bibitem[\protect\citeauthoryear{{Reiners} et~al.,}{{Reiners}
  et~al.}{2022}]{Reiners22}
{Reiners} A.,  et~al., 2022, \mn@doi [\aap] {10.1051/0004-6361/202243251},
  \href {https://ui.adsabs.harvard.edu/abs/2022A&A...662A..41R} {662, A41}

\bibitem[\protect\citeauthoryear{{Ryabchikova}, {Piskunov}, {Kurucz},
  {Stempels}, {Heiter}, {Pakhomov}  \& {Barklem}}{{Ryabchikova}
  et~al.}{2015}]{Ryabchikova15}
{Ryabchikova} T.,  {Piskunov} N.,  {Kurucz} R.~L.,  {Stempels} H.~C.,  {Heiter}
  U.,  {Pakhomov} Y.,   {Barklem} P.~S.,  2015, \mn@doi [\physscr]
  {10.1088/0031-8949/90/5/054005}, \href
  {https://ui.adsabs.harvard.edu/abs/2015PhyS...90e4005R} {90, 054005}

\bibitem[\protect\citeauthoryear{{Siess}, {Dufour}  \& {Forestini}}{{Siess}
  et~al.}{2000}]{Siess00}
{Siess} L.,  {Dufour} E.,   {Forestini} M.,  2000, \aap, 358, 593

\bibitem[\protect\citeauthoryear{{Skilling} \& {Bryan}}{{Skilling} \&
  {Bryan}}{1984}]{Skilling84}
{Skilling} J.,  {Bryan} R.~K.,  1984, \mn@doi [\mnras]
  {10.1093/mnras/211.1.111}, \href
  {https://ui.adsabs.harvard.edu/abs/1984MNRAS.211..111S} {211, 111}

\bibitem[\protect\citeauthoryear{{Strugarek}, {Brun}, {Matt}  \&
  {R{\'e}ville}}{{Strugarek} et~al.}{2015}]{Strugarek15}
{Strugarek} A.,  {Brun} A.~S.,  {Matt} S.~P.,   {R{\'e}ville} V.,  2015,
  \mn@doi [\apj] {10.1088/0004-637X/815/2/111}, \href
  {https://ui.adsabs.harvard.edu/abs/2015ApJ...815..111S} {815, 111}

\bibitem[\protect\citeauthoryear{{Su{\'a}rez Mascare{\~n}o}
  et~al.,}{{Su{\'a}rez Mascare{\~n}o} et~al.}{2020}]{Suarez20}
{Su{\'a}rez Mascare{\~n}o} A.,  et~al., 2020, \mn@doi [\aap]
  {10.1051/0004-6361/202037745}, \href
  {https://ui.adsabs.harvard.edu/abs/2020A&A...639A..77S} {639, A77}

\bibitem[\protect\citeauthoryear{{Su{\'a}rez Mascare{\~n}o}
  et~al.,}{{Su{\'a}rez Mascare{\~n}o} et~al.}{2021}]{Suarez22}
{Su{\'a}rez Mascare{\~n}o} A.,  et~al., 2021, \mn@doi [Nature Astronomy]
  {10.1038/s41550-021-01533-7}, \href
  {https://ui.adsabs.harvard.edu/abs/2022NatAs...6..232S} {6, 232}

\bibitem[\protect\citeauthoryear{{Szab{\'o}} et~al.,}{{Szab{\'o}}
  et~al.}{2021}]{Szabo21}
{Szab{\'o}} G.~M.,  et~al., 2021, \mn@doi [\aap] {10.1051/0004-6361/202140345},
  \href {https://ui.adsabs.harvard.edu/abs/2021A&A...654A.159S} {654, A159}

\bibitem[\protect\citeauthoryear{{Szab{\'o}} et~al.,}{{Szab{\'o}}
  et~al.}{2022}]{Szabo22}
{Szab{\'o}} G.~M.,  et~al., 2022, \mn@doi [\aap] {10.1051/0004-6361/202243076},
  \href {https://ui.adsabs.harvard.edu/abs/2022A&A...659L...7S} {659, L7}

\bibitem[\protect\citeauthoryear{{Van Eylen} et~al.,}{{Van Eylen}
  et~al.}{2019}]{vanEylen19}
{Van Eylen} V.,  et~al., 2019, \mn@doi [\aj] {10.3847/1538-3881/aaf22f}, \href
  {https://ui.adsabs.harvard.edu/abs/2019AJ....157...61V} {157, 61}

\bibitem[\protect\citeauthoryear{{Wittrock} et~al.,}{{Wittrock}
  et~al.}{2023}]{Wittrock23}
{Wittrock} J.~M.,  et~al., 2023, \mn@doi [arXiv e-prints]
  {10.48550/arXiv.2302.04922}, \href
  {https://ui.adsabs.harvard.edu/abs/2023arXiv230204922W} {p. arXiv:2302.04922}

\bibitem[\protect\citeauthoryear{{Zanni} \& {Ferreira}}{{Zanni} \&
  {Ferreira}}{2013}]{Zanni13}
{Zanni} C.,  {Ferreira} J.,  2013, \mn@doi [\aap]
  {10.1051/0004-6361/201220168}, \href
  {http://adsabs.harvard.edu/abs/2013A%26A...550A..99Z} {550, A99}

\bibitem[\protect\citeauthoryear{{Zeng}, {Sasselov}  \& {Jacobsen}}{{Zeng}
  et~al.}{2016}]{Zeng16}
{Zeng} L.,  {Sasselov} D.~D.,   {Jacobsen} S.~B.,  2016, \mn@doi [\apj]
  {10.3847/0004-637X/819/2/127}, \href
  {https://ui.adsabs.harvard.edu/abs/2016ApJ...819..127Z} {819, 127}

\bibitem[\protect\citeauthoryear{{Zeng} et~al.,}{{Zeng} et~al.}{2019}]{Zeng19}
{Zeng} L.,  et~al., 2019, \mn@doi [Proceedings of the National Academy of
  Science] {10.1073/pnas.1812905116}, \href
  {https://ui.adsabs.harvard.edu/abs/2019PNAS..116.9723Z} {116, 9723}

\bibitem[\protect\citeauthoryear{{Zicher} et~al.,}{{Zicher}
  et~al.}{2022}]{Zicher22}
{Zicher} N.,  et~al., 2022, \mn@doi [\mnras] {10.1093/mnras/stac614}, \href
  {https://ui.adsabs.harvard.edu/abs/2022MNRAS.512.3060Z} {512, 3060}

\makeatother
\end{thebibliography}
\bibliographystyle{mnras}

\appendix
\section{Observation log}
\label{sec:appA}
Table~\ref{tab:log} outlines the full observation log, as well as all individual measurements carried out from the collected spectra at each epoch.  

\begin{table*}
\small
\caption[]{Observation log.  All exposures consist of 4 sub-exposures of equal length, except that marked with an `x' for which only 2 of the 4 sub-exposures could be used.  
For each visit, we list the corresponding barycentric Julian date BJD, rotation cycle c and phase $\phi$ (computed as indicated in Sec.~\ref{sec:obs}), the total observing time 
t$_{\rm exp}$, the peak SNR in the spectrum, the noise level in the LSD Stokes $V$ profile, the estimated \Bl\ and <$B$> with error bars, the RV and error bar estimated by LBL whenever 
relevant (once per night) and available, the wPCA component $W_1$ derived from the LBL dLW proxy (see text), and the EWVs as measured in the \hei\ and \pab\ lines.  
Exposures affected with a flare (detected in the \hei\ line) are marked with a `+'.  } 
\hspace{-6mm}
\begin{tabular}{ccccccccccc}
\hline
BJD        & c / $\phi$ & t$_{\rm exp}$ & SNR & $\sigma_V$            & \Bl\  &  <$B$>  & RV    & $W_1$    & EWV \hei & EWV \pab \\ 
(2459000+) &            &   (s)         &     & ($10^{-4} I_c$)       & (G)   &  (kG)   & (\ms) & (\kmsss) & (\kms)   & (\kms)   \\  
\hline
-399.8747695 & -83 / 0.721 & 378.9 & 427 & 1.53 & -68.6$\pm$8.2 & & 2.7$\pm$4.9 & -1.32$\pm$0.13 & 0.749$\pm$0.077 & 0.269$\pm$0.033 \\
+ -350.8732586 & -73 / 0.804 & 763.4 & 752 & 1.19 & -63.3$\pm$6.4 & 2.89$\pm$0.03 & -55.2$\pm$4.6 & -1.24$\pm$0.11 & 3.111$\pm$1.045 & 0.218$\pm$0.019 \\
-349.8781098 & -72 / 0.009 & 741.1 & 753 & 1.14 & -27.4$\pm$6.0 & 2.59$\pm$0.02 & -24.0$\pm$3.5 & 1.15$\pm$0.12 & 0.405$\pm$0.043 & -0.032$\pm$0.019 \\
-348.9179834 & -72 / 0.206 & 735.5 & 753 & 1.36 & 123.9$\pm$7.3 & 2.69$\pm$0.02 & 31.7$\pm$3.7 & 0.43$\pm$0.12 & 0.438$\pm$0.042 & -0.231$\pm$0.018 \\
-348.0685108 & -72 / 0.381 & 490.3 & 407 & 1.73 & 82.8$\pm$9.6 & & & & -0.407$\pm$0.073 & 0.138$\pm$0.031 \\
-348.0607403 & -72 / 0.383 & 490.3 & 451 & 1.60 & 67.7$\pm$8.9 & & & & -0.233$\pm$0.068 & 0.020$\pm$0.027 \\
-348.0539511 & -72 / 0.384 & 490.3 & 511 & 1.36 & 74.2$\pm$7.6 & & & & -0.387$\pm$0.066 & -0.086$\pm$0.028 \\
-348.0466565 & -72 / 0.385 & 490.3 & 532 & 1.33 & 67.2$\pm$7.4 & & & & -0.252$\pm$0.068 & 0.071$\pm$0.029 \\
-348.0392277 & -72 / 0.387 & 490.3 & 519 & 1.31 & 76.5$\pm$7.2 & & & & -0.196$\pm$0.069 & 0.047$\pm$0.029 \\
-348.0319186 & -72 / 0.388 & 490.3 & 523 & 1.36 & 85.0$\pm$7.5 & & & & -0.216$\pm$0.067 & 0.073$\pm$0.029 \\
-348.0247451 & -72 / 0.390 & 490.3 & 509 & 1.37 & 70.6$\pm$7.5 & & & & -0.290$\pm$0.081 & 0.029$\pm$0.033 \\
-348.0172993 & -72 / 0.392 & 490.3 & 523 & 1.33 & 64.8$\pm$7.2 & & & & -0.079$\pm$0.065 & 0.097$\pm$0.027 \\
-348.0099247 & -72 / 0.393 & 490.3 & 545 & 1.29 & 64.5$\pm$7.0 & & & & -0.227$\pm$0.069 & -0.026$\pm$0.028 \\
-348.0028446 & -72 / 0.394 & 490.3 & 545 & 1.27 & 69.5$\pm$6.9 & & & & -0.159$\pm$0.075 & 0.081$\pm$0.032 \\
-347.9953218 & -72 / 0.396 & 490.3 & 523 & 1.31 & 54.1$\pm$7.1 & & & & -0.409$\pm$0.071 & 0.088$\pm$0.030 \\
-347.9880112 & -72 / 0.398 & 490.3 & 551 & 1.35 & 83.1$\pm$7.2 & 2.73$\pm$0.02 & 18.5$\pm$0.8 & -0.34$\pm$0.12 & -0.300$\pm$0.069 & 0.024$\pm$0.029 \\
-347.9807843 & -72 / 0.399 & 490.3 & 546 & 1.29 & 73.2$\pm$6.9 & & & & -0.173$\pm$0.070 & -0.050$\pm$0.029 \\
-347.9734398 & -72 / 0.401 & 490.3 & 518 & 1.31 & 62.0$\pm$7.0 & & & & -0.009$\pm$0.063 & 0.077$\pm$0.027 \\
-347.9660598 & -72 / 0.402 & 490.3 & 527 & 1.32 & 57.6$\pm$7.1 & & & & -0.124$\pm$0.066 & 0.029$\pm$0.028 \\
-347.9589404 & -72 / 0.404 & 490.3 & 539 & 1.34 & 52.8$\pm$7.2 & & & & -0.207$\pm$0.067 & 0.066$\pm$0.029 \\
-347.9515016 & -72 / 0.405 & 490.3 & 536 & 1.34 & 58.5$\pm$7.2 & & & & -0.049$\pm$0.088 & -0.007$\pm$0.036 \\
-347.9442715 & -72 / 0.407 & 490.3 & 506 & 1.36 & 64.9$\pm$7.2 & & & & -0.311$\pm$0.067 & 0.040$\pm$0.028 \\
-347.9368452 & -72 / 0.408 & 490.3 & 497 & 1.39 & 42.4$\pm$7.4 & & & & -0.063$\pm$0.066 & 0.057$\pm$0.028 \\
-347.9298065 & -72 / 0.410 & 490.3 & 499 & 1.45 & 48.4$\pm$7.7 & & & & -0.093$\pm$0.063 & 0.041$\pm$0.027 \\
-347.9221049 & -72 / 0.411 & 490.3 & 462 & 1.48 & 52.7$\pm$7.9 & & & & -0.172$\pm$0.062 & -0.071$\pm$0.026 \\
-347.9149019 & -72 / 0.413 & 490.3 & 469 & 1.45 & 45.2$\pm$7.7 & & & & -0.292$\pm$0.062 & -0.037$\pm$0.026 \\
-347.9074044 & -72 / 0.414 & 490.3 & 481 & 1.44 & 54.6$\pm$7.7 & & & & -0.278$\pm$0.062 & 0.038$\pm$0.026 \\
-347.9002434 & -72 / 0.416 & 490.3 & 501 & 1.45 & 46.2$\pm$7.8 & & & & -0.277$\pm$0.062 & 0.038$\pm$0.026 \\
-347.8927767 & -72 / 0.417 & 490.3 & 505 & 1.42 & 51.8$\pm$7.5 & & & & -0.157$\pm$0.062 & 0.048$\pm$0.027 \\
-347.8854737 & -72 / 0.419 & 490.3 & 518 & 1.38 & 53.7$\pm$7.4 & & & & -0.434$\pm$0.065 & -0.005$\pm$0.027 \\
-347.8782414 & -72 / 0.420 & 490.3 & 531 & 1.33 & 43.7$\pm$7.1 & & & & -0.176$\pm$0.065 & -0.069$\pm$0.028 \\
-347.8710231 & -72 / 0.422 & 490.3 & 535 & 1.38 & 52.4$\pm$7.3 & & & & -0.218$\pm$0.067 & 0.011$\pm$0.028 \\
-347.8636629 & -72 / 0.423 & 490.3 & 522 & 1.35 & 34.4$\pm$7.2 & & & & -0.286$\pm$0.062 & -0.036$\pm$0.026 \\
-345.9614621 & -72 / 0.815 & 752.2 & 748 & 1.10 & -55.9$\pm$6.0 & 2.87$\pm$0.03 & -40.3$\pm$2.9 & -1.07$\pm$0.13 & 0.259$\pm$0.043 & 0.136$\pm$0.019 \\
-255.1779946 & -53 / 0.494 & 780.1 & 731 & 1.00 & -66.8$\pm$5.6 & 2.77$\pm$0.01 & 51.7$\pm$4.3 & -0.47$\pm$0.10 & 0.062$\pm$0.044 & -0.179$\pm$0.019 \\
-249.2450248 & -52 / 0.715 & 780.1 & 765 & 1.16 & 42.3$\pm$6.7 & 3.01$\pm$0.03 & -12.7$\pm$5.0 & -1.69$\pm$0.11 & 0.658$\pm$0.041 & 0.387$\pm$0.017 \\
-248.2538870 & -52 / 0.919 & 780.1 & 769 & 1.07 & 12.3$\pm$5.7 & 2.74$\pm$0.03 & -60.8$\pm$4.2 & -0.40$\pm$0.10 & 0.568$\pm$0.042 & -0.169$\pm$0.018 \\
-247.2094110 & -51 / 0.134 & 780.1 & 763 & 1.00 & 75.5$\pm$5.3 & 2.57$\pm$0.03 & 12.6$\pm$4.3 & 1.40$\pm$0.10 & 0.114$\pm$0.041 & -0.156$\pm$0.017 \\
-241.2704220 & -50 / 0.356 & 780.1 & 706 & 0.96 & 16.4$\pm$5.1 & 2.65$\pm$0.03 & 20.7$\pm$3.7 & 0.36$\pm$0.09 & -0.638$\pm$0.046 & -0.165$\pm$0.020 \\
-240.1938818 & -50 / 0.577 & 780.1 & 648 & 1.03 & 21.6$\pm$5.6 & 2.93$\pm$0.03 & 34.6$\pm$3.7 & -1.41$\pm$0.10 & 0.417$\pm$0.054 & 0.028$\pm$0.022 \\
-239.2714096 & -50 / 0.767 & 780.1 & 580 & 1.17 & 45.6$\pm$6.3 & 2.94$\pm$0.03 & -3.6$\pm$3.8 & -1.50$\pm$0.11 & 0.781$\pm$0.064 & 0.106$\pm$0.026 \\
-238.2687216 & -50 / 0.974 & 780.1 & 757 & 1.01 & -1.4$\pm$5.2 & 2.68$\pm$0.03 & -80.4$\pm$4.3 & 0.18$\pm$0.10 & 0.647$\pm$0.044 & 0.079$\pm$0.018 \\
-237.2676540 & -49 / 0.179 & 780.1 & 749 & 0.94 & 79.6$\pm$4.9 & 2.63$\pm$0.03 & -4.0$\pm$4.0 & 0.96$\pm$0.10 & 0.247$\pm$0.044 & -0.109$\pm$0.019 \\
-235.2421162 & -49 / 0.596 & 780.1 & 759 & 1.26 & 31.2$\pm$7.0 & 2.96$\pm$0.03 & 42.6$\pm$4.5 & -1.71$\pm$0.11 & 0.245$\pm$0.041 & 0.226$\pm$0.018 \\
-234.2298202 & -49 / 0.805 & 780.1 & 764 & 1.04 & 13.8$\pm$5.8 & 2.97$\pm$0.03 & -25.3$\pm$4.3 & -1.59$\pm$0.10 & 0.713$\pm$0.042 & -0.011$\pm$0.018 \\
-230.2553866 & -48 / 0.622 & 780.1 & 767 & 1.02 & 42.8$\pm$5.6 & 2.99$\pm$0.03 & 29.8$\pm$4.3 & -1.71$\pm$0.10 & -0.002$\pm$0.044 & 0.158$\pm$0.018 \\
-229.2585186 & -48 / 0.827 & 780.1 & 702 & 1.11 & 20.7$\pm$6.0 & 2.90$\pm$0.03 & -16.6$\pm$4.2 & -1.21$\pm$0.10 & 0.338$\pm$0.047 & 0.070$\pm$0.020 \\
-228.2780477 & -47 / 0.029 & 780.1 & 697 & 1.03 & 5.1$\pm$5.6 & 2.65$\pm$0.03 & -26.0$\pm$4.1 & 0.50$\pm$0.10 & 0.958$\pm$0.046 & -0.182$\pm$0.020 \\
-227.2576342 & -47 / 0.239 & 780.1 & 665 & 1.12 & 82.2$\pm$6.3 & 2.65$\pm$0.02 & 26.9$\pm$4.3 & 0.41$\pm$0.10 & -0.084$\pm$0.051 & -0.144$\pm$0.021 \\
-212.2836795 & -44 / 0.320 & 780.1 & 650 & 1.09 & -6.9$\pm$5.9 & 2.74$\pm$0.02 & 20.1$\pm$4.0 & 0.22$\pm$0.10 & -0.625$\pm$0.051 & -0.194$\pm$0.021 \\
-211.2946856 & -44 / 0.524 & 780.1 & 673 & 1.06 & -7.8$\pm$5.9 & 2.90$\pm$0.03 & 40.3$\pm$4.2 & -0.95$\pm$0.11 & -0.769$\pm$0.050 & 0.014$\pm$0.021 \\
-210.2625258 & -44 / 0.736 & 780.1 & 740 & 1.00 & 106.4$\pm$5.6 & 2.98$\pm$0.03 & & -1.87$\pm$0.10 & 0.248$\pm$0.043 & 0.121$\pm$0.018 \\
-209.2981790 & -44 / 0.935 & 780.1 & 757 & 1.21 & -13.2$\pm$6.7 & 2.76$\pm$0.02 & & -0.38$\pm$0.11 & 0.811$\pm$0.042 & 0.077$\pm$0.018 \\
-208.3008722 & -43 / 0.140 & 780.1 & 747 & 0.96 & 58.3$\pm$5.1 & 2.55$\pm$0.02 & 9.3$\pm$4.2 & 1.02$\pm$0.10 & 0.734$\pm$0.042 & -0.013$\pm$0.018 \\
-207.3015996 & -43 / 0.345 & 780.1 & 778 & 1.08 & -42.1$\pm$5.8 & 2.69$\pm$0.02 & 36.2$\pm$4.0 & 0.34$\pm$0.10 & -0.910$\pm$0.041 & -0.212$\pm$0.018 \\
-203.3133195 & -42 / 0.166 & 780.1 & 586 & 1.23 & 59.1$\pm$6.7 & 2.64$\pm$0.02 & -12.9$\pm$4.3 & 0.86$\pm$0.11 & -0.251$\pm$0.058 & 0.016$\pm$0.023 \\
x -202.3142043 & -42 / 0.372 & 390.0 & 424 & 1.73 & -59.8$\pm$10.1 & & & & -1.426$\pm$0.059 & 0.078$\pm$0.024 \\
-202.2674475 & -42 / 0.381 & 780.1 & 593 & 1.19 & -70.2$\pm$6.9 & 2.72$\pm$0.02 & 27.0$\pm$4.3 & 0.37$\pm$0.11 & -1.282$\pm$0.082 & 0.059$\pm$0.033 \\
-201.3118745 & -42 / 0.578 & 780.1 & 631 & 1.07 & 85.8$\pm$6.0 & 2.88$\pm$0.02 & 29.8$\pm$3.8 & -1.51$\pm$0.10 & -0.540$\pm$0.054 & 0.111$\pm$0.022 \\
\hline
\end{tabular}
\label{tab:log}
\end{table*}

\setcounter{table}{0}
\begin{table*}
\caption[]{continued} 
\hspace{-6mm}
\begin{tabular}{ccccccccccc}
\hline
BJD        & c / $\phi$ & t$_{\rm exp}$ & SNR & $\sigma_V$            & \Bl\  &  <$B$>  & RV    & $W_1$    & EWV \hei & EWV \pab \\ 
(2459000+) &            &   (s)         &     & ($10^{-4} I_c$)       & (G)   &  (kG)   & (\ms) & (\kmsss) & (\kms)   & (\kms)   \\  
\hline
-200.3108791 & -42 / 0.784 & 780.1 & 579 & 1.19 & 91.7$\pm$6.8 & 2.93$\pm$0.03 & -13.2$\pm$4.3 & -1.59$\pm$0.11 & 0.044$\pm$0.060 & 0.280$\pm$0.025 \\
-199.3096404 & -42 / 0.990 & 780.1 & 746 & 1.00 & -16.9$\pm$5.5 & 2.68$\pm$0.03 & -46.4$\pm$4.1 & -0.04$\pm$0.10 & 0.010$\pm$0.044 & 0.031$\pm$0.019 \\
-198.3118516 & -41 / 0.195 & 780.1 & 721 & 1.12 & 67.2$\pm$6.1 & 2.66$\pm$0.03 & 6.1$\pm$4.6 & 0.90$\pm$0.11 & 0.252$\pm$0.045 & -0.200$\pm$0.019 \\
\hline
-15.8798518 & -4 / 0.733 & 780.1 & 701 & 0.94 & 196.7$\pm$5.2 & 2.77$\pm$0.03 & 2.2$\pm$3.6 & -0.21$\pm$0.09 & -0.062$\pm$0.045 & 0.174$\pm$0.019 \\
-14.8996031 & -4 / 0.934 & 780.1 & 736 & 0.90 & -5.5$\pm$4.8 & 2.81$\pm$0.03 & 22.9$\pm$3.8 & -0.74$\pm$0.09 & 0.343$\pm$0.042 & 0.153$\pm$0.018 \\
+ 0.0859254 & 0 / 0.018 & 780.1 & 785 & 0.92 & 33.0$\pm$5.0 & 2.83$\pm$0.03 & -11.5$\pm$4.2 & -0.58$\pm$0.10 & 4.757$\pm$1.040 & 0.038$\pm$0.017 \\
1.0109417 & 0 / 0.208 & 735.5 & 787 & 0.93 & -75.8$\pm$4.9 & 2.72$\pm$0.03 & -45.7$\pm$4.4 & 0.67$\pm$0.10 & 0.333$\pm$0.039 & -0.173$\pm$0.017 \\
3.0657407 & 0 / 0.631 & 780.1 & 764 & 1.14 & 211.0$\pm$6.4 & 2.71$\pm$0.03 & -22.1$\pm$4.2 & 0.22$\pm$0.10 & -0.349$\pm$0.039 & 0.016$\pm$0.017 \\
4.0703036 & 0 / 0.838 & 780.1 & 760 & 0.94 & 83.5$\pm$5.1 & 2.82$\pm$0.03 & 21.6$\pm$4.1 & -0.94$\pm$0.10 & 0.088$\pm$0.040 & -0.074$\pm$0.017 \\
5.0685152 & 1 / 0.043 & 774.5 & 774 & 0.87 & 31.9$\pm$4.7 & 2.79$\pm$0.03 & -26.0$\pm$4.0 & -0.53$\pm$0.09 & 0.266$\pm$0.039 & -0.009$\pm$0.017 \\
6.0994963 & 1 / 0.255 & 780.1 & 777 & 0.85 & -149.0$\pm$4.4 & 2.70$\pm$0.02 & -26.7$\pm$3.9 & 1.17$\pm$0.10 & -0.369$\pm$0.040 & 0.060$\pm$0.017 \\
7.0806725 & 1 / 0.457 & 757.8 & 785 & 0.85 & -37.0$\pm$4.3 & 2.72$\pm$0.03 & 8.9$\pm$4.0 & -0.60$\pm$0.10 & -0.224$\pm$0.039 & -0.392$\pm$0.016 \\
8.0635561 & 1 / 0.659 & 774.5 & 777 & 0.90 & 221.5$\pm$4.7 & 2.72$\pm$0.03 & 0.2$\pm$3.6 & 0.41$\pm$0.09 & -0.410$\pm$0.040 & 0.046$\pm$0.017 \\
+ 9.0639632 & 1 / 0.865 & 780.1 & 740 & 0.95 & 55.8$\pm$5.1 & 2.82$\pm$0.03 & 55.0$\pm$4.0 & -0.75$\pm$0.10 & 2.526$\pm$1.043 & 0.240$\pm$0.018 \\
10.0304082 & 2 / 0.064 & 780.1 & 705 & 1.05 & 33.8$\pm$5.8 & 2.72$\pm$0.03 & -39.1$\pm$3.8 & -0.64$\pm$0.10 & 0.060$\pm$0.042 & 0.060$\pm$0.018 \\
11.0027820 & 2 / 0.264 & 780.1 & 747 & 0.92 & -164.4$\pm$5.1 & 2.64$\pm$0.03 & -23.4$\pm$4.0 & 1.04$\pm$0.10 & 1.847$\pm$0.042 & -0.051$\pm$0.018 \\
31.0278512 & 6 / 0.384 & 707.6 & 789 & 0.98 & -201.9$\pm$5.1 & 2.69$\pm$0.03 & 19.4$\pm$4.1 & 0.61$\pm$0.10 & -0.991$\pm$0.035 & -0.088$\pm$0.016 \\
32.0334169 & 6 / 0.591 & 796.8 & 781 & 1.01 & 225.2$\pm$5.5 & 2.66$\pm$0.03 & -18.0$\pm$4.0 & 0.26$\pm$0.09 & -0.452$\pm$0.035 & -0.047$\pm$0.017 \\
33.0034293 & 6 / 0.791 & 802.4 & 797 & 0.84 & 106.7$\pm$4.5 & 2.76$\pm$0.03 & 33.9$\pm$3.9 & -0.23$\pm$0.09 & 0.057$\pm$0.035 & 0.109$\pm$0.016 \\
34.0637925 & 7 / 0.009 & 802.4 & 793 & 0.96 & 36.8$\pm$5.0 & 2.80$\pm$0.03 & 13.5$\pm$3.9 & -0.85$\pm$0.09 & 0.054$\pm$0.036 & 0.016$\pm$0.016 \\
34.9930744 & 7 / 0.200 & 802.4 & 787 & 0.96 & -91.3$\pm$5.0 & 2.64$\pm$0.03 & -48.1$\pm$3.7 & 0.56$\pm$0.09 & 0.556$\pm$0.039 & -0.079$\pm$0.017 \\
36.9557369 & 7 / 0.604 & 802.4 & 801 & 1.19 & 247.1$\pm$6.4 & 2.65$\pm$0.02 & 15.6$\pm$3.9 & 0.23$\pm$0.09 & -0.200$\pm$0.035 & 0.158$\pm$0.017 \\
37.9853735 & 7 / 0.816 & 802.4 & 809 & 0.93 & 108.5$\pm$4.9 & 2.72$\pm$0.03 & 56.9$\pm$3.7 & -0.41$\pm$0.09 & 0.199$\pm$0.035 & 0.201$\pm$0.016 \\
38.9902282 & 8 / 0.023 & 802.4 & 813 & 0.97 & 49.6$\pm$5.0 & 2.80$\pm$0.03 & 34.6$\pm$4.0 & -0.87$\pm$0.10 & 0.059$\pm$0.035 & 0.088$\pm$0.016 \\
40.0452587 & 8 / 0.240 & 802.4 & 797 & 0.97 & -140.6$\pm$4.9 & 2.61$\pm$0.02 & -12.6$\pm$4.0 & 0.85$\pm$0.10 & 0.279$\pm$0.034 & 0.180$\pm$0.017 \\
40.9999808 & 8 / 0.436 & 802.4 & 731 & 0.88 & -119.6$\pm$4.7 & 2.73$\pm$0.03 & 33.4$\pm$3.8 & -0.16$\pm$0.09 & -0.834$\pm$0.039 & -0.065$\pm$0.018 \\
55.8565090 & 11 / 0.493 & 802.4 & 816 & 0.85 & 56.6$\pm$4.5 & 2.72$\pm$0.02 & -14.3$\pm$4.0 & -0.42$\pm$0.09 & -0.139$\pm$0.034 & -0.088$\pm$0.016 \\
58.9526368 & 12 / 0.130 & 802.4 & 776 & 0.81 & -12.7$\pm$4.1 & 2.72$\pm$0.03 & -47.5$\pm$3.3 & -0.28$\pm$0.08 & 1.165$\pm$0.037 & 0.149$\pm$0.017 \\
59.9590901 & 12 / 0.337 & 802.4 & 800 & 0.90 & -240.0$\pm$5.0 & 2.68$\pm$0.03 & 0.8$\pm$3.7 & 0.55$\pm$0.09 & -0.451$\pm$0.034 & -0.067$\pm$0.017 \\
60.9540595 & 12 / 0.542 & 802.4 & 736 & 0.88 & 168.6$\pm$4.6 & 2.68$\pm$0.03 & -31.9$\pm$3.4 & 0.07$\pm$0.09 & -0.427$\pm$0.037 & 0.036$\pm$0.018 \\
61.9319175 & 12 / 0.743 & 802.4 & 812 & 0.89 & 142.0$\pm$4.6 & 2.67$\pm$0.03 & -6.7$\pm$3.7 & 0.31$\pm$0.09 & 1.748$\pm$0.035 & 0.130$\pm$0.016 \\
62.9758774 & 12 / 0.958 & 802.4 & 822 & 0.92 & 98.7$\pm$4.7 & 2.77$\pm$0.03 & 12.7$\pm$3.9 & -0.37$\pm$0.09 & 1.225$\pm$0.034 & 0.335$\pm$0.016 \\
64.9688429 & 13 / 0.368 & 802.4 & 817 & 0.98 & -233.1$\pm$5.3 & 2.72$\pm$0.02 & -11.8$\pm$4.0 & 0.43$\pm$0.09 & -0.551$\pm$0.033 & -0.037$\pm$0.015 \\
65.8979758 & 13 / 0.559 & 802.4 & 824 & 1.01 & 185.7$\pm$5.2 & 2.67$\pm$0.03 & -31.4$\pm$3.9 & 0.42$\pm$0.09 & -0.728$\pm$0.034 & -0.109$\pm$0.015 \\
68.8956491 & 14 / 0.176 & 802.4 & 472 & 1.38 & -71.5$\pm$7.4 & & & & 0.214$\pm$0.045 & -0.053$\pm$0.021 \\
68.9014223 & 14 / 0.177 & 802.4 & 621 & 1.09 & -69.0$\pm$5.8 & 2.68$\pm$0.03 & & & 0.346$\pm$0.062 & -0.046$\pm$0.028 \\
70.9359245 & 14 / 0.596 & 802.4 & 754 & 0.95 & 236.2$\pm$5.3 & 2.64$\pm$0.03 & -3.1$\pm$3.6 & 0.55$\pm$0.08 & -0.453$\pm$0.036 & 0.043$\pm$0.017 \\
71.9158815 & 14 / 0.798 & 802.4 & 798 & 0.87 & 122.2$\pm$4.7 & 2.69$\pm$0.03 & 16.7$\pm$3.6 & -0.13$\pm$0.08 & 0.435$\pm$0.035 & 0.126$\pm$0.016 \\
72.9451954 & 15 / 0.009 & 802.4 & 814 & 0.86 & 66.4$\pm$4.7 & 2.79$\pm$0.03 & 28.8$\pm$3.9 & -0.58$\pm$0.09 & -0.151$\pm$0.034 & -0.167$\pm$0.016 \\
87.8489695 & 18 / 0.076 & 802.4 & 484 & 1.35 & -7.5$\pm$7.3 & & & & 0.110$\pm$0.048 & 0.112$\pm$0.022 \\
87.8596173 & 18 / 0.078 & 802.4 & 588 & 1.10 & 5.2$\pm$5.9 & 2.74$\pm$0.03 & -36.8$\pm$2.6 & -0.45$\pm$0.10 & 0.257$\pm$0.060 & 0.087$\pm$0.027 \\
88.8540279 & 18 / 0.283 & 802.4 & 613 & 1.12 & -173.5$\pm$6.1 & 2.67$\pm$0.03 & -50.3$\pm$3.1 & 0.38$\pm$0.09 & -0.382$\pm$0.047 & 0.161$\pm$0.021 \\
89.8517444 & 18 / 0.488 & 802.4 & 829 & 1.08 & 41.5$\pm$5.5 & 2.68$\pm$0.03 & -38.8$\pm$3.7 & -0.18$\pm$0.09 & -0.306$\pm$0.033 & -0.066$\pm$0.015 \\
90.8053844 & 18 / 0.684 & 802.4 & 852 & 0.92 & 180.6$\pm$4.8 & 2.59$\pm$0.03 & -18.3$\pm$3.7 & 0.72$\pm$0.09 & -0.199$\pm$0.032 & 0.030$\pm$0.015 \\
91.8503861 & 18 / 0.899 & 802.4 & 838 & 1.00 & 115.0$\pm$5.1 & 2.79$\pm$0.03 & -4.3$\pm$3.7 & -0.33$\pm$0.09 & -0.158$\pm$0.031 & 0.060$\pm$0.015 \\
92.8230413 & 19 / 0.099 & 802.4 & 876 & 0.95 & -3.4$\pm$5.0 & 2.71$\pm$0.03 & -55.0$\pm$3.9 & -0.62$\pm$0.09 & 0.304$\pm$0.030 & -0.158$\pm$0.014 \\
93.9083146 & 19 / 0.323 & 802.4 & 807 & 0.92 & -208.2$\pm$4.8 & 2.66$\pm$0.03 & -27.6$\pm$3.4 & 0.23$\pm$0.08 & 0.387$\pm$0.034 & 0.088$\pm$0.016 \\
94.8964898 & 19 / 0.526 & 802.4 & 854 & 0.99 & 127.0$\pm$5.1 & 2.67$\pm$0.03 & -45.0$\pm$3.9 & 0.16$\pm$0.09 & 0.141$\pm$0.031 & -0.075$\pm$0.015 \\
95.9362467 & 19 / 0.740 & 802.4 & 871 & 1.02 & 126.6$\pm$5.3 & 2.66$\pm$0.03 & 0.6$\pm$4.4 & -0.05$\pm$0.10 & 0.268$\pm$0.030 & 0.136$\pm$0.014 \\
96.9073483 & 19 / 0.940 & 802.4 & 831 & 0.97 & 101.0$\pm$5.5 & 2.77$\pm$0.03 & 7.7$\pm$4.2 & -0.25$\pm$0.09 & 1.308$\pm$0.031 & -0.021$\pm$0.015 \\
97.9143932 & 20 / 0.147 & 802.4 & 586 & 1.14 & -25.4$\pm$6.2 & 2.76$\pm$0.03 & -50.1$\pm$4.1 & -0.28$\pm$0.10 & 0.164$\pm$0.046 & 0.133$\pm$0.021 \\
98.8671179 & 20 / 0.343 & 802.4 & 839 & 0.99 & -211.5$\pm$5.1 & 2.70$\pm$0.03 & -23.0$\pm$3.8 & 0.07$\pm$0.09 & -0.216$\pm$0.032 & -0.009$\pm$0.014 \\
99.8825843 & 20 / 0.552 & 802.4 & 662 & 1.01 & 177.9$\pm$5.5 & 2.65$\pm$0.03 & -26.6$\pm$3.5 & 0.32$\pm$0.08 & 0.163$\pm$0.042 & 0.094$\pm$0.019 \\
100.9216089 & 20 / 0.766 & 802.4 & 782 & 0.93 & 125.8$\pm$4.9 & 2.64$\pm$0.03 & -18.7$\pm$3.3 & -0.07$\pm$0.08 & 0.233$\pm$0.035 & 0.114$\pm$0.016 \\
101.8507189 & 20 / 0.957 & 802.4 & 867 & 0.97 & 98.7$\pm$5.2 & 2.74$\pm$0.03 & 9.5$\pm$4.2 & -0.33$\pm$0.10 & -0.220$\pm$0.030 & -0.034$\pm$0.014 \\
102.8595043 & 21 / 0.165 & 802.4 & 849 & 0.85 & -4.5$\pm$4.5 & 2.68$\pm$0.03 & -55.2$\pm$3.8 & -0.24$\pm$0.09 & 0.588$\pm$0.031 & 0.184$\pm$0.014 \\
110.8131877 & 22 / 0.801 & 802.4 & 681 & 1.23 & 103.2$\pm$6.9 & & & & 0.342$\pm$0.061 & -0.022$\pm$0.029 \\
110.8244244 & 22 / 0.803 & 802.4 & 434 & 1.65 & 109.6$\pm$9.2 & 2.73$\pm$0.03 & 15.8$\pm$3.1 & -0.27$\pm$0.11 & 0.226$\pm$0.038 & 0.057$\pm$0.018 \\
111.8027093 & 23 / 0.005 & 802.4 & 748 & 0.91 & 67.9$\pm$5.0 & 2.76$\pm$0.03 & 25.6$\pm$3.4 & -0.46$\pm$0.08 & -0.270$\pm$0.035 & 0.112$\pm$0.017 \\
112.7898810 & 23 / 0.208 & 802.4 & 718 & 0.96 & -61.3$\pm$5.1 & 2.75$\pm$0.03 & -22.8$\pm$3.8 & -0.39$\pm$0.09 & 0.555$\pm$0.037 & 0.042$\pm$0.017 \\
113.8113221 & 23 / 0.418 & 802.4 & 661 & 0.99 & -137.9$\pm$5.2 & 2.74$\pm$0.03 & 3.6$\pm$3.7 & -0.23$\pm$0.09 & -0.511$\pm$0.043 & 0.077$\pm$0.018 \\
114.8048469 & 23 / 0.622 & 802.4 & 813 & 0.95 & 203.2$\pm$5.1 & 2.65$\pm$0.03 & 8.8$\pm$3.9 & 0.68$\pm$0.09 & -0.054$\pm$0.033 & 0.063$\pm$0.015 \\
115.7991501 & 23 / 0.827 & 802.4 & 725 & 0.95 & 85.6$\pm$5.1 & 2.74$\pm$0.03 & 23.3$\pm$3.4 & -0.35$\pm$0.08 & 0.167$\pm$0.039 & 0.105$\pm$0.017 \\
\hline
\end{tabular}
\end{table*}

\setcounter{table}{0}
\begin{table*}
\caption[]{continued} 
\hspace{-6mm}
\begin{tabular}{ccccccccccc}
\hline
BJD        & c / $\phi$ & t$_{\rm exp}$ & SNR & $\sigma_V$            & \Bl\  &  <$B$>  & RV    & $W_1$    & EWV \hei & EWV \pab \\ 
(2459000+) &            &   (s)         &     & ($10^{-4} I_c$)       & (G)   &  (kG)   & (\ms) & (\kmsss) & (\kms)   & (\kms)   \\  
\hline
117.8244814 & 24 / 0.244 & 802.4 & 853 & 0.83 & -108.8$\pm$4.4 & 2.76$\pm$0.03 & -50.0$\pm$3.5 & -0.25$\pm$0.08 & 0.787$\pm$0.030 & 0.182$\pm$0.014 \\
118.7992235 & 24 / 0.444 & 802.4 & 843 & 0.84 & -79.7$\pm$4.3 & 2.72$\pm$0.03 & -10.0$\pm$3.3 & -0.28$\pm$0.08 & 1.411$\pm$0.030 & 0.135$\pm$0.015 \\
119.7937402 & 24 / 0.649 & 802.4 & 840 & 1.03 & 207.0$\pm$5.5 & 2.62$\pm$0.03 & 5.4$\pm$4.0 & 0.72$\pm$0.09 & 1.343$\pm$0.033 & 0.248$\pm$0.015 \\
120.7913733 & 24 / 0.854 & 802.4 & 884 & 0.90 & 77.7$\pm$4.8 & 2.75$\pm$0.03 & 6.4$\pm$4.3 & -0.33$\pm$0.10 & 0.901$\pm$0.030 & -0.181$\pm$0.013 \\
121.7909067 & 25 / 0.060 & 802.4 & 856 & 0.94 & 93.2$\pm$5.1 & 2.79$\pm$0.03 & -2.0$\pm$3.9 & -0.59$\pm$0.09 & 0.455$\pm$0.031 & 0.186$\pm$0.014 \\
122.7558637 & 25 / 0.258 & 802.4 & 885 & 1.00 & -130.9$\pm$5.4 & 2.74$\pm$0.03 & -31.4$\pm$4.3 & -0.14$\pm$0.10 & -0.048$\pm$0.029 & 0.084$\pm$0.014 \\
123.7949656 & 25 / 0.472 & 802.4 & 866 & 0.90 & 1.1$\pm$4.8 & 2.74$\pm$0.03 & -20.0$\pm$3.9 & -0.22$\pm$0.09 & -0.680$\pm$0.030 & -0.054$\pm$0.014 \\
125.7947743 & 25 / 0.884 & 802.4 & 816 & 0.86 & 79.6$\pm$4.6 & 2.75$\pm$0.03 & 22.1$\pm$3.4 & -0.28$\pm$0.08 & 0.229$\pm$0.034 & 0.059$\pm$0.015 \\
126.8009561 & 26 / 0.091 & 802.4 & 720 & 0.94 & 71.1$\pm$5.1 & 2.80$\pm$0.03 & -14.0$\pm$3.2 & -0.60$\pm$0.08 & 0.232$\pm$0.038 & -0.027$\pm$0.018 \\
127.7893997 & 26 / 0.294 & 802.4 & 802 & 0.95 & -172.9$\pm$5.2 & 2.74$\pm$0.03 & -32.9$\pm$3.6 & -0.16$\pm$0.08 & 0.139$\pm$0.033 & 0.143$\pm$0.015 \\
129.7872490 & 26 / 0.705 & 802.4 & 844 & 0.88 & 162.5$\pm$4.8 & 2.67$\pm$0.03 & -3.2$\pm$3.6 & 0.44$\pm$0.08 & 0.054$\pm$0.031 & 0.016$\pm$0.015 \\
130.7305717 & 26 / 0.899 & 802.4 & 834 & 0.85 & 101.9$\pm$4.5 & 2.72$\pm$0.03 & 17.1$\pm$3.6 & -0.32$\pm$0.08 & 0.230$\pm$0.032 & 0.039$\pm$0.015 \\
153.7188640 & 31 / 0.629 & 802.4 & 916 & 0.81 & 206.3$\pm$4.4 & 2.59$\pm$0.03 & -15.4$\pm$3.8 & 1.13$\pm$0.09 & 0.389$\pm$0.027 & -0.042$\pm$0.013 \\
154.7176220 & 31 / 0.835 & 802.4 & 910 & 0.83 & 91.4$\pm$4.5 & 2.74$\pm$0.03 & 13.6$\pm$3.8 & -0.25$\pm$0.09 & 1.560$\pm$0.027 & -0.249$\pm$0.013 \\
156.8153501 & 32 / 0.267 & 802.4 & 881 & 1.01 & -139.7$\pm$5.7 & 2.82$\pm$0.03 & -16.4$\pm$5.0 & -0.42$\pm$0.09 & 1.341$\pm$0.028 & -0.118$\pm$0.014 \\
157.7715397 & 32 / 0.463 & 802.4 & 544 & 1.56 & -1.6$\pm$8.7 & & & & 0.165$\pm$0.029 & -0.232$\pm$0.014 \\
157.7835742 & 32 / 0.466 & 802.4 & 870 & 0.87 & 11.4$\pm$4.8 & 2.74$\pm$0.03 & -35.7$\pm$3.9 & -0.16$\pm$0.08 & 0.491$\pm$0.046 & -0.240$\pm$0.022 \\
158.7220851 & 32 / 0.659 & 802.4 & 910 & 0.92 & 196.6$\pm$5.1 & 2.62$\pm$0.03 & -1.1$\pm$3.8 & 0.97$\pm$0.09 & 0.611$\pm$0.027 & -0.188$\pm$0.013 \\
\hline
385.0563840 & 79 / 0.230 & 802.4 & 494 & 1.36 & -57.8$\pm$7.3 & & & & -0.074$\pm$0.057 & 0.115$\pm$0.027 \\
385.0673174 & 79 / 0.232 & 802.4 & 451 & 1.48 & -49.7$\pm$7.9 & 2.87$\pm$0.03 & 22.0$\pm$3.1 & -1.04$\pm$0.11 & 0.154$\pm$0.052 & 0.080$\pm$0.025 \\
386.0605961 & 79 / 0.436 & 802.4 & 765 & 0.91 & -91.4$\pm$4.8 & 2.71$\pm$0.03 & -49.4$\pm$3.8 & -0.09$\pm$0.09 & -0.099$\pm$0.032 & 0.033$\pm$0.016 \\
387.0534251 & 79 / 0.641 & 802.4 & 820 & 0.87 & 220.5$\pm$4.6 & 2.52$\pm$0.03 & 12.8$\pm$3.5 & 1.53$\pm$0.09 & 0.034$\pm$0.031 & -0.130$\pm$0.015 \\
388.0576673 & 79 / 0.847 & 802.4 & 663 & 1.08 & 136.4$\pm$5.7 & 2.66$\pm$0.02 & 23.7$\pm$3.8 & 0.29$\pm$0.09 & 0.034$\pm$0.038 & -0.144$\pm$0.019 \\
389.0548333 & 80 / 0.052 & 802.4 & 707 & 0.96 & 33.6$\pm$5.1 & 2.73$\pm$0.03 & -4.3$\pm$3.4 & -0.03$\pm$0.08 & -0.111$\pm$0.036 & -0.200$\pm$0.017 \\
390.0934567 & 80 / 0.266 & 802.4 & 868 & 0.81 & -44.5$\pm$4.2 & 2.89$\pm$0.03 & 18.4$\pm$3.7 & -1.12$\pm$0.09 & -0.016$\pm$0.028 & 0.232$\pm$0.014 \\
391.0825145 & 80 / 0.470 & 802.4 & 890 & 0.99 & -67.3$\pm$5.2 & 2.71$\pm$0.03 & -41.3$\pm$4.0 & 0.39$\pm$0.10 & 0.226$\pm$0.028 & 0.093$\pm$0.014 \\
392.1045181 & 80 / 0.680 & 802.4 & 912 & 0.96 & 214.1$\pm$5.0 & 2.52$\pm$0.02 & 7.1$\pm$3.7 & 1.84$\pm$0.09 & -0.519$\pm$0.028 & -0.038$\pm$0.014 \\
393.0787054 & 80 / 0.880 & 802.4 & 897 & 0.89 & 134.4$\pm$4.7 & 2.69$\pm$0.03 & 30.8$\pm$3.8 & -0.23$\pm$0.09 & -0.369$\pm$0.027 & 0.025$\pm$0.014 \\
393.9688957 & 81 / 0.064 & 802.4 & 934 & 1.05 & 35.9$\pm$5.7 & 2.78$\pm$0.03 & 19.1$\pm$4.1 & 0.02$\pm$0.09 & -0.644$\pm$0.026 & 0.079$\pm$0.013 \\
394.9871872 & 81 / 0.273 & 802.4 & 666 & 1.18 & -36.5$\pm$6.2 & 2.90$\pm$0.03 & 23.0$\pm$4.0 & -1.14$\pm$0.10 & -0.479$\pm$0.038 & 0.084$\pm$0.019 \\
396.0914947 & 81 / 0.500 & 802.4 & 822 & 0.85 & -28.3$\pm$4.3 & 2.63$\pm$0.03 & -35.8$\pm$3.4 & 0.76$\pm$0.08 & 1.004$\pm$0.032 & -0.013$\pm$0.015 \\
397.0124331 & 81 / 0.690 & 802.4 & 791 & 0.90 & 223.1$\pm$4.8 & 2.50$\pm$0.03 & 7.8$\pm$3.5 & 1.75$\pm$0.09 & 1.597$\pm$0.034 & -0.152$\pm$0.016 \\
398.0083393 & 81 / 0.895 & 802.4 & 508 & 1.33 & 143.1$\pm$7.1 & 2.71$\pm$0.03 & 13.6$\pm$3.9 & -0.25$\pm$0.10 & 0.319$\pm$0.054 & 0.132$\pm$0.025 \\
412.9575971 & 84 / 0.971 & 802.4 & 890 & 0.89 & 140.6$\pm$4.5 & 2.76$\pm$0.03 & 40.8$\pm$3.7 & -0.39$\pm$0.09 & 0.171$\pm$0.029 & 0.036$\pm$0.015 \\
413.9187473 & 85 / 0.168 & 802.4 & 820 & 0.86 & -79.4$\pm$4.5 & 2.83$\pm$0.03 & 43.6$\pm$3.4 & -0.84$\pm$0.08 & 1.429$\pm$0.033 & 0.399$\pm$0.016 \\
414.8852716 & 85 / 0.367 & 802.4 & 785 & 0.88 & -83.5$\pm$4.8 & 2.81$\pm$0.03 & -19.4$\pm$3.9 & -0.80$\pm$0.09 & 0.149$\pm$0.033 & -0.073$\pm$0.016 \\
415.9648538 & 85 / 0.589 & 802.4 & 873 & 0.85 & 124.4$\pm$4.6 & 2.53$\pm$0.03 & -8.3$\pm$3.8 & 1.61$\pm$0.09 & -0.363$\pm$0.029 & -0.280$\pm$0.014 \\
417.9378291 & 85 / 0.995 & 802.4 & 604 & 1.40 & 120.2$\pm$7.4 & 2.77$\pm$0.03 & -5.6$\pm$2.1 & -0.40$\pm$0.11 & 0.637$\pm$0.033 & 0.117$\pm$0.016 \\
417.9526248 & 85 / 0.998 & 802.4 & 784 & 0.87 & 112.8$\pm$4.5 & & & & 0.462$\pm$0.043 & 0.088$\pm$0.021 \\
422.9456638 & 87 / 0.026 & 802.4 & 821 & 0.94 & 67.7$\pm$5.0 & 2.77$\pm$0.03 & -30.3$\pm$3.3 & -0.38$\pm$0.08 & 0.947$\pm$0.030 & 0.178$\pm$0.018 \\
439.8521184 & 90 / 0.505 & 802.4 & 877 & 0.82 & -46.8$\pm$4.2 & 2.59$\pm$0.03 & -38.8$\pm$3.8 & 1.32$\pm$0.09 & -0.556$\pm$0.030 & -0.072$\pm$0.014 \\
440.8775457 & 90 / 0.716 & 802.4 & 875 & 0.97 & 221.3$\pm$5.1 & 2.49$\pm$0.02 & 20.7$\pm$4.1 & 1.93$\pm$0.10 & -0.157$\pm$0.030 & -0.118$\pm$0.014 \\
441.8563799 & 90 / 0.917 & 802.4 & 891 & 0.95 & 175.6$\pm$5.0 & 2.74$\pm$0.03 & 35.0$\pm$4.0 & -0.47$\pm$0.09 & 0.593$\pm$0.029 & 0.185$\pm$0.014 \\
442.8235780 & 91 / 0.116 & 802.4 & 798 & 0.83 & -40.5$\pm$4.3 & 2.81$\pm$0.03 & 33.1$\pm$3.3 & -0.59$\pm$0.08 & 0.384$\pm$0.032 & 0.142$\pm$0.016 \\
443.8226988 & 91 / 0.322 & 802.4 & 847 & 0.88 & -71.2$\pm$4.6 & 2.84$\pm$0.03 & -6.8$\pm$3.3 & -0.93$\pm$0.08 & 0.309$\pm$0.031 & 0.188$\pm$0.015 \\
444.9453386 & 91 / 0.553 & 802.4 & 929 & 1.09 & 31.2$\pm$5.8 & 2.55$\pm$0.03 & -13.1$\pm$4.3 & 1.80$\pm$0.10 & -0.169$\pm$0.026 & -0.382$\pm$0.013 \\
445.8079924 & 91 / 0.730 & 802.4 & 906 & 0.92 & 214.0$\pm$4.9 & 2.56$\pm$0.02 & 65.9$\pm$4.2 & 2.03$\pm$0.10 & -0.173$\pm$0.028 & -0.151$\pm$0.014 \\
446.8413365 & 91 / 0.943 & 802.4 & 832 & 0.90 & 161.9$\pm$4.8 & 2.75$\pm$0.03 & 62.6$\pm$3.7 & -0.53$\pm$0.09 & 1.046$\pm$0.031 & -0.033$\pm$0.015 \\
447.8771338 & 92 / 0.156 & 802.4 & 808 & 0.86 & -62.2$\pm$4.7 & 2.83$\pm$0.03 & 30.7$\pm$3.6 & -0.81$\pm$0.08 & -0.022$\pm$0.033 & 0.060$\pm$0.015 \\
448.8642996 & 92 / 0.359 & 802.4 & 672 & 0.97 & -87.9$\pm$5.2 & 2.80$\pm$0.03 & -29.2$\pm$3.2 & -0.62$\pm$0.08 & -0.553$\pm$0.041 & 0.004$\pm$0.019 \\
449.8496155 & 92 / 0.562 & 802.4 & 887 & 0.87 & 41.7$\pm$4.5 & 2.41$\pm$0.03 & -74.2$\pm$4.0 & 1.80$\pm$0.09 & 0.196$\pm$0.029 & -0.088$\pm$0.014 \\
450.8177319 & 92 / 0.761 & 802.4 & 832 & 0.92 & 222.4$\pm$5.2 & 2.48$\pm$0.03 & 15.9$\pm$4.0 & 1.52$\pm$0.09 & 0.357$\pm$0.030 & 0.036$\pm$0.015 \\
451.8200490 & 92 / 0.967 & 802.4 & 789 & 0.88 & 139.2$\pm$4.8 & 2.75$\pm$0.03 & -11.3$\pm$3.4 & -0.59$\pm$0.08 & 0.373$\pm$0.033 & 0.064$\pm$0.015 \\
452.8143475 & 93 / 0.172 & 802.4 & 870 & 0.89 & -71.3$\pm$4.6 & 2.79$\pm$0.03 & 47.7$\pm$3.8 & -1.02$\pm$0.09 & 1.115$\pm$0.030 & 0.194$\pm$0.014 \\
472.7817257 & 97 / 0.280 & 802.4 & 835 & 0.83 & -71.1$\pm$4.4 & 2.86$\pm$0.03 & -12.4$\pm$3.5 & -1.36$\pm$0.08 & 1.367$\pm$0.031 & 0.285$\pm$0.015 \\
474.8452436 & 97 / 0.705 & 802.4 & 767 & 0.99 & 261.3$\pm$5.5 & 2.51$\pm$0.03 & 27.8$\pm$3.7 & 1.86$\pm$0.09 & 0.709$\pm$0.033 & -0.150$\pm$0.016 \\
475.7741501 & 97 / 0.896 & 802.4 & 858 & 0.85 & 199.5$\pm$4.6 & 2.72$\pm$0.03 & 30.9$\pm$3.7 & -0.24$\pm$0.09 & -0.258$\pm$0.030 & -0.108$\pm$0.014 \\
476.7556882 & 98 / 0.098 & 802.4 & 859 & 0.80 & -55.9$\pm$4.2 & 2.88$\pm$0.03 & 21.9$\pm$3.8 & -0.72$\pm$0.10 & 0.363$\pm$0.030 & -0.007$\pm$0.014 \\
477.7814701 & 98 / 0.309 & 802.4 & 893 & 0.76 & -67.0$\pm$4.0 & 2.83$\pm$0.03 & -30.5$\pm$3.3 & -1.29$\pm$0.08 & -0.001$\pm$0.028 & 0.005$\pm$0.013 \\
478.7906948 & 98 / 0.517 & 802.4 & 869 & 0.80 & -49.8$\pm$4.1 & 2.52$\pm$0.03 & -25.3$\pm$3.6 & 1.53$\pm$0.08 & 0.448$\pm$0.029 & -0.164$\pm$0.014 \\
479.7890571 & 98 / 0.722 & 802.4 & 954 & 0.87 & 254.7$\pm$4.7 & 2.51$\pm$0.02 & 8.6$\pm$3.9 & 2.05$\pm$0.09 & 0.029$\pm$0.026 & -0.086$\pm$0.013 \\
480.7740582 & 98 / 0.925 & 802.4 & 913 & 0.85 & 179.3$\pm$4.6 & 2.77$\pm$0.03 & 11.7$\pm$3.7 & -0.45$\pm$0.09 & 1.144$\pm$0.027 & -0.044$\pm$0.013 \\
481.7763948 & 99 / 0.131 & 802.4 & 868 & 0.78 & -87.0$\pm$4.2 & 2.83$\pm$0.03 & 29.0$\pm$3.5 & -0.73$\pm$0.08 & -0.197$\pm$0.028 & 0.099$\pm$0.014 \\
\hline
\end{tabular}
\end{table*}

\setcounter{table}{0}
\begin{table*}
\caption[]{continued} 
\hspace{-6mm}
\begin{tabular}{ccccccccccc}
\hline
BJD        & c / $\phi$ & t$_{\rm exp}$ & SNR & $\sigma_V$            & \Bl\  &  <$B$>  & RV    & $W_1$    & EWV \hei & EWV \pab \\ 
(2459000+) &            &   (s)         &     & ($10^{-4} I_c$)       & (G)   &  (kG)   & (\ms) & (\kmsss) & (\kms)   & (\kms)   \\  
\hline
501.8155848 & 103 / 0.254 & 802.4 & 910 & 0.81 & -78.2$\pm$4.5 & 2.89$\pm$0.03 & -15.4$\pm$3.8 & -1.71$\pm$0.09 & 0.168$\pm$0.027 & 0.060$\pm$0.013 \\
502.7430654 & 103 / 0.445 & 802.4 & 910 & 0.82 & -83.5$\pm$4.4 & 2.65$\pm$0.03 & -58.4$\pm$4.2 & 0.84$\pm$0.10 & 0.012$\pm$0.027 & -0.191$\pm$0.013 \\
503.7409031 & 103 / 0.650 & 802.4 & 560 & 1.18 & 165.1$\pm$6.1 & & & & -0.362$\pm$0.030 & -0.253$\pm$0.015 \\
503.7918633 & 103 / 0.661 & 802.4 & 838 & 0.87 & 181.3$\pm$4.5 & 2.45$\pm$0.03 & 7.4$\pm$2.5 & 1.49$\pm$0.09 & -0.299$\pm$0.050 & -0.217$\pm$0.023 \\
504.7752355 & 103 / 0.863 & 802.4 & 861 & 0.82 & 198.8$\pm$4.4 & 2.61$\pm$0.03 & 50.4$\pm$3.5 & 0.37$\pm$0.08 & 0.423$\pm$0.029 & -0.194$\pm$0.014 \\
505.7435779 & 104 / 0.062 & 802.4 & 787 & 0.90 & -7.5$\pm$4.9 & 2.79$\pm$0.03 & 35.2$\pm$3.7 & -0.45$\pm$0.09 & -0.109$\pm$0.031 & 0.104$\pm$0.015 \\
506.7361818 & 104 / 0.267 & 802.4 & 855 & 0.92 & -82.3$\pm$5.0 & 2.97$\pm$0.03 & -4.4$\pm$4.0 & -1.86$\pm$0.09 & 0.171$\pm$0.029 & 0.011$\pm$0.014 \\
507.7452686 & 104 / 0.474 & 802.4 & 866 & 0.83 & -71.9$\pm$4.5 & 2.56$\pm$0.03 & -39.2$\pm$3.9 & 1.45$\pm$0.09 & 0.251$\pm$0.027 & -0.122$\pm$0.013 \\
508.6910043 & 104 / 0.669 & 802.4 & 784 & 0.94 & 192.0$\pm$5.1 & 2.47$\pm$0.03 & -21.5$\pm$4.1 & 1.75$\pm$0.09 & 1.489$\pm$0.031 & 0.070$\pm$0.015 \\
509.7227087 & 104 / 0.881 & 802.4 & 744 & 0.95 & 199.8$\pm$5.2 & 2.70$\pm$0.03 & 31.6$\pm$3.8 & -0.04$\pm$0.09 & -0.080$\pm$0.033 & -0.102$\pm$0.016 \\
510.7570249 & 105 / 0.094 & 802.4 & 827 & 1.07 & -49.1$\pm$5.8 & 2.89$\pm$0.03 & 32.5$\pm$4.3 & -0.52$\pm$0.10 & 0.540$\pm$0.029 & -0.328$\pm$0.014 \\
511.6891490 & 105 / 0.286 & 802.4 & 849 & 0.81 & -63.0$\pm$4.4 & 2.87$\pm$0.03 & -50.1$\pm$3.6 & -1.68$\pm$0.08 & 0.586$\pm$0.029 & 0.204$\pm$0.014 \\
512.7317607 & 105 / 0.500 & 802.4 & 866 & 0.73 & -51.2$\pm$3.7 & 2.51$\pm$0.03 & -24.3$\pm$3.3 & 1.54$\pm$0.08 & 0.313$\pm$0.029 & -0.170$\pm$0.014 \\
513.7533589 & 105 / 0.711 & 802.4 & 938 & 0.82 & 214.2$\pm$4.5 & 2.51$\pm$0.03 & 10.3$\pm$3.8 & 2.01$\pm$0.09 & -0.400$\pm$0.026 & -0.346$\pm$0.013 \\
514.7142377 & 105 / 0.908 & 802.4 & 897 & 0.84 & 177.2$\pm$4.6 & 2.71$\pm$0.03 & 23.6$\pm$3.8 & -0.35$\pm$0.09 & -0.442$\pm$0.026 & -0.026$\pm$0.013 \\
515.7261028 & 106 / 0.116 & 802.4 & 892 & 0.82 & -68.7$\pm$4.3 & 2.83$\pm$0.03 & 38.3$\pm$3.9 & -0.84$\pm$0.09 & 0.370$\pm$0.028 & 0.051$\pm$0.013 \\
+ 531.7132915 & 109 / 0.406 & 802.4 & 847 & 0.89 & -59.2$\pm$4.8 & 2.61$\pm$0.03 & -110.9$\pm$4.0 & 0.60$\pm$0.10 & 2.967$\pm$1.030 & 0.365$\pm$0.015 \\
537.7172157 & 110 / 0.641 & 802.4 & 871 & 0.80 & 134.5$\pm$4.2 & 2.45$\pm$0.03 & 0.3$\pm$3.7 & 1.79$\pm$0.09 & -0.421$\pm$0.028 & -0.306$\pm$0.014 \\
538.7109023 & 110 / 0.846 & 802.4 & 823 & 0.84 & 196.6$\pm$4.6 & 2.58$\pm$0.03 & 35.0$\pm$3.4 & 0.65$\pm$0.08 & -0.677$\pm$0.031 & -0.159$\pm$0.015 \\
539.7110514 & 111 / 0.052 & 802.4 & 884 & 0.84 & 28.9$\pm$4.5 & 2.77$\pm$0.03 & 42.4$\pm$3.8 & -0.45$\pm$0.09 & -0.095$\pm$0.028 & 0.055$\pm$0.014 \\
\hline
711.0801811 & 146 / 0.313 & 802.4 & 835 & 0.85 & -87.7$\pm$4.4 & 2.85$\pm$0.03 & -3.1$\pm$3.7 & -1.06$\pm$0.09 & 0.740$\pm$0.031 & 0.258$\pm$0.015 \\
712.0952528 & 146 / 0.522 & 802.4 & 738 & 0.89 & 39.6$\pm$4.6 & 2.56$\pm$0.03 & -66.4$\pm$3.6 & 0.74$\pm$0.09 & 0.104$\pm$0.035 & -0.073$\pm$0.017 \\
713.1104824 & 146 / 0.731 & 802.4 & 834 & 0.83 & 82.3$\pm$4.2 & 2.40$\pm$0.03 & -4.0$\pm$3.8 & 2.42$\pm$0.09 & -0.967$\pm$0.030 & -0.200$\pm$0.015 \\
714.1174816 & 146 / 0.938 & 802.4 & 849 & 1.19 & 138.7$\pm$6.1 & 2.58$\pm$0.03 & 86.2$\pm$4.1 & 1.27$\pm$0.10 & 1.206$\pm$0.031 & -0.086$\pm$0.015 \\
715.1316342 & 147 / 0.146 & 802.4 & 816 & 0.85 & 6.2$\pm$4.3 & 2.89$\pm$0.03 & 20.1$\pm$3.5 & -1.27$\pm$0.08 & -0.011$\pm$0.032 & 0.135$\pm$0.015 \\
716.1256920 & 147 / 0.351 & 802.4 & 894 & 0.87 & -91.3$\pm$4.5 & 2.82$\pm$0.03 & -14.4$\pm$4.0 & -0.95$\pm$0.09 & 0.505$\pm$0.028 & 0.198$\pm$0.013 \\
717.1270682 & 147 / 0.557 & 802.4 & 877 & 0.92 & 74.2$\pm$4.7 & 2.55$\pm$0.03 & -54.9$\pm$4.0 & 1.33$\pm$0.10 & -0.089$\pm$0.028 & -0.113$\pm$0.014 \\
718.1146161 & 147 / 0.760 & 802.4 & 578 & 1.13 & 111.5$\pm$6.0 & 2.40$\pm$0.03 & 2.9$\pm$3.9 & 2.08$\pm$0.10 & -0.208$\pm$0.046 & -0.102$\pm$0.022 \\
733.0771500 & 150 / 0.839 & 802.4 & 913 & 0.90 & 160.8$\pm$4.8 & 2.44$\pm$0.03 & 50.7$\pm$4.2 & 2.38$\pm$0.10 & 0.068$\pm$0.028 & -0.058$\pm$0.013 \\
734.0673419 & 151 / 0.043 & 802.4 & 888 & 0.97 & 45.9$\pm$5.2 & 2.84$\pm$0.03 & 29.6$\pm$4.0 & -0.84$\pm$0.09 & 0.706$\pm$0.028 & -0.068$\pm$0.013 \\
735.1196227 & 151 / 0.259 & 802.4 & 844 & 1.08 & -80.0$\pm$5.8 & 2.89$\pm$0.03 & 11.5$\pm$3.8 & -1.19$\pm$0.09 & 1.092$\pm$0.030 & -0.155$\pm$0.015 \\
736.1041074 & 151 / 0.462 & 802.4 & 798 & 0.87 & -41.5$\pm$4.6 & 2.71$\pm$0.03 & -64.3$\pm$3.6 & -0.07$\pm$0.09 & -0.510$\pm$0.032 & -0.096$\pm$0.016 \\
737.1112682 & 151 / 0.669 & 802.4 & 841 & 0.84 & 92.8$\pm$4.3 & 2.44$\pm$0.03 & -22.0$\pm$3.6 & 1.94$\pm$0.09 & 0.072$\pm$0.031 & -0.304$\pm$0.015 \\
738.1066549 & 151 / 0.874 & 802.4 & 857 & 0.82 & 168.6$\pm$4.3 & & & & -0.492$\pm$0.030 & -0.041$\pm$0.014 \\
739.0936458 & 152 / 0.077 & 802.4 & 839 & 0.80 & 13.6$\pm$4.0 & 2.84$\pm$0.03 & 15.5$\pm$3.4 & -1.18$\pm$0.08 & 0.199$\pm$0.031 & -0.152$\pm$0.015 \\
740.0401410 & 152 / 0.272 & 802.4 & 888 & 0.97 & -77.3$\pm$4.9 & 2.92$\pm$0.03 & -9.8$\pm$3.8 & -1.18$\pm$0.09 & 0.715$\pm$0.029 & 0.180$\pm$0.014 \\
741.0854219 & 152 / 0.487 & 802.4 & 867 & 0.83 & -8.4$\pm$4.1 & 2.68$\pm$0.03 & -63.0$\pm$3.6 & 0.15$\pm$0.09 & -1.115$\pm$0.029 & -0.048$\pm$0.015 \\
742.0225617 & 152 / 0.680 & 802.4 & 895 & 0.78 & 80.0$\pm$3.9 & 2.44$\pm$0.03 & -4.0$\pm$3.7 & 2.21$\pm$0.09 & 0.793$\pm$0.029 & -0.284$\pm$0.014 \\
743.0773300 & 152 / 0.897 & 802.4 & 617 & 1.07 & 162.4$\pm$5.5 & 2.52$\pm$0.03 & 53.7$\pm$3.6 & 1.50$\pm$0.09 & -0.294$\pm$0.045 & 0.000$\pm$0.021 \\
744.0024264 & 153 / 0.087 & 802.4 & 717 & 0.93 & 23.1$\pm$5.0 & 2.85$\pm$0.03 & 22.4$\pm$3.4 & -1.12$\pm$0.08 & 1.848$\pm$0.038 & -0.006$\pm$0.018 \\
\hline
\end{tabular}
\end{table*}

\section{ZDI fit to Stokes $V$ LSD profiles}
\label{sec:appB}

\begin{figure*}
\centerline{\includegraphics[scale=0.6,angle=-90]{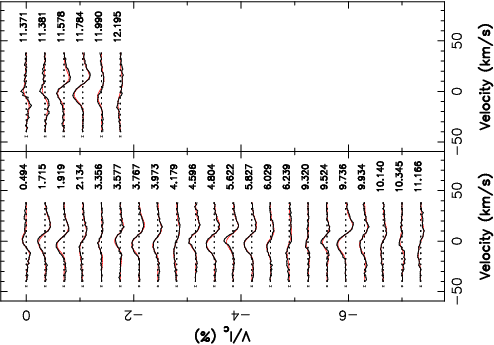}\hspace{2mm}
            \includegraphics[scale=0.6,angle=-90]{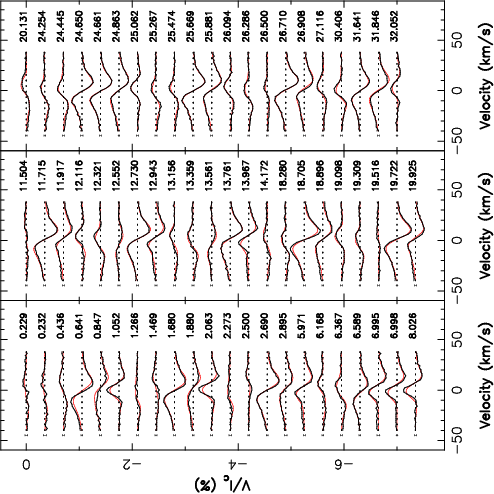}\vspace{2mm}}
\centerline{\includegraphics[scale=0.6,angle=-90]{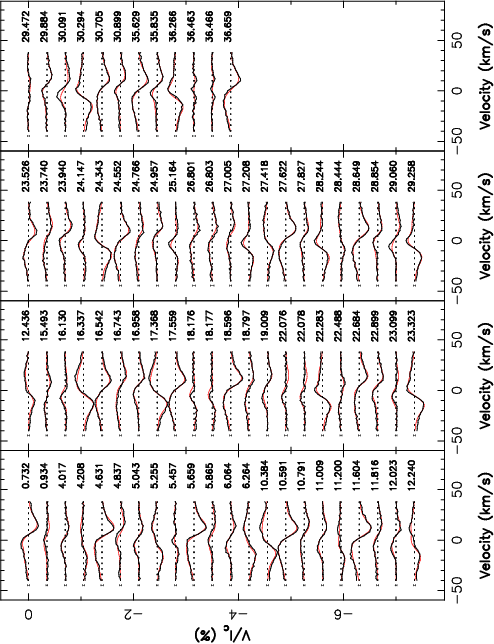}\hspace{2mm}
            \includegraphics[scale=0.6,angle=-90]{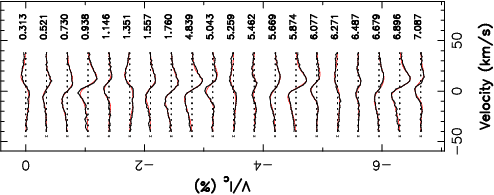}} 
\caption[]{Observed (thick black line) and modelled (thin red line) LSD Stokes $V$ profiles of the photospheric lines 
of AU~Mic, for seasons 2019 Sep-Nov (top left), 2020 Apr-Nov (bottom left), 2021 Jun-Nov (top right) and 2022 May-Jun (bottom right).  The ZDI modeling of these profiles 
is described in Sec.~\ref{sec:zd1}.  Rotation cycles (counting from $-53$, $-4$, 79 and 146 for 2019, 2020, 2021 and 2022 respectively, see 
Table~\ref{tab:log}) and $\pm$1$\sigma$ error bars are indicated to the right and left of each profile.  } 
\label{fig:fit}
\end{figure*}

We show in Fig.~\ref{fig:fit} the collected Stokes $V$ LSD profiles along with the ZDI fit described in the Stokes $V$ analysis of Sec.~\ref{sec:zd1}.  

\section{Simulated Stokes $Q$ and $U$ LSD profiles}
\label{sec:appC}

\begin{figure*}
\includegraphics[scale=0.6,angle=-90]{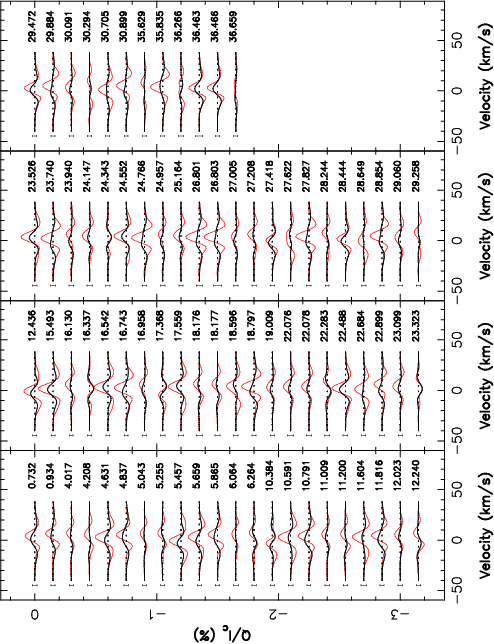}\vspace{2mm}
\includegraphics[scale=0.6,angle=-90]{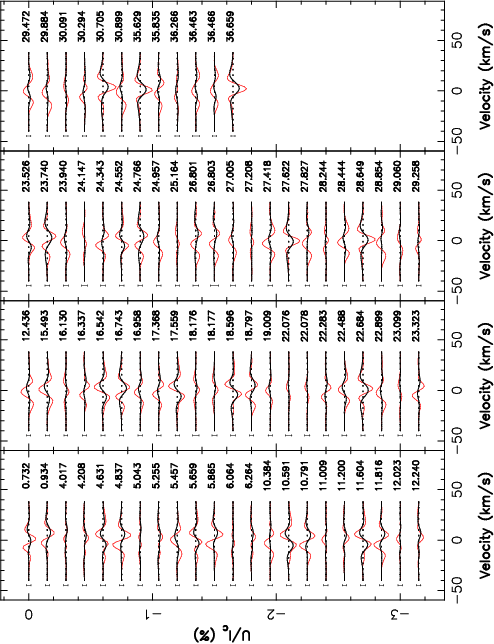}
\caption[]{Calculated LSD Stokes $Q$ (top panel) and $U$ (bottom panel) signatures associated with the reconstructed magnetic topologies of AU~Mic for season 2020 Apr-Nov, 
derived with the Stokes $V$ (black line) and Stokes $I$ \& $V$ (red line) analyses outlined in Secs.~\ref{sec:zd1} and \ref{sec:zd2} and shown in Figs.~\ref{fig:map} 
(second top row) and \ref{fig:mapI}.  Error bars to the left of each profile, depicting the achievable photon noise with exposure times listed in Table~\ref{tab:log}, 
demonstrate that such signatures are detectable and can be used to distinguish between both magnetic configurations, whose Stokes $V$ signatures are similar.  } 
\label{fig:lqu}
\end{figure*}

We present in Fig.~\ref{fig:lqu} the simulated $Q$ and $U$ LSD profiles corresponding to our ZDI images of AU~Mic for season 2020 Apr-Nov, shown in 
Figs.~\ref{fig:map} and Figs.~\ref{fig:mapI} and derived with the Stokes $V$ and Stokes $I$ \& $V$ analyses detailed in Secs.~\ref{sec:zd1} and \ref{sec:zd3}.  
It shows in particular that Stokes $Q$ and $U$ observations of AU~Mic can be used to distinguish between both magnetic configurations, whose Stokes $V$ signatures are similar. 

\section{The \hei\ and \pab\ line profiles of AU~Mic}
\label{sec:appD}

\begin{figure*}
\centerline{\includegraphics[scale=0.4]{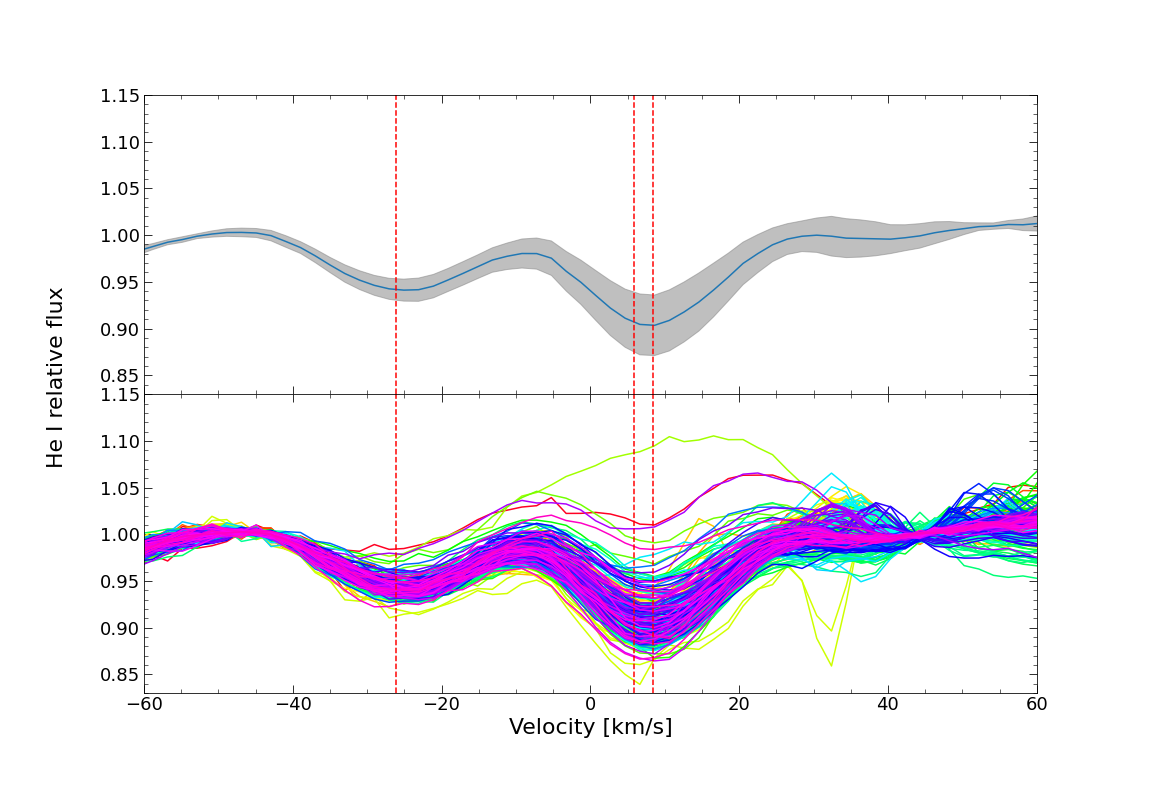}}\vspace{2mm}
\centerline{\includegraphics[scale=0.4]{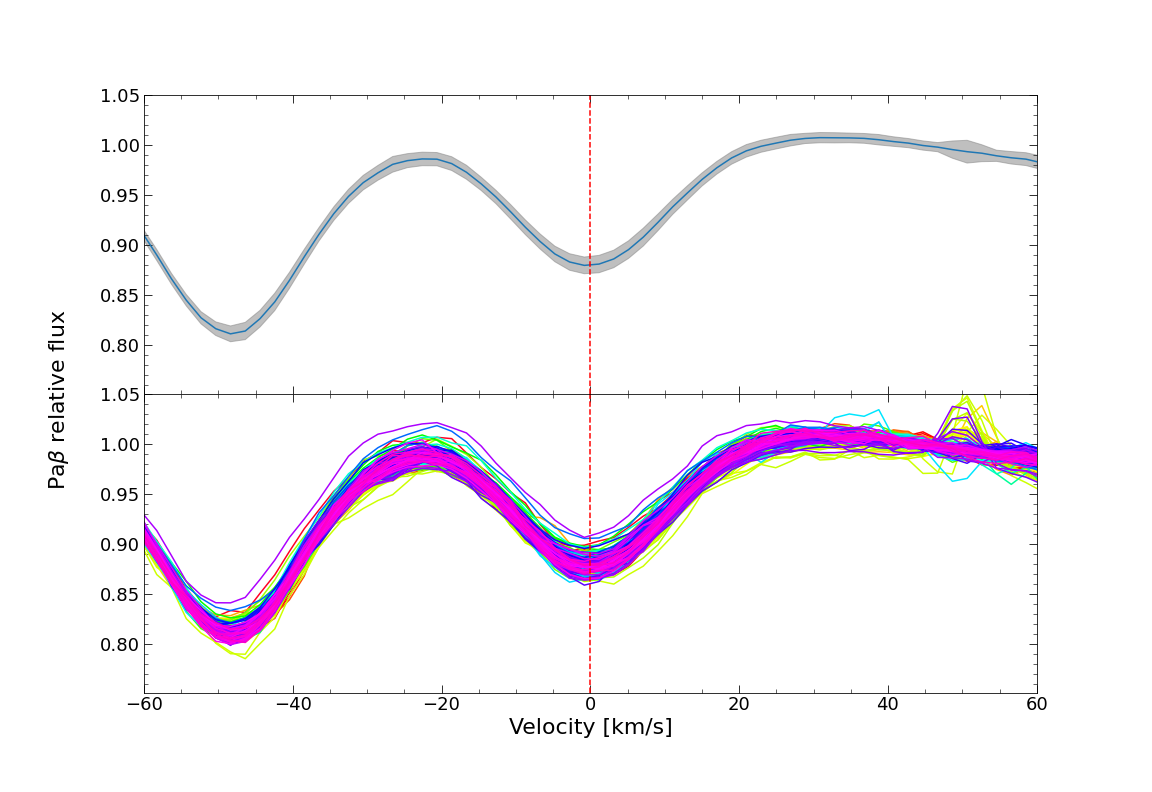}} 
\caption[]{Spectra of AU~Mic in the region of the \hei\ triplet (top panel) and \pab\ line (bottom panel), in the stellar rest frame.  In each panel, the bottom 
curves show the superposition of individual spectra, whereas the top curve shows the median profile and dispersion (in grey shade).  The red vertical dotted 
lines depict the location of the \hei\ triplet end of the \pab\ line.  } 
\label{fig:hpl}
\end{figure*}

We show in Fig.~\ref{fig:hpl} the spectra of AU~Mic in the region of the \hei\ triplet and \pab\ line, in the stellar rest frame.  

\section{Stability and predicted TTVs of the planetary system}
\label{sec:appE}

\begin{figure*}
\centerline{\includegraphics[scale=0.5,bb=40 10 800 300]{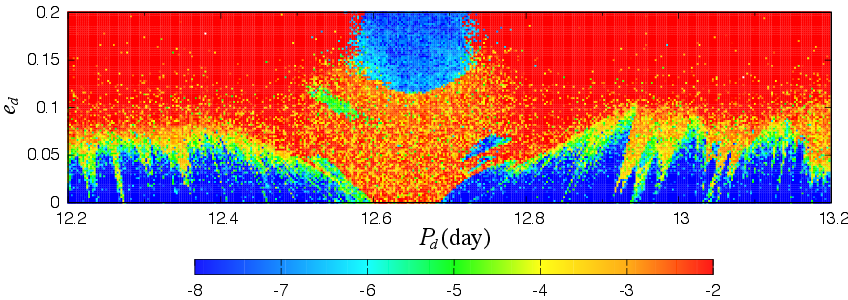}}\vspace{2mm}
\centerline{\includegraphics[scale=0.5,bb=40 10 800 300]{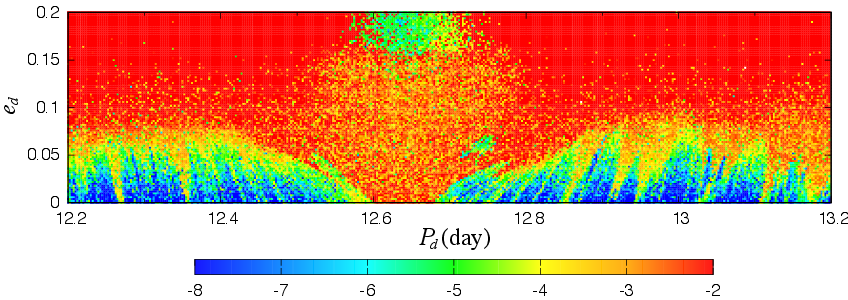}} 
\caption[]{\emr Same as Fig.~\ref{fig:stab} for candidate planet d in the case of the b+c+d (top panel) and b+c+d+e (bottom panel) systems.  In both cases, the system is stable for the 12.74-d period 
and low eccentricity orbit of candidate planet d suggested by \citet[][]{Wittrock23}, though slightly less for the 4-planet than for the 3-planet system.  The 3:2 resonances between d-b and c-d 
are strongest around $P_d\simeq12.66$~d, with b+c+d being only stable for $e_d=$~0.15--0.20 and b+c+d+e becoming unstable on a Gyr timescale. } 
\label{fig:stab2}
\end{figure*}

\begin{figure*}
\centerline{\includegraphics[scale=1.2,bb=40 10 300 220]{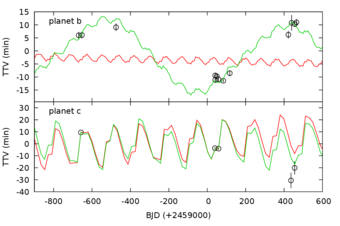}}
\caption[]{\ems Predicted TTVs for transiting planets b (top) and c (bottom), in the b+c+e (red line) and b+c+d+e (green line, with the mass of candidate planet d set to 0.68~\me) cases, compared with 
the transit timings derived by TESS, Spitzer and CHEOPS \citep[black open circles,][]{Szabo22}. }  
\label{fig:ttv}
\end{figure*}

We show in Fig.~\ref{fig:stab2} the stability of the planetary system in cases b+c+d and b+c+d+e, for a range of periods and eccentricities of candidate planet d.  
It confirms that b+c+d is stable for the 12.74-d period and low eccentricity orbit of planet d suggested by \citet[][]{Wittrock23}, whereas b+c+d+e, though less stable, should 
still be able to survive on long timescales.  {\ems Besides, we show in Fig.~\ref{fig:ttv} the predicted TTVs for transiting planets b and c in the b+c+d+e and b+d+e cases, 
compared with the TESS, Spitzer and CHEOPS timing measurements.  } 

\bsp	
\label{lastpage}
\end{document}